\documentclass[aps,pra,10pt,showpacs,tightenlines,notitlepage,nofootinbib,longbibliography,superscriptaddress,twocolumn,floatfix]{revtex4-1}

\usepackage{graphicx,amssymb,amstext}
\usepackage{amsmath,bbold}
\usepackage{hyperref}
\usepackage{epsfig}
\usepackage{verbatim}
\usepackage [utf8]{inputenc}

\begin{document}


\title{Fundamental noisy multiparameter quantum bounds}

\author{Shibdas Roy}
\email{Email: roy\_shibdas@yahoo.co.in}
\affiliation{Department of Physics, University of Warwick, Coventry CV4 7AL, United Kingdom.}



\date{\today}

\begin{abstract}
Quantum multiparameter estimation involves estimating multiple parameters simultaneously and can be more precise than estimating them individually. Our interest here is to determine fundamental quantum limits to the achievable multiparameter estimation precision in the presence of noise. We present a lower bound to the estimation error covariance for a noisy initial probe state evolving via a noiseless quantum channel. We then present a lower bound to the estimation error covariance in the most general form for a noisy initial probe state evolving via a noisy quantum channel. We show conditions and accordingly measurements to attain these estimation precision limits for noisy systems. We see that the Heisenberg precision scaling of $1/N$ can be achieved with a probe comprising $N$ particles even in the presence of noise. In fact, some noise in the initial probe state or the quantum channel can serve as a feature rather than a bug, since the estimation precision scaling achievable in the presence of noise in the initial state or the channel in some situations is impossible in the absence of noise in the initial state or the channel. However, a lot of noise harms the quantum advantage achievable with $N$ parallel resources, and allows for a best precision scaling of $1/\sqrt{N}$. Moreover, the Heisenberg precision limit can be beaten with noise in the channel, and we present a super-Heisenberg precision limit with scaling of $1/N^2$ for optimal amount of noise in the channel, characterized by one-particle evolution operators. Further, using $\gamma$-particle evolution operators for the noisy channel, where $\gamma>1$, the best precision scaling attainable is $1/N^{2\gamma}$, which is otherwise known to be only possible using $2\gamma$-particle evolution operators for a noiseless channel.
\end{abstract}


\maketitle

\section{Introduction}\label{sec:intro}
Studying quantum multiparameter estimation has recently been of significant interest \cite{GPANKK,VDGJKKDBW,CDBW,HBDW,BD,SBD,PCSHDWBSS,GBD,NLKA,LY,RGMSSGB,CSVPSS,PKD,GPS}. While quantum resources allow for surpassing measurement limits set by classical physics \cite{GLM1,GLM2,FOSSDBD}, it is important to consider fundamental measurement limits set by quantum mechanics. Although quantum estimation of a single parameter captures many scenarios \cite{GLM3}, the practically more relevant problem of estimating multiple parameters simultaneously has started drawing more attention, mainly because unlike in quantum single-parameter estimation case, quantum measurements required to attain multiparameter bounds do not necessarily commute \cite{CWH,MGAP,BD,LC}.

Multiparameter estimation using a pure (i.e.~noiseless) probe state under unitary (i.e.~noiseless) evolution has been studied, e.g.~in Ref.~\cite{BD}. This work, like most in the literature, used symmetric logarithmic derivatives (SLDs), as used by Helstrom \cite{CWH}, to define the quantum Fisher information matrix (QFIM) \cite{YZF}. Then, the estimation error covariance (that is the multiparameter counterpart to the mean-squared estimation error in single parameter estimation) is lower-bounded by the inverse of the QFIM and the bound is called a quantum Cram\'{e}r-Rao bound (QCRB) \cite{WM}. Such a QFIM for a probe with multiple particles under unitary evolution via one particle Hamiltonians \cite{BD,WM,NC} was shown to depend only on the one- and two-particle reduced density operators \cite{NC} of the probe state. However, when the initial probe state is mixed (i.e.~noisy) but the quantum channel is unitary, even for single parameter estimation, only an upper bound to such an SLD-based QFIM (and therefore, a lower bound to the corresponding QCRB) can be explicitly established in general \cite{BCM,EFD}. Although noiseless quantum parameter estimation has been studied extensively and is well understood, it is important to study and better understand fundamental quantum estimation limits in more practical noisy situations \cite{VDGJKKDBW,SBD1,DRFG,KDD,EFD,YZF,DDKG,DDM,DDCS,NBCA,YGXWS,RCSGRMGRB,JCH,SSD,SSKD,FAAMLOGO,HMM,ZZPJ, CASZZHTXKLG,DDJK,BAR,BdC,AMR,YF,MT,ARTG,HSKDDH}.

In this article, we present a multiparameter QCRB for a noisy initial state evolving unitarily, based on anti-symmetric logarithmic derivatives (ALDs) \cite{TWC,FN}, that lend a convenient way to study noisy quantum metrology. Moreover, we use a similar ALD-approach to present an upper bound to the QFIM (like in Refs.~\cite{EFD,YZF}) for the case of impure initial states under arbitrary evolution. That is, we consider a noisy quantum channel and a mixed intial probe state and define a quantum lower bound for the estimation error covariance in this general-most case. Such bounds for an $N$-particle probe state depend on the one- and two-particle reduced density operators only, similar to the case of pure state evolving unitarily in Ref.~\cite{BD}. We also provide conditions and accordingly measurements that allow to attain these bounds.

Our results here are fundamentally profound because of several reasons. Firstly, the tight bounds presented here are explicitly computable (e.g.~in terms of the Kraus operators of a noisy channel), without any knowledge of the eigenvalues and the eigenvectors of the evolved probe state \cite{MGAP,SBD1} and are not known to be possible for these most general noisy cases using the conventional SLD-approach. A similar bound with SLDs was obtained for single parameter estimation earlier \cite{FI,EFD,DDM}, but it was not considered tight being an upper bound to the SLD QFI, and accordingly a tighter bound, linear in the number $N$ of resources was considered. Secondly, our bounds are such that the quantum enhancement to the estimation precision is provided by the two-particle reduced density matrices of the probe state and the attainability of the quantum enhancement is determined solely by the one-particle reduced density matrices of the probe state, when the channel is characterized by one-particle evolution operators, even in the presence of noise, similar to the noiseless case from Ref.~\cite{BD}. Thirdly, the results here suggest that the Heisenberg scaling of $1/N$ in the estimation precision, with $N$ number of resources, is achievable even in the presence of noise. Moreover, some noise in the quantum channel or the initial probe state can act as a feature rather than a bug, since we see that there are situations when it is not possible to attain the Heisenberg limit in the absence of noise in the channel or the initial state, but it is possible in the presence of noise in the channel or the initial state. However, too much noise in the initial state or the channel harms the quantum advantage achievable with $N$ parallel resources.

Furthermore, we show that the Heisenberg precision limit can be beaten with noise in the quantum channel. The best achievable precision limit for non-unitary channel is then determined by two-particle reduced density operators of the evolved probe state being maximally entangled and one-particle reduced density operators being maximally mixed, and corresponds to a precision scaling of $1/N^2$, attained with one-particle evolution operators for the channel. Further, using $\gamma$-particle (instead of one-particle) evolution operators for a noisy channel, where $\gamma>1$, the best precision scaling achievable is $1/N^{2\gamma}$, that is otherwise known as achievable with $2\gamma$-particle evolution operators of a noiseless channel.

Before we proceed, it is important to explicitly point out why the non-standard ALD-approach instead of the standard SLD-approach is adopted in this paper. The way we choose the ALDs in this article, it turns out that the ALD-based QFIM is an upper bound to the standard SLD-based QFIM for the noiseless channel case. As already pointed out, such an upper bound to the SLD QFIM for single parameter estimation has been obtained earlier, but it was not considered a tight bound, since beating the SLD QFIM would mean that the Heisenberg limit can be beaten. However, we show here that such an upper bound to the SLD QFIM can be tight too, but the use of ALD-approach indicates that the Heisenberg limit is not beaten for the noiseless channel case. Thus, the QFIM obtained here for the noiseless channel case cannot be obtained using the SLD-approach and the corresponding equivalent bound obtained using SLDs would seem to beat the Heisenberg limit. Moreover, for the multiparameter noisy channel case considered here, the upper bound to the ALD QFIM we obtained cannot be obtained using the SLD-approach, since it would be an upper bound to the aforementioned upper bound to the SLD QFIM. We show that such an upper bound to the ALD QFIM can also be tight, implying that the Heisenberg limit can be beaten. It is unlikely that there exists some other logarithmic derivative for which the QFIM would be the upper bound to the ALD QFIM, suggesting that the Heisenberg limit is still not beaten.

\vspace*{-3mm}
\section{Multiparameter Quantum Cram\'{e}r-Rao Bound}\label{sec:qcrb}
An experiment for estimation of some unknown parameters corresponding to a quantum process involves three stages. First, a probe state is prepared in an initial state, comprising $N$ number of resources, and evolves under the action of the quantum process. The second stage involves choosing a suitable measurement, applied to the evolved probe state. The final step involves associating, through an estimator, each experimental result with an estimation of the parameters \cite{EFD}. The Heisenberg limit to the estimation precision is then the precision scaling of $1/N$.

Consider that a probe state $\hat{\rho}$ acquires $q$ number of parameters $\boldsymbol{\theta} = \left[\begin{array}{cccc} \theta_1 & \theta_2 & \hdots & \theta_q \end{array}\right]^T$ via a unitary transformation $\hat{U}(\boldsymbol{\theta})$, and we seek the best quantum strategy to estimate the parameters from the evolved probe state, $\hat{\rho}(\boldsymbol{\theta})=\hat{U}(\boldsymbol{\theta})\hat{\rho}\hat{U}^\dagger(\boldsymbol{\theta})$. Let a measurement performed on the evolved state $\hat{\rho}(\boldsymbol{\theta})$ be given by some positive operator valued measure (POVM) $\{\hat{P}_m\}$. The conditional probability to obtain the outcome $m$ given the parameters have the value $\boldsymbol{\theta}$ is $p(m|\boldsymbol{\theta}) = \mathrm{Tr}\left(\hat{P}_m \hat{\rho}(\boldsymbol{\theta})\right)$. The estimates $\boldsymbol{\tilde{\theta}}(m) = \left[\begin{array}{cccc} \tilde{\theta}_1(m) & \tilde{\theta}_2(m) & \hdots & \tilde{\theta}_q(m) \end{array}\right]^T$ are unbiased if

\small\vspace*{-2mm}
\begin{equation}\label{eq:unbiased_est}
\sum_m p(m|\boldsymbol{\theta})\tilde{\theta}_j(m) = \theta_j \qquad \forall j.
\end{equation}\normalsize
Then, the estimation error covariance is
\small
\begin{equation}\label{eq:est_err_cov}
V\left[\boldsymbol{\tilde{\theta}}(m)\right] = \sum_m p(m|\boldsymbol{\theta}) \left(\boldsymbol{\tilde{\theta}}(m)-\boldsymbol{\theta}\right)\left(\boldsymbol{\tilde{\theta}}(m)-\boldsymbol{\theta}\right)^T.
\end{equation}\normalsize
Then, for unbiased estimators, the above covariance satisfies the Cram\'{e}r-Rao inequality:

\small\vspace*{-2mm}
\begin{equation}
\nu V\left[\boldsymbol{\tilde{\theta}}(m)\right] \geq \left[J_C(\boldsymbol{\theta})\right]^{-1},
\end{equation}\normalsize
where where $\nu$ is the number of times the overall experiment is repeated and $J_C(\boldsymbol{\theta})$ is the classical Fisher Information Matrix (FIM), given by

\small\vspace*{-2mm}
\begin{equation}\label{eq:fim1}
J_C^{jk} = \sum_m \frac{1}{p(m|\boldsymbol{\theta})}\frac{\partial}{\partial\theta_j}p(m|\boldsymbol{\theta})\frac{\partial}{\partial\theta_k}p(m|\boldsymbol{\theta}).
\end{equation}\normalsize
Further, the maximisation of the FIM over all possible POVMs yields the quantum Fisher Information Matrix (QFIM), $J_Q(\boldsymbol{\theta})$, which is determined from \cite{TWC,FN}:

\small\vspace*{-2mm}
\begin{equation}\label{eq:ald_diffeqn1}
\frac{1}{2}\left(\hat{L}_k\hat{\rho}(\boldsymbol{\theta})+\hat{\rho}(\boldsymbol{\theta})\hat{L}_k^\dagger\right)=\frac{\partial}{\partial\theta_k}\hat{\rho}(\boldsymbol{\theta}).
\end{equation}\normalsize
where $\hat{L}_k$ is an operator. The QFIM $J_Q(\boldsymbol{\theta})$ is then \cite{TWC}:
\small
\begin{equation}\label{eq:qfim1}
J_Q^{jk} = \frac{1}{2}{\rm Tr}\left[\left(\hat{L}_j^\dagger \hat{L}_k+\hat{L}_k^\dagger \hat{L}_j\right)\hat{\rho}(\boldsymbol{\theta})\right],
\end{equation}\normalsize
Then, we have
\small
\begin{equation}\label{eq:ald_qcrb}
\nu V\left[\boldsymbol{\tilde{\theta}}(m)\right] \geq \left[J_C(\boldsymbol{\theta})\right]^{-1} \geq \left[J_Q(\boldsymbol{\theta})\right]^{-1},
\end{equation}\normalsize
where, $\hat{L}_k$ was taken to be Hermitian by Helstrom \cite{CWH}, in which case it is called the symmetric logarithmic derivative (SLD). In general, $\hat{L}_k$ need not be Hermitian. We assume that $\hat{L}_k$ is anti-Hermitian, such that $\hat{L}_k^\dagger=-\hat{L}_k$ \cite{TWC,FN}, in which case it is called the anti-symmetric logarithmic derivative (ALD). Thus, (\ref{eq:qfim1}) defines a certain family of logarithmic derivatives, satisfying ${\rm Tr}\left[\hat{\rho}(\boldsymbol{\theta})\hat{L}_k\right]=0$, such that a Hermitian $\hat{L}_k$ is an SLD and an anti-Hermitian $\hat{L}_k$ is an ALD \cite{FN}. Although Ref.~\cite{TWC} considered a different (Bayesian waveform-) estimation problem, (\ref{eq:ald_qcrb}) can be similarly proven here. See Appendix \ref{sec:app1}.

Although the classical Cram\'{e}r-Rao bound (i.e.~the first inequality in (\ref{eq:ald_qcrb})) can always be saturated, e.g.~by a maximum likelihood estimator \cite{SLB}, the QCRB (i.e.~the second inequality in (\ref{eq:ald_qcrb})) for SLDs are not known to be attainable in general. We claim that an ALD-based QCRB of the form (\ref{eq:ald_qcrb}) can be saturated (i.e.~attained), when the QFIM is not rank deficient and the expectation of the commutator of every pair of the ALDs vanishes, similar to the case of SLD-based QCRB \cite{KM,RJD,BD}:

\small\vspace*{-2mm}
\begin{equation}\label{eq:qcrb_saturate}
{\rm Tr}\left[\left(\hat{L}_j^\dagger \hat{L}_k-\hat{L}_k^\dagger \hat{L}_j\right)\hat{\rho}(\boldsymbol{\theta})\right]={\rm Tr}\left(\left[\hat{L}_j,\hat{L}_k\right]\hat{\rho}(\boldsymbol{\theta})\right)= 0.
\end{equation}\normalsize
See Appendix \ref{sec:app2}. The above condition is trivially true for single parameter estimation. Then, the set of POVMs of cardinality $q+2$, comprising the following $q+1$ elements,

\small\vspace*{-2mm}
\begin{equation}\label{eq:qcrb_measure1}
\begin{split}
\hat{P}_0&=\hat{\rho}(\boldsymbol{\theta})=\hat{U}(\boldsymbol{\theta})\hat{\rho}\hat{U}^\dagger(\boldsymbol{\theta}),\\
\hat{P}_m&=\frac{\partial\hat{\rho}(\boldsymbol{\theta})}{\partial\theta_m}=\frac{\partial\hat{U}(\boldsymbol{\theta})}{\partial\theta_m}\hat{\rho}\hat{U}^\dagger(\boldsymbol{\theta})+\hat{U}(\boldsymbol{\theta})\hat{\rho}\frac{\partial\hat{U}^\dagger(\boldsymbol{\theta})}{\partial\theta_m} \, \forall m=1,\ldots,q,
\end{split}
\end{equation}\normalsize
along with one normalising element, saturates the QCRB (see Appendix \ref{sec:app7}). For pure states $|\psi\rangle$, the $q+1$ projectors

\small\vspace*{-2mm}
\begin{equation}
\begin{split}
\hat{P}_0&=\hat{\rho}(\boldsymbol{\theta})=\hat{U}(\boldsymbol{\theta})|\psi\rangle\langle\psi|\hat{U}^\dagger(\boldsymbol{\theta}),\\
\hat{P}_m&=\frac{\partial\hat{U}(\boldsymbol{\theta})}{\partial\theta_m}|\psi\rangle\langle\psi|\frac{\partial\hat{U}^\dagger(\boldsymbol{\theta})}{\partial\theta_m} \quad \forall m=1,\ldots,q,
\end{split}
\end{equation}\normalsize
along with one normalising element, saturates the QCRB. This follows from Refs.~\cite{HBDW,BD} (see Appendix \ref{sec:app6}).

\vspace*{-3mm}
\section{The QFIM for One-Particle Hamiltonians}\label{sec:qfim}
Let us now consider that the unitary evolution is $\hat{U}(\boldsymbol{\theta}) = e^{-i\hat{H}(\boldsymbol{\theta})}$ and that the probe state $\hat{\rho}$ comprises $N$ particles evolving under the one-particle Hamiltonian $\hat{h}^{[n]} = \sum_{k=1}^q \theta_k\hat{h}^{[n]}_k$ for $n=1,\hdots,N$, such that \cite{BD}:

\small\vspace*{-2mm}
\begin{equation}
\hat{H}(\boldsymbol{\theta}) = \sum_{n=1}^N\hat{h}^{[n]} = \sum_{k=1}^q\theta_k\sum_{n=1}^N\hat{h}_k^{[n]} \equiv \sum_{k=1}^q\theta_k\hat{H}_k.
\end{equation}\normalsize
The generators $\hat{H}_k$ are assumed to not depend on $\boldsymbol{\theta}$ and do not generally commute with each other. Then, as employed by Ref.~\cite{BD}, we have \cite{RMW}:

\small\vspace*{-2mm}
\begin{equation}\label{eq:noncommuting_generators}
\frac{\partial\hat{U}(\boldsymbol{\theta})}{\partial\theta_k}=-i\int_0^1 d\alpha e^{-i(1-\alpha)\hat{H}(\boldsymbol{\theta})}\frac{\partial\hat{H}(\boldsymbol{\theta})}{\partial\theta_k}e^{-i\alpha\hat{H}(\boldsymbol{\theta})}.
\end{equation}\normalsize
Then, we have

\small\vspace*{-2mm}
\begin{equation}\label{eq:unitary_delrho}
\frac{\partial}{\partial\theta_k}\hat{\rho}(\boldsymbol{\theta})=\frac{\partial}{\partial\theta_k}\left(\hat{U}(\boldsymbol{\theta})\hat{\rho}\hat{U}^\dagger(\boldsymbol{\theta})\right)=-i\left[\hat{M}_k(\boldsymbol{\theta}),\hat{\rho}(\boldsymbol{\theta})\right],
\end{equation}
\normalsize
where

\small\vspace*{-2mm}
\begin{equation}\label{eq:unitary_mk}
\hat{M}_k(\boldsymbol{\theta}) = i\frac{\partial\hat{U}(\boldsymbol{\theta})}{\partial\theta_k}\hat{U}^\dagger(\boldsymbol{\theta}) = \hat{U}(\boldsymbol{\theta})\hat{A}_k(\boldsymbol{\theta})\hat{U}^\dagger(\boldsymbol{\theta}),
\end{equation}\normalsize
with $\hat{A}_k(\boldsymbol{\theta}) = \int_0^1 d\alpha e^{i\alpha\hat{H}(\boldsymbol{\theta})}\hat{H}_ke^{-i\alpha\hat{H}(\boldsymbol{\theta})}$. We choose the operator $\hat{L}_k$ as the anti-Hermitian, $\hat{L}_k = -2i\Delta \hat{M}_k$, where

\small\vspace*{-2mm}
\begin{equation}\label{eq:unitary_delmk}
\Delta \hat{M}_k \equiv \hat{M}_k(\boldsymbol{\theta}) - {\rm Tr}\left(\hat{M}_k(\boldsymbol{\theta})\hat{\rho}(\boldsymbol{\theta})\right).
\end{equation}\normalsize
The QFIM from (\ref{eq:qfim1}) then takes the form:

\small\vspace*{-2mm}
\begin{equation}\label{eq:qfim3}
J_Q^{jk} = 2{\rm Tr}\left[\left(\Delta\hat{A}_j\Delta\hat{A}_k+\Delta\hat{A}_k\Delta\hat{A}_j\right)\hat{\rho}\right],
\end{equation}\normalsize
where

\small\vspace*{-2mm}
\begin{equation}
\begin{split}
\Delta\hat{A}_k &= \hat{A}_k(\boldsymbol{\theta}) - {\rm Tr}\left(\hat{A}_k(\boldsymbol{\theta})\hat{\rho}\right)\\
&= \sum_n \left(\hat{b}_k^{[n]} - {\rm Tr}\left(\hat{b}_k^{[n]}\hat{\rho}^{[n]}\right) \right) \equiv \sum_n \hat{c}_k^{[n]},
\end{split}
\end{equation}\normalsize
with $\hat{b}_k^{[n]} = \int_0^1 d\alpha e^{i\alpha\hat{h}^{[n]}}\hat{h}_k^{[n]}e^{-i\alpha\hat{h}^{[n]}}$. Thus, (\ref{eq:qfim3}) becomes:

\small\vspace*{-2mm}
\begin{equation*}
\begin{split}
J_Q^{jk}&=2\sum_n{\rm Tr}\left[ \left(\hat{c}_j^{[n]}\hat{c}_k^{[n]}+\hat{c}_k^{[n]}\hat{c}_j^{[n]}\right)\hat{\rho}^{[n]} \right]\\
&+2\sum_{n\neq m}{\rm Tr}\left[ \left(\hat{c}_j^{[n]}\otimes\hat{c}_k^{[m]}+\hat{c}_k^{[n]}\otimes\hat{c}_j^{[m]}\right)\hat{\rho}^{[n,m]} \right]\\
&=4\sum_n{\rm Re}\left[{\rm Tr}\left[\hat{\rho}^{[n]}\hat{b}_j^{[n]}\hat{b}_k^{[n]}\right]-{\rm Tr}\left[\hat{\rho}^{[n]}\hat{b}_j^{[n]}\right]{\rm Tr}\left[\hat{\rho}^{[n]}\hat{b}_k^{[n]}\right]\right]\\
&+4\sum_{n\neq m}{\rm Re}\left[{\rm Tr}\left[\hat{\rho}^{[n,m]}\left(\hat{b}_j^{[n]}\otimes\hat{b}_k^{[m]}\right)\right]\right.\\
&\left.-{\rm Tr}\left[\hat{\rho}^{[n]}\hat{b}_j^{[n]}\right]{\rm Tr}\left[\hat{\rho}^{[m]}\hat{b}_k^{[m]}\right]\right]\\
&=\sum_n J_Q^{jk,[1]}\left(\hat{\rho}^{[n]}\right) + \sum_{n\neq m} J_Q^{jk,[2]}\left(\hat{\rho}^{[n,m]}\right),
\end{split}
\end{equation*}
\normalsize
where $J_Q^{jk,[1]}$ depends only on one-particle reduced density matrix on subsystem $n$ and $J_Q^{jk,[2]}$ depends on two-particle reduced density matrix on subsystems $n$, $m$.

We now restrict to permutationally invariant states \cite{BD}, i.e.~states that are invariant under any permutation of its constituents: $\hat{\rho}=\hat{O}_\pi\hat{\rho}\hat{O}_\pi^\dagger$ for all possible $\pi$, where $\hat{O}_\pi$ is the unitary operator for the permutation $\pi$. Then,

\small\vspace*{-3mm}
\begin{equation}\label{eq:qfim_1p2p}
J_Q^{jk} = NJ_Q^{jk,[1]}\left(\hat{\rho}^{[1]}\right)+N(N-1)J_Q^{jk,[2]}\left(\hat{\rho}^{[1]},\hat{\rho}^{[2]}\right),
\end{equation}\normalsize
where
\small
\begin{equation*}
J_Q^{jk,[1]}=4{\rm Re}\left[{\rm Tr}\left[\hat{\rho}^{[1]}\hat{b}_j\hat{b}_k\right]-{\rm Tr}\left[\hat{\rho}^{[1]}\hat{b}_j\right]{\rm Tr}\left[\hat{\rho}^{[1]}\hat{b}_k\right]\right]
\end{equation*}\normalsize
only depends on the first order reduced density matrix,
\small
\begin{equation*}
J_Q^{jk,[2]}=4{\rm Re}\left[{\rm Tr}\left[\hat{\rho}^{[2]}\left(\hat{b}_j\otimes\hat{b}_k\right)\right]-{\rm Tr}\left[\hat{\rho}^{[1]}\hat{b}_j\right]{\rm Tr}\left[\hat{\rho}^{[1]}\hat{b}_k\right]\right]
\end{equation*}\normalsize
also depends on the second order reduced density matrix.

Then, similar observations can be made as were made in Ref.~\cite{BD} for pure state. For example, if the probe state is a product state, i.e.~$\hat{\rho}=\bigotimes_{n=1}^N\hat{\rho}^{[n]}$, and permutationally invariant, then $\hat{\rho}^{[2]}=\hat{\rho}^{[1]}\otimes\hat{\rho}^{[1]}$, such that $J_Q^{jk,[2]}=0$, and so $J_Q^{jk}=NJ_Q^{jk,[1]}$. This implies that quantum correlations are necessary for achieving the Heisenberg scaling $1/N$, which is evidently attainable even when the initial probe state is mixed. However, if both $\hat{\rho}^{[1]}$ and $\hat{\rho}^{[2]}$ are maximally mixed, the Heisenberg scaling is lost, i.e.~too much quantum correlations harm the quantum advantage with $N$ parallel resources \cite{BD,HGS}. Thus, any quantum enhancement to the estimation precision is provided by the two-particle reduced density matrices of the probe state.

Moreover, from (\ref{eq:qcrb_measure1}), the set of POVMs, comprising

\small\vspace*{-3mm}
\begin{equation}
\begin{split}
\hat{P}_0 &= \hat{\rho}(\boldsymbol{\theta}) = \hat{U}(\boldsymbol{\theta})\hat{\rho}\hat{U}^\dagger(\boldsymbol{\theta}),\\
\hat{P}_m &=\frac{\partial\hat{\rho}(\boldsymbol{\theta})}{\partial\theta_m} = -i\left[\hat{M}_m(\boldsymbol{\theta}),\hat{\rho}(\boldsymbol{\theta})\right] \quad \forall m=1,\ldots,q,
\end{split}
\end{equation}\normalsize
along with one element accounting for normalisation, saturates the QCRB for (\ref{eq:qfim3}), provided we have (\ref{eq:qcrb_saturate}), i.e.~here

\small\vspace*{-3mm}
\begin{equation}
\begin{split}
&2{\rm Tr}\left[\left(\Delta\hat{A}_j\Delta\hat{A}_k-\Delta\hat{A}_k\Delta\hat{A}_j\right)\hat{\rho}\right]=0 \quad \forall j,k\\
\Rightarrow &4\sum_n{\rm Im}{\rm Tr}\left[\hat{\rho}^{[n]}\hat{b}_j^{[n]}\hat{b}_k^{[n]}\right]\\
+&4\sum_{n\neq m}{\rm Im}{\rm Tr}\left[\hat{\rho}^{[n,m]}\left(\hat{b}_j^{[n]}\otimes\hat{b}_k^{[m]}\right)\right]=0\\
\Rightarrow &4\sum_n{\rm Im}{\rm Tr}\left[\hat{\rho}^{[n]}\hat{b}_j^{[n]}\hat{b}_k^{[n]}\right]=0,
\end{split}
\end{equation}\normalsize
since ${\rm Tr}\left[\hat{\rho}^{[n,m]}\left(\hat{b}_j^{[n]}\otimes\hat{b}_k^{[m]}\right)\right]\in\mathbb{R}$. Hence, the attainability of the quantum enhancement to the estimation precision is determined solely by the one-particle reduced density matrices of the probe state.

\vspace*{-3mm}
\section{Estimating a Magnetic Field in Three Dimensions}\label{sec:mag_fld}
Now consider the task of estimating the components of a magnetic field in three dimensions simultaneously using two-level systems. The Hamilton operator for this system is given by $\hat{h}=\boldsymbol{\hat{\mu}}\cdot\mathbf{B}=\sum_{k=1}^3\hat{\mu}_kB_k=\sum_{k=1}^3(\mu/2)B_k\hat{\sigma}_k:=\sum_{k=1}^3\theta_k\hat{\sigma}_k$, where the magnetic moment $\hat{\mu}_k=\mu\hat{\sigma}_k/2$ is proportional to the spin, $\{\hat{\sigma}_k\}$ are the unnormalized Pauli operators, and $\theta_k=\mu B_k/2$ \citep{BD}.

Start with a Greenberger-Horne-Zeilinger (GHZ) type pure state $|\Phi_k\rangle = \left(|\phi_k^{+}\rangle^{\otimes N}+|\phi_k^{-}\rangle^{\otimes N}\right)/\sqrt{2}$, where $|\phi_k^{\pm}\rangle$ is the eigenvector of the Pauli operator $\hat{\sigma}_k$ corresponding to the eigenvalue $\pm 1$ ($k=1$, $2$, $3$ corresponding to the $X$, $Y$, and $Z$ directions). These states are permutationally invariant with first and second order reduced density matrices $\hat{\rho}_k^{[1]}=\mathbb{1}_2/2$ and $\hat{\rho}_k^{[2]}=(|\phi_k^{+},\phi_k^{+}\rangle\langle\phi_k^{+},\phi_k^{+}|+|\phi_k^{-},\phi_k^{-}\rangle\langle\phi_k^{-},\phi_k^{-}|)/2=(\mathbb{1}_2\otimes\mathbb{1}_2+\hat{\sigma}_k\otimes\hat{\sigma}_k)/4$, respectively \cite{BD}. Now, Ref.~\cite{BD} used the pure state $|\psi\rangle = \mathcal{N}\left(e^{i\delta_1}|\Phi_1\rangle+e^{i\delta_2}|\Phi_2\rangle+e^{i\delta_3}|\Phi_3\rangle\right)$, where $\mathcal{N}$ is the normalization constant and $\{\delta_k\}$ are adjustable local phases. We here intend to estimate the three components of the magnetic field using a mixed state $\hat{\rho}^N$, obtained from the above pure state in the presence of local dephasing, described using two single-particle Kraus operators \cite{NC},

\vspace*{-3mm}\small
\begin{equation}
\hat{E}_0 = \left[\begin{array}{cc}
1 & 0\\
0 & e^{-\lambda}
\end{array}\right], \qquad \hat{E}_1 = \left[\begin{array}{cc}
0 & 0\\
0 & \sqrt{1-e^{-2\lambda}}
\end{array}\right],
\end{equation}\normalsize
where $\lambda$ is some constant causing the phase damping, such that the off-diagonal elements of the density matrix decay exponentially to zero with time. Considering that all particles dephase uniformly, the $N$-particle density matrix of the desired mixed state is then \cite{JD}:

\vspace*{-4mm}
\small
\begin{equation*}
\hat{\rho}^N = \sum_{g=0}^N\sum_{\pi_g^N}\pi_g^N\left[\hat{E}_1^{\otimes g}\otimes \hat{E}_0^{\otimes N-g}\right]|\psi\rangle\langle\psi|\pi_g^N\left[\hat{E}_1^{\dagger\otimes g}\otimes \hat{E}_0^{\dagger\otimes N-g}\right],
\end{equation*}\normalsize
where $\pi_g^N$ represents different permutations of $g$ and $N-g$ copies of the $\hat{E}_1$ and $\hat{E}_0$ operators, respectively. Note that the operators $\hat{E}_0$ and $\hat{E}_1$ are Hermitian, so the $\dagger$s can be dropped. Clearly, these states $\hat{\rho}^N$ are permutationally invariant as well, with now first and second order reduced density matrices $\hat{\rho}^{[1]}=\mathbb{1}_2/2$ and $\hat{\rho}_k^{[2]}=[\mathbb{1}_2\otimes\mathbb{1}_2+(\sum_{r=0}^1\hat{E}_r\hat{\sigma}_k\hat{E}_r)\otimes(\sum_{s=0}^1\hat{E}_s\hat{\sigma}_k\hat{E}_s)]/4$, respectively. This is shown in Appendix \ref{sec:app3}.

For $N=8n$, $n \in \mathbb{N}$ (and $\delta_k=0$ for all $k$), the two-body reduced density matrix of $\hat{\rho}^N$ is an equal mixture of those in all directions (as in the pure state case in Ref.~\cite{BD}), given by $\hat{\rho}^{[2]}=\frac{1}{3}\sum_{k=1}^3\hat{\rho}_k^{[2]}=\frac{1}{4}\mathbb{1}_2\otimes\mathbb{1}_2+\frac{1}{12}\sum_{k=1}^3\left[\left(\sum_{r=0}^1\hat{E}_r\hat{\sigma}_k\hat{E}_r\right)\otimes\left(\sum_{s=0}^1\hat{E}_s\hat{\sigma}_k\hat{E}_s\right)\right]$. For any other $N$, the difference from the form of $\hat{\rho}^{[2]}$ is exponentially small in $N$. This directly follows from the way it was shown in Ref.~\cite{BD} for the pure state case. Hence, we consider the probe state to have marginals $\hat{\rho}^{[1]}=\mathbb{1}_2/2$ and $\hat{\rho}^{[2]}$ as above and calculate the QFIM. We get

\vspace*{-2mm}\small
\begin{equation}\label{eq:qfim_1p}
J_Q^{jk,[1]}=2{\rm Tr}\left[\hat{b}_j\hat{b}_k\right],
\end{equation}\normalsize\vspace*{-3mm}
and

\vspace*{-3mm}\small
\begin{equation}\label{eq:qfim_2p}
\begin{split}
J_Q^{jk,[2]}&=\frac{1}{3}\sum_{t=1}^3{\rm Tr}\left[\left(\sum_{r=0}^1\hat{E}_r\hat{\sigma}_t\hat{E}_r\otimes\sum_{s=0}^1\hat{E}_s\hat{\sigma}_t\hat{E}_s\right)\left(\hat{b}_j\otimes\hat{b}_k\right)\right]\\
&=\frac{1}{3}\sum_{t=1}^3{\rm Tr}\left[\sum_{r=0}^1\hat{E}_r\hat{\sigma}_t\hat{E}_r\hat{b}_j\right]{\rm Tr}\left[\sum_{s=0}^1\hat{E}_s\hat{\sigma}_t\hat{E}_s\hat{b}_k\right]\\
&=\frac{1}{3}\sum_{t=1}^3{\rm Tr}\left[\hat{\sigma}_t\sum_{r=0}^1\hat{E}_r\hat{b}_j\hat{E}_r\right]{\rm Tr}\left[\hat{\sigma}_t\sum_{s=0}^1\hat{E}_s\hat{b}_k\hat{E}_s\right]\\
&=\frac{2}{3}{\rm Tr}\left[\sum_{t=1}^3{\rm Tr}\left[\hat{\sigma}_t\left(\sum_{r=0}^1\hat{E}_r\hat{b}_j\hat{E}_r\right)\right]\hat{\sigma}_t\left(\sum_{s=0}^1\hat{E}_s\hat{b}_k\hat{E}_s\right)\right]\\
&=\frac{2}{3}{\rm Tr}\left[\left(\sum_{r=0}^1\hat{E}_r\hat{b}_j\hat{E}_r\right)\left(\sum_{s=0}^1\hat{E}_s\hat{b}_k\hat{E}_s\right)\right].
\end{split}
\end{equation}
\normalsize

Define $\hat{f}_j=\sum_{r=0}^1\hat{E}_r\hat{b}_j\hat{E}_r$ and $\hat{f}_k=\sum_{s=0}^1\hat{E}_s\hat{b}_k\hat{E}_s$. Also, let $\xi=\sqrt{\theta_1^2+\theta_2^2+\theta_3^2}, \eta_k=\frac{\theta_k}{\sqrt{\theta_1^2+\theta_2^2+\theta_3^2}}$ for all $k=1$, $2$, $3$ (corresponding to the $X$, $Y$ and $Z$ directions). Here, (\ref{eq:qfim_1p}) is found to be (following Ref.~\cite{BD}):
\vspace*{-2mm}
\begin{equation}\label{eq:qfim_1p_x}
J_Q^{jk,[1]}=4\left[\left(1-{\rm sinc}^2[\xi]\right)\eta_j\eta_k+\delta_{jk}{\rm sinc}^2[\xi]\right],
\end{equation}
where ${\rm sinc}[\xi]={\rm sin}[\xi]/\xi$. From (\ref{eq:qfim_1p2p}), (\ref{eq:qfim_1p_x}), (\ref{eq:qfim_2p}), we get:
\vspace*{-2mm}
\begin{equation}\label{eq:qfim_mgfld}
\begin{split}
J_Q^{jk}&=4N\left[\left(1-{\rm sinc}^2[\xi]\right)\eta_j\eta_k+\delta_{jk}{\rm sinc}^2[\xi]\right]\\
&+\frac{2N(N-1)}{3}{\rm Tr}\left[\hat{f}_j\hat{f}_k\right],
\end{split}
\end{equation}
where the terms ${\rm Tr}\left[\hat{f}_j\hat{f}_k\right]$ can be explicitly calculated.

Since some or all of the terms ${\rm Tr}\left[\hat{b}_j\hat{b}_k\right]$ are non-zero, we can have the terms ${\rm Tr}\left[\hat{f}_j\hat{f}_k\right]$ as non-zero, such that the scaling $1/N$ can be achieved, as the parallel scheme bound without ancillas from Ref.~\cite{DDM} can be tight even for $\beta\neq 0$. Even when ${\rm Tr}\left[\hat{b}_j\hat{b}_k\right]=0$, the terms ${\rm Tr}\left[\hat{f}_j\hat{f}_k\right]$ in general (i.e.~when $\hat{E}_0$ and $\hat{E}_1$ need not be local dephasing operators), can be non-zero. This implies that it is possible to achieve the Heisenberg scaling with the presence of noise in the initial probe state, even when such a scaling cannot be achieved in the absence of noise in the initial state. This is because mixed separable states can be as nonclassical as entangled pure states \cite{PGACHW}. Thus, noise in the initial probe state can act as a feature rather than a bug in attaining the Heisenberg limit. Note though that it is unlikely for all the terms ${\rm Tr}\left[\hat{b}_j\hat{b}_k\right]$ to be zero, since that would mean that the QFIM $J_Q$ is zero for the pure state case from Ref.~\cite{BD}. However, even when some or all of the terms ${\rm Tr}\left[\hat{b}_j\hat{b}_k\right]$ are non-zero, it may be possible for the terms ${\rm Tr}\left[\hat{f}_j\hat{f}_k\right]$ to be such that the QFIM $J_Q$ for the mixed state case considered here is larger than that for the pure state case from Ref.~\cite{BD}. This is because mixed entangled states can be more nonclassical than pure entangled states \cite{PGACHW}. Thus, noise in the initial probe state can allow for better estimation precision than the case of no noise in the initial state.

Although noise is known to reduce quantum correlations in a system in most cases \cite{HHHH,NC}, noise can also introduce or increase quantum correlations in a system \cite{DB,BFP,SKB,OCMBRM}. For example, local dephasing considered in this section is a local unital noise \cite{SKB}, that mostly decreases quantum correlations. Instead, if local non-unital noise, such as local dissipation \cite{OCMBRM} as represented by the following single-particle Kraus operators \cite{NC}, is used to obtain the initial mixed probe state from a classically correlated separable state, the mixed state so obtained can have quantum correlations, that may be activated into entanglement, allowing for better estimation precision \cite{BIWH,SC,HWD,MCWV}:

\vspace*{-3mm}\small
\begin{equation}
\hat{E}_0 = \left[\begin{array}{cc}
1 & 0\\
0 & \sqrt{1-e^{-2\kappa}}
\end{array}\right], \qquad \hat{E}_1 = \left[\begin{array}{cc}
0 & e^{-\kappa}\\
0 & 0
\end{array}\right],
\end{equation}\normalsize
where $\kappa$ is a constant causing amplitude damping. This is why ancilla-assisted schemes of Ref.~\cite{DDM} yielded scaling better than that without ancillas for amplitude damping.

Nonetheless, if $\hat{\rho}^{[2]}=\hat{\rho}^{[1]}\otimes\hat{\rho}^{[1]}=\mathbb{1}_4/4$, i.e.~both $\hat{\rho}^{[1]}$ and $\hat{\rho}^{[2]}$ are maximally mixed, then ${\rm Tr}\left[\hat{f}_j\hat{f}_k\right]=0$, since $\sum_{t=1}^3\left(\sum_{r=0}^1\hat{E}_r\hat{\sigma}_t\hat{E}_r\otimes\sum_{s=0}^1\hat{E}_s\hat{\sigma}_t\hat{E}_s\right)$ would be zero in (\ref{eq:qfim_2p}), such that the best scaling achievable is $1/\sqrt{N}$. Thus, unlike the conventional wisdom that any amount of noise is harmful, we see that some amount of noise in the initial probe state can be useful and provides a quantum advantage through its quantum correlations, but a lot of noise is harmful because of too much quantum correlations in the state.

\vspace*{-3mm}
\section{Noisy Quantum Channel}\label{sec:noisy}
We consider a general noisy quantum channel that allows the state $\hat{\rho}$ to evolve not necessarily unitarily. Let $\hat{\Pi}_l(\boldsymbol{\theta})$ be the Kraus operators that describe the dynamical evolution of the system. The state of the system after the evolution is \cite{EFD,YZF}

\small\vspace*{-2mm}
\begin{equation}\label{eq:kraus_evolution}
\hat{\rho}(\boldsymbol{\theta}) = \sum_l \hat{\Pi}_l(\boldsymbol{\theta})\hat{\rho}\hat{\Pi}_l^\dagger(\boldsymbol{\theta}),
\end{equation}\normalsize
where $\sum_l\hat{\Pi}_l^\dagger(\boldsymbol{\theta})\hat{\Pi}_l(\boldsymbol{\theta}) = \mathbb{1}$.
Even when the transformation (\ref{eq:kraus_evolution}) is non-unitary, it may be described by a unitary evolution $\hat{U}_{SB}(\boldsymbol{\theta})$ in a bigger space, comprising the system $S$ and some vacuum state ancillary bath $B$. The evolved state in $S+B$ space is given by

\small\vspace*{-2mm}
\begin{equation*}
\begin{split}
\hat{\rho}_{SB}(\boldsymbol{\theta})&=\hat{U}_{SB}(\boldsymbol{\theta})\left(\hat{\rho}\otimes|0\rangle\langle 0|\right)\hat{U}_{SB}^\dagger(\boldsymbol{\theta})\\
&=\sum_{l,v}\hat{\Pi}_l(\boldsymbol{\theta})\hat{\rho}\hat{\Pi}_v^\dagger(\boldsymbol{\theta})\otimes|l\rangle\langle v|
\end{split}
\end{equation*}\normalsize
Then, following from (\ref{eq:unitary_delrho}), (\ref{eq:unitary_mk}), (\ref{eq:unitary_delmk}) for the noiseless $S+B$ space, we get

\small\vspace*{-2mm}
\begin{equation*}
\frac{\partial}{\partial\theta_k}\hat{\rho}_{SB}(\boldsymbol{\theta})=-i\left[\hat{M}_k(\boldsymbol{\theta}),\hat{\rho}_{SB}(\boldsymbol{\theta})\right]
=\frac{1}{2}\left(\hat{L}_k\hat{\rho}_{SB}(\boldsymbol{\theta})+\hat{\rho}_{SB}(\boldsymbol{\theta})\hat{L}_k^\dagger\right),
\end{equation*}\normalsize
where $\hat{M}_k(\boldsymbol{\theta}) \equiv i\frac{\partial\hat{U}_{SB}(\boldsymbol{\theta})}{\partial\theta_k}\hat{U}_{SB}^\dagger(\boldsymbol{\theta})$, and $\hat{L}_k = -2i\Delta\hat{M}_k$ is anti-Hermitian, $\Delta\hat{M}_k \equiv \hat{M}_k(\boldsymbol{\theta}) - {\rm Tr}\left(\hat{M}_k(\boldsymbol{\theta})\hat{\rho}_{SB}(\boldsymbol{\theta})\right)$. Then, the QFIM from (\ref{eq:qfim1}) for $\hat{\rho}_{SB}(\boldsymbol{\theta})$ takes the form:
\small
\begin{equation}\label{eq:noisy_qfim1}
\begin{split}
J_Q^{jk}&=4{\rm Re}\left[{\rm Tr}\left(\hat{H}^{jk}_1(\boldsymbol{\theta})\left(\hat{\rho}\otimes|0\rangle\langle 0|\right)\right)\right.\\
&\left.-{\rm Tr}\left(\hat{H}^j_2(\boldsymbol{\theta})\left(\hat{\rho}\otimes|0\rangle\langle 0|\right)\right){\rm Tr}\left(\hat{H}^k_2(\boldsymbol{\theta})\left(\hat{\rho}\otimes|0\rangle\langle 0|\right)\right)\right],
\end{split}
\end{equation}
\normalsize
where
\small
\begin{equation*}
\hat{H}^{jk}_1(\boldsymbol{\theta})=\frac{\partial\hat{U}_{SB}^\dagger(\boldsymbol{\theta})}{\partial\theta_j}\frac{\partial\hat{U}_{SB}(\boldsymbol{\theta})}{\partial\theta_k}, \, 
\hat{H}^k_2(\boldsymbol{\theta})=i\frac{\partial\hat{U}_{SB}^\dagger(\boldsymbol{\theta})}{\partial\theta_k}\hat{U}_{SB}(\boldsymbol{\theta}).
\end{equation*}\normalsize
However, when only the system $S$ is monitored but the bath $B$ is not monitored, we recover (\ref{eq:kraus_evolution}) by taking a partial trace with respect to $B$: ${\rm Tr}_B\left(\hat{\rho}_{SB}(\boldsymbol{\theta})\right) = \hat{\rho}(\boldsymbol{\theta})$. Then, if we trace out the bath $B$ before having the traces in (\ref{eq:noisy_qfim1}), we obtain an upper bound (like those obtained in Refs.~\cite{EFD,YZF}) to the QFIM in (\ref{eq:qfim1}) for $\hat{\rho}(\boldsymbol{\theta})$:

\small\vspace*{-2mm}
\begin{equation}\label{eq:noisy_qfim2}
C_Q^{jk}=4{\rm Re}\left[{\rm Tr}\left(\hat{K}^{jk}_1(\boldsymbol{\theta})\hat{\rho}\right)-{\rm Tr}\left(\hat{K}^j_2(\boldsymbol{\theta})\hat{\rho}\right){\rm Tr}\left(\hat{K}^k_2(\boldsymbol{\theta})\hat{\rho}\right)\right],
\end{equation}\normalsize
where
\small
\begin{equation*}
\hat{K}^{jk}_1(\boldsymbol{\theta})=\sum_l\frac{\partial\hat{\Pi}_l^\dagger(\boldsymbol{\theta})}{\partial\theta_j}\frac{\partial\hat{\Pi}_l(\boldsymbol{\theta})}{\partial\theta_k}, \, 
\hat{K}^k_2(\boldsymbol{\theta})=i\sum_p\frac{\partial\hat{\Pi}_p^\dagger(\boldsymbol{\theta})}{\partial\theta_k}\hat{\Pi}_p(\boldsymbol{\theta}),
\end{equation*}\normalsize
such that
\small
\begin{equation}
\begin{split}
\hat{K}^{jk}_1(\boldsymbol{\theta})\hat{\rho}&={\rm Tr}_B\left[\hat{H}^{jk}_1(\boldsymbol{\theta})\left(\hat{\rho}\otimes|0\rangle\langle 0|\right)\right],\\
\hat{K}^k_2(\boldsymbol{\theta})\hat{\rho}&={\rm Tr}_B\left[\hat{H}^k_2(\boldsymbol{\theta})\left(\hat{\rho}\otimes|0\rangle\langle 0|\right)\right].
\end{split}
\end{equation}\normalsize
We prove in Appendix \ref{sec:app5} that $C_Q$ from (\ref{eq:noisy_qfim2}) is an upper bound to the QFIM $J_Q$ from (\ref{eq:qfim1}) for $\hat{\rho}(\boldsymbol{\theta})$.

One may compare these results with those in Ref.~\cite{YZF}, where initially pure states in different modes were assumed to evolve independently. We made no such assumption and our initial state is mixed, and so our results are more general. Also, we consider estimation of multiple parameters, as opposed to single parameter estimation studied in Ref.~\cite{EFD}. Our upper bound to the QFIM is relevant, since there are an infinitude of Kraus representations $\hat{\Pi}_l(\boldsymbol{\theta})$ of the channel that make the bound to equal the QFIM \cite{EFD}.

Now, we claim that (\ref{eq:noisy_qfim2}) is saturated, when the following condition is satisfied:

\small\vspace*{-2mm}
\begin{equation}\label{eq:qfim_bound_saturate}
{\rm Im}\left[\sum_l{\rm Tr}\left\lbrace\left(\frac{\partial\hat{\Pi}_l^\dagger(\boldsymbol{\theta})}{\partial\theta_j}\frac{\partial\hat{\Pi}_l(\boldsymbol{\theta})}{\partial\theta_k}\right)\hat{\rho}\right\rbrace\right]=0 \quad \forall j,k,
\end{equation}\normalsize
which is obtained from (\ref{eq:qcrb_saturate}) for $S+B$ space, by tracing out $B$ (see Appendix \ref{sec:app8}). That is, the bound (\ref{eq:noisy_qfim2}) is saturated, when the expectation, with respect to the initial probe state, of the commutator of every pair of the derivatives of the channel Kraus operator and its adjoint vanishes. Clearly, when the above condition is satisfied, it is possible to attain the elusive Heisenberg limit even in the most general noisy estimation scenario. The above condition is trivially true for single parameter estimation.

Then, the set of POVMs of cardinality $q+2$, comprising the following $q+1$ elements,

\small
\begin{equation}\label{eq:qfim_bound_measure1}
\begin{split}
\hat{P}_0&=\hat{\rho}(\boldsymbol{\theta})=\sum_l\hat{\Pi}_l(\boldsymbol{\theta})\hat{\rho}\hat{\Pi}_l^\dagger(\boldsymbol{\theta}),\\
\hat{P}_m&=\frac{\partial\hat{\rho}(\boldsymbol{\theta})}{\partial\theta_m}=\sum_l\left[\frac{\partial\hat{\Pi}_l(\boldsymbol{\theta})}{\partial\theta_m}\hat{\rho}\hat{\Pi}_l^\dagger(\boldsymbol{\theta})\right.\\
&\left.+\hat{\Pi}_l(\boldsymbol{\theta})\hat{\rho}\frac{\partial\hat{\Pi}_l^\dagger(\boldsymbol{\theta})}{\partial\theta_m}\right] \qquad \forall m=1,\ldots,q,
\end{split}
\end{equation}\normalsize
along with one element accounting for normalisation, saturates (\ref{eq:noisy_qfim2}) (See Appendices \ref{sec:app9} and \ref{sec:app10}).

\vspace*{3mm}
\section{Upper Bound to the QFIM for $N$ Particles Evolving via Noisy Channel}\label{sec:qfim_bound}
Consider that the probe comprising $N$ particles evolves not necessarily unitarily. Then, the QFIM (\ref{eq:qfim3}) is for unitary evolution of a probe comprising more than $N$ particles in $S+B$ space. The evolution of the probe comprising $N$ particles in $S$ space alone is described here by some unital Kraus operators $\hat{\Pi}_l(\boldsymbol{\theta})=\frac{1}{\sqrt{L}}e^{-i\hat{G}_l(\boldsymbol{\theta})}$, where $l=1,\ldots,L$, $\sum_l\hat{\Pi}_l^\dagger(\boldsymbol{\theta})\hat{\Pi}_l(\boldsymbol{\theta})=\sum_l\hat{\Pi}_l(\boldsymbol{\theta})\hat{\Pi}_l^\dagger(\boldsymbol{\theta})=\mathbb{1}$, and
\begin{equation}
\hat{G}_l(\boldsymbol{\theta}) = \sum_{n=1}^N\hat{\pi}_{l_n}^{[n]} = \sum_{k=1}^q\theta_k\sum_{n=1}^N\hat{\pi}_{l_nk}^{[n]} \equiv \sum_{k=1}^q\theta_k\hat{G}_{lk}.
\end{equation}
The generators $\hat{G}_{lk}$ do not depend on $\boldsymbol{\theta}$ and do not generally commute with each other. Then, as in Section \ref{sec:qfim},
\small
\begin{equation}
\frac{\partial\hat{\Pi}_l(\boldsymbol{\theta})}{\partial\theta_k}=\frac{-i}{L\sqrt{L}}\int_0^1 d\alpha e^{-i(1-\alpha)\hat{G}_l(\boldsymbol{\theta})}\frac{\partial\hat{G}_l(\boldsymbol{\theta})}{\partial\theta_k}e^{-i\alpha\hat{G}_l(\boldsymbol{\theta})}.
\end{equation}\normalsize
So, ${\small{\rm Tr}_B\left[\hat{M}_k(\boldsymbol{\theta})\right]=i\sum_l\frac{\partial\hat{\Pi}_l(\boldsymbol{\theta})}{\partial\theta_k}\hat{\Pi}_l^\dagger(\boldsymbol{\theta})=\sum_l\hat{\Pi}_l(\boldsymbol{\theta})\hat{B}_{lk}(\boldsymbol{\theta})\hat{\Pi}_l^\dagger(\boldsymbol{\theta})}$\normalsize, where $\hat{M}_k(\boldsymbol{\theta})$ is from (\ref{eq:unitary_mk}) for $S+B$ space, $\sum_l\hat{B}_{lk}(\boldsymbol{\theta})={\rm Tr}_B\left[\hat{A}_k(\boldsymbol{\theta})\right]=\frac{1}{L}\sum_l\int_0^1 d\alpha e^{i\alpha\hat{G}_l(\boldsymbol{\theta})}\hat{G}_{lk}e^{-i\alpha\hat{G}_l(\boldsymbol{\theta})}$, since we have $\frac{\partial\hat{G}_l(\boldsymbol{\theta})}{\partial\theta_k}=\hat{G}_{lk}$. Then, we have in $S+B$ space
\small
\begin{equation}
\frac{\partial\hat{U}_{SB}^\dagger(\boldsymbol{\theta})}{\partial\theta_j}\frac{\partial\hat{U}_{SB}(\boldsymbol{\theta})}{\partial\theta_k}=\hat{A}_j(\boldsymbol{\theta})\hat{U}_{SB}^\dagger(\boldsymbol{\theta})\hat{U}_{SB}(\boldsymbol{\theta})\hat{A}_k(\boldsymbol{\theta}).
\end{equation}\normalsize
Tracing out the bath $B$, we get (see Appendix \ref{sec:app11} to understand why an extra $1/L$ does not arise below):
\begin{widetext}
\small
\begin{equation}
\begin{split}
\sum_l\frac{\partial\hat{\Pi}_l^\dagger(\boldsymbol{\theta})}{\partial\theta_j}\frac{\partial\hat{\Pi}_l(\boldsymbol{\theta})}{\partial\theta_k}&=\sum_l\hat{B}_{lj}(\boldsymbol{\theta})\hat{\Pi}_l^\dagger(\boldsymbol{\theta})\hat{\Pi}_l(\boldsymbol{\theta})\hat{B}_{lk}(\boldsymbol{\theta})\\
\Rightarrow{\rm Tr}\left[\sum_l\frac{\partial\hat{\Pi}_l^\dagger(\boldsymbol{\theta})}{\partial\theta_j}\frac{\partial\hat{\Pi}_l(\boldsymbol{\theta})}{\partial\theta_k}\right]&={\rm Tr}\left[\sum_l\frac{\partial\hat{\Pi}_l^\dagger(\boldsymbol{\theta})}{\partial\theta_j}\hat{\Pi}_l(\boldsymbol{\theta})\hat{\Pi}_l^\dagger(\boldsymbol{\theta})\frac{\partial\hat{\Pi}_l(\boldsymbol{\theta})}{\partial\theta_k}\right]={\rm Tr}\left[\sum_l\hat{B}_{lj}(\boldsymbol{\theta})\hat{\Pi}_l^\dagger(\boldsymbol{\theta})\hat{\Pi}_l(\boldsymbol{\theta})\hat{\Pi}_l^\dagger(\boldsymbol{\theta})\hat{\Pi}_l(\boldsymbol{\theta})\hat{B}_{lk}(\boldsymbol{\theta})\right],
\end{split}
\end{equation}\normalsize
\end{widetext}
since $\sum_l\hat{\Pi}_l(\boldsymbol{\theta})\hat{\Pi}_l^\dagger(\boldsymbol{\theta})=\mathbb{1}$. Also, we have in $S+B$ space
\begin{equation}
i\frac{\partial\hat{U}_{SB}^\dagger(\boldsymbol{\theta})}{\partial\theta_k}\hat{U}_{SB}(\boldsymbol{\theta})=-\hat{A}_k(\boldsymbol{\theta})\hat{U}_{SB}^\dagger(\boldsymbol{\theta})\hat{U}_{SB}(\boldsymbol{\theta}).
\end{equation}

Again, tracing out the bath $B$, we get:
\begin{equation}
i\sum_l\frac{\partial\hat{\Pi}_l^\dagger(\boldsymbol{\theta})}{\partial\theta_k}\hat{\Pi}_l(\boldsymbol{\theta})=-\sum_l\hat{B}_{lk}(\boldsymbol{\theta})\hat{\Pi}_l^\dagger(\boldsymbol{\theta})\hat{\Pi}_l(\boldsymbol{\theta}).
\end{equation}

Then, we get the desired upper bound $C_Q$ to the QFIM from (\ref{eq:qfim3}) as follows:
\begin{equation}\label{eq:qfim4}
\begin{split}
C_Q^{jk}&=4{\rm Re}\left[{\rm Tr}\left(\sum_l\hat{B}_{lj}(\boldsymbol{\theta})\hat{B}_{lk}(\boldsymbol{\theta})\hat{\rho}\right)\right.\\
&\left.-{\rm Tr}\left(\sum_p\hat{B}_{pj}(\boldsymbol{\theta})\hat{\rho}\right){\rm Tr}\left(\sum_r\hat{B}_{rk}(\boldsymbol{\theta})\hat{\rho}\right)\right],
\end{split}
\end{equation}
since $\sum_l\hat{\Pi}_l^\dagger(\boldsymbol{\theta})\hat{\Pi}_l(\boldsymbol{\theta})=\mathbb{1}$. Also, here we have
\small
\begin{equation*}
\sum_l\hat{B}_{lk}(\boldsymbol{\theta})=\sum_n\sum_{l_n}\hat{d}_{l_nk}^{[n]}=\frac{1}{L}\sum_n\sum_{l_n}\int_0^1 d\alpha e^{i\alpha\hat{\pi}_{l_n}^{[n]}}\hat{\pi}_{l_nk}^{[n]}e^{-i\alpha\hat{\pi}_{l_n}^{[n]}}.
\end{equation*}\normalsize

Thus, (\ref{eq:qfim4}) becomes:
\vspace*{-3mm}\small
\begin{equation*}
\begin{split}
C_Q^{jk}&=4\sum_n{\rm Re}\left[{\rm Tr}\left[\sum_{l_n}\hat{\rho}^{[n]}\hat{d}_{l_nj}^{[n]}\hat{d}_{l_nk}^{[n]}\right]\right.\\
&\left.-{\rm Tr}\left[\sum_{l_{n_1}}\hat{\rho}^{[n]}\hat{d}_{l_{n_1}j}^{[n]}\right]{\rm Tr}\left[\sum_{l_{n_2}}\hat{\rho}^{[n]}\hat{d}_{l_{n_2}k}^{[n]}\right]\right]\\
&+4\sum_{n\neq m}\sum_{l_n,l_m}{\rm Re}\left[{\rm Tr}\left[\hat{\rho}^{[n,m]}\left(\hat{d}_{l_nj}^{[n]}\otimes\hat{d}_{l_mk}^{[m]}\right)\right]\right.\\
&\left.-{\rm Tr}\left[\hat{\rho}^{[n]}\hat{d}_{l_nj}^{[n]}\right]{\rm Tr}\left[\hat{\rho}^{[m]}\hat{d}_{l_mk}^{[m]}\right]\right]\\
&=\sum_n C_Q^{jk,[1]}\left(\hat{\rho}^{[n]}\right) + \sum_{n\neq m} C_Q^{jk,[2]}\left(\hat{\rho}^{[n,m]}\right),
\end{split}
\end{equation*}\normalsize
where $C_Q^{jk,[1]}$ depends only on one-particle reduced density matrix on subsystem $n$ and $C_Q^{jk,[2]}$ depends on two-particle reduced density matrix on subsystems $n$, $m$.

Further, if we restrict ourselves to only permutationally invariant states, the upper bound to the QFIM from (\ref{eq:qfim_1p2p}) is as follows:
\small
\begin{equation}\label{eq:qfim_bound_1p2p}
C_Q^{jk} = NC_Q^{jk,[1]}\left(\hat{\rho}^{[1]}\right)+N(N-1)C_Q^{jk,[2]}\left(\hat{\rho}^{[1]},\hat{\rho}^{[2]}\right),
\end{equation}\normalsize
where
\small
\begin{equation*}
C_Q^{jk,[1]}=4\sum_{p,r}{\rm Re}\left[{\rm Tr}\left[\hat{\rho}^{[1]}\hat{d}_{pj}\hat{d}_{pk}\right]-{\rm Tr}\left[\hat{\rho}^{[1]}\hat{d}_{pj}\right]{\rm Tr}\left[\hat{\rho}^{[1]}\hat{d}_{rk}\right]\right]
\end{equation*}\normalsize
only depends on the first order reduced density matrix,
\small
\begin{equation*}
\begin{split}
C_Q^{jk,[2]}&=4\sum_{p,r}{\rm Re}\left[{\rm Tr}\left[\hat{\rho}^{[2]}\left(\hat{d}_{pj}\otimes\hat{d}_{rk}\right)\right]\right.\\
&\left.-{\rm Tr}\left[\hat{\rho}^{[1]}\hat{d}_{pj}\right]{\rm Tr}\left[\hat{\rho}^{[1]}\hat{d}_{rk}\right]\right]
\end{split}
\end{equation*}\normalsize
also depends on the second order reduced density matrix.

Clearly, when the two-particle reduced density matrix of the initial probe state is a product state, we get $C_Q^{jk,[2]}=0$. When both the one- and two-particle reduced density matrices of the initial probe state are maximally mixed, we again get $C_Q^{jk,[2]}=0$. Thus, a precision scaling of $1/N$ cannot be achieved, when there are no correlations or too much quantum correlations in the initial state, like in unitary channel case. Thus, any quantum enhancement to the estimation precision is provided by the two-particle reduced density matrices of the probe state.

Now, from (\ref{eq:qfim_bound_measure1}), the set of POVMs comprising
\small
\begin{equation}
\begin{split}
\hat{P}_0&=\hat{\rho}(\boldsymbol{\theta})=\sum_l\hat{\Pi}_l(\boldsymbol{\theta})\hat{\rho}\hat{\Pi}_l^\dagger(\boldsymbol{\theta}),\\
\hat{P}_m&=\frac{\partial\hat{\rho}(\boldsymbol{\theta})}{\partial\theta_m}=\sum_l\left[\hat{\Pi}_l(\boldsymbol{\theta})\hat{B}_{lm}(\boldsymbol{\theta})\hat{\rho}\hat{\Pi}_l^\dagger(\boldsymbol{\theta})\right.\\
&\left.+\hat{\Pi}_l(\boldsymbol{\theta})\hat{\rho}\hat{B}_{lm}(\boldsymbol{\theta})\hat{\Pi}_l^\dagger(\boldsymbol{\theta})\right] \qquad \forall m=1,\ldots,q,
\end{split}
\end{equation}\normalsize
along with one element accounting for normalisation, saturates the upper bound (\ref{eq:qfim4}) to the QFIM, provided we have (\ref{eq:qfim_bound_saturate}), i.e.~here

\small\vspace*{-3mm}
\begin{equation}
\begin{split}
&4\sum_l{\rm Im}\left[{\rm Tr}\left(\hat{B}_{lj}(\boldsymbol{\theta})\hat{B}_{lk}(\boldsymbol{\theta})\hat{\rho}\right)\right]=0 \quad \forall j,k\\
\Rightarrow &4\sum_n{\rm Im}{\rm Tr}\left[\sum_{l_n}\hat{\rho}^{[n]}\hat{d}_{l_nj}^{[n]}\hat{d}_{l_nk}^{[n]}\right]\\
+&4\sum_{n\neq m}\sum_{l_n,l_m}{\rm Im}{\rm Tr}\left[\hat{\rho}^{[n,m]}\left(\hat{d}_{l_nj}^{[n]}\otimes\hat{d}_{l_mk}^{[m]}\right)\right]=0\\
\Rightarrow &4\sum_n{\rm Im}{\rm Tr}\left[\sum_{l_n}\hat{\rho}^{[n]}\hat{d}_{l_nj}^{[n]}\hat{d}_{l_nk}^{[n]}\right]=0,
\end{split}
\end{equation}\normalsize
since $\sum_{l_n,l_m}{\rm Tr}\left[\hat{\rho}^{[n,m]}\left(\hat{d}_{l_nj}^{[n]}\otimes\hat{d}_{l_mk}^{[m]}\right)\right]\in\mathbb{R}$. Hence, the attainability of the quantum enhancement to the estimation precision is determined solely by the one-particle reduced density matrices of the probe state.

Consider the magnetic field example again here in the context of noisy channel. The same permutationally invariant mixed input probe state is used. Thus, the first and second order marginals are the same. Moreover, for the purposes of this example here, each Pauli operator $\hat{\sigma}_k$ for $k=1,2,3$ (corresponding to $X$, $Y$ and $Z$ directions) can be split into a sum of two single particle Kraus operators as $\hat{\sigma}_k=\sum_{l=1}^2\hat{\pi}_{lk}$, so that $\hat{\pi}_l=\sum_{k=1}^3\theta_k\hat{\pi}_{lk}$, e.g.
\begin{equation}
\begin{split}
\hat{\sigma}_1&=\left[\begin{array}{cc}0 & 1\\1 & 0\end{array}\right]=\left[\begin{array}{cc}0 & 1\\0 & 0\end{array}\right]+\left[\begin{array}{cc}0 & 0\\1 & 0\end{array}\right],\\
\hat{\sigma}_2&=\left[\begin{array}{cc}0 & -i\\i & 0\end{array}\right]=\left[\begin{array}{cc}0 & -i\\0 & 0\end{array}\right]+\left[\begin{array}{cc}0 & 0\\i & 0\end{array}\right],\\
\hat{\sigma}_3&=\left[\begin{array}{cc}1 & 0\\0 & -1\end{array}\right]=\left[\begin{array}{cc}1 & 0\\0 & 0\end{array}\right]+\left[\begin{array}{cc}0 & 0\\0 & -1\end{array}\right].
\end{split}
\end{equation}
One can verify that such a decomposition for each Pauli operator $\hat{\sigma}_k$ satisfies $\sum_l\hat{\pi}_{lk}^\dagger\hat{\pi}_{lk}=\sum_l\hat{\pi}_{lk}\hat{\pi}_{lk}^\dagger=\mathbb{1}_2$. Then,
\begin{equation}
\hat{d}_{lk}=\frac{1}{2}\int_0^1 d\alpha e^{i\alpha\hat{\pi}_l}\hat{\pi}_{lk}e^{-i\alpha\hat{\pi}_l}.
\end{equation}
Then, we get:
\small
\begin{equation}\label{eq:qfim_bound_1p}
C_Q^{jk,[1]}=\sum_{l,p}{\rm Re}\left[2{\rm Tr}\left[\hat{d}_{lj}\hat{d}_{lk}\right]-{\rm Tr}\left[\hat{d}_{lj}\right]{\rm Tr}\left[\hat{d}_{pk}\right]\right],
\end{equation}\normalsize
and
\small
\begin{equation}\label{eq:qfim_bound_2p}
\begin{split}
C_Q^{jk,[2]}&=\frac{1}{3}\sum_{l,p}\sum_{t=1}^3{\rm Re}\left[{\rm Tr}\left[\left(\sum_{r=0}^1\hat{E}_r\hat{\sigma}_t\hat{E}_r\otimes\sum_{s=0}^1\hat{E}_s\hat{\sigma}_t\hat{E}_s\right)\right.\right.\\
&\left.\left.\times\left(\hat{d}_{lj}\otimes\hat{d}_{pk}\right)\right]\right]=\frac{2}{3}\sum_{l,p}{\rm Re}\left[{\rm Tr}\left[\left(\sum_{r=0}^1\hat{E}_r\hat{d}_{lj}\hat{E}_r\right)\right.\right.\\
&\left.\left.\times\left(\sum_{s=0}^1\hat{E}_s\hat{d}_{pk}\hat{E}_s\right)\right]\right].
\end{split}
\end{equation}\normalsize

Define $\hat{g}_{lj}=\sum_{r=0}^1\hat{E}_r\hat{d}_{lj}\hat{E}_r$ and $\hat{g}_{pk}=\sum_{s=0}^1\hat{E}_s\hat{d}_{pk}\hat{E}_s$.

Thus, from (\ref{eq:qfim_bound_1p2p}), (\ref{eq:qfim_bound_1p}) and (\ref{eq:qfim_bound_2p}), we get:
\small
\begin{equation*}
\begin{split}
C_Q^{jk}&=N\sum_{l,p}{\rm Re}\left[2{\rm Tr}\left[\hat{d}_{lj}\hat{d}_{lk}\right]-{\rm Tr}\left[\hat{d}_{lj}\right]{\rm Tr}\left[\hat{d}_{pk}\right]\right]\\
&+\frac{2N(N-1)}{3}\sum_{l,p}{\rm Re}\left[{\rm Tr}\left[\hat{g}_{lj}\hat{g}_{pk}\right]\right],
\end{split}
\end{equation*}\normalsize
where all the quantities may be explicitly calculated. 

Again, note that when the terms ${\rm Tr}\left[\hat{d}_{lj}\hat{d}_{pk}\right]$ are all zero, the terms ${\rm Tr}\left[\hat{g}_{lj}\hat{g}_{pk}\right]$ in general (i.e.~when $\hat{E}_0$ and $\hat{E}_1$ need not be local dephasing operators) can be non-zero, such that it is possible to achieve the Heisenberg limit with the presence of noise in the initial probe state, even when it cannot be achieved in the absence of noise in the initial probe state. Moreover, when the terms ${\rm Tr}\left[\hat{d}_{lj}\hat{d}_{pk}\right]$ are not all zero, the terms ${\rm Tr}\left[\hat{g}_{lj}\hat{g}_{pk}\right]$ can be such that $C_Q$ with noise in the initial probe state, such as by means of $\hat{E}_0$ and $\hat{E}_1$ for local dissipation, is larger than that without noise in the initial probe state, so that the estimation precision can be better with noise in the initial probe state than that without noise in the initial state.

We next consider the more general situation, where the noisy channel need not be necessarily unital, and illustrate that the presence of noise in the channel can actually serve as a feature rather than a bug, since even when the Heisenberg precision scaling cannot be achieved with a unitary channel, it is possible to attain the Heisenberg scaling, and in fact, even beat it with a noisy channel.

\section{Noise in Channel as a Feature rather than a Bug}\label{sec:noise_feature}
We now look at the utility of the presence of noise in a general channel in achieving or even beating the Heisenberg precision limit.

Consider first the case of a mixed probe state, comprising $N$ particles, evolving through a unitary channel, and that the $N$ particles of the probe undergo $N$ independent $\boldsymbol{\theta}$-dependent unitary evolutions, i.e.~the unitary operator of the channel is a product of $N$ independent unitary operators $\hat{U}(\boldsymbol{\theta})=\bigotimes_{n=1}^N\hat{U}_{(n)}(\boldsymbol{\theta})$.

Then, the QFIM takes the form as in (\ref{eq:noisy_qfim1}) as follows:
\begin{widetext}\vspace*{-3mm}
\small
\begin{equation*}
\begin{split}
J_Q^{jk}&=4{\rm Re}\sum_n\left[{\rm Tr}\left(\frac{\partial\hat{U}_{(n)}^{\dagger}(\boldsymbol{\theta})}{\partial\theta_j}\frac{\partial\hat{U}_{(n)}(\boldsymbol{\theta})}{\partial\theta_k}\hat{\rho}^{[n]}\right)-{\rm Tr}\left(i\frac{\partial\hat{U}_{(n)}^\dagger(\boldsymbol{\theta})}{\partial\theta_j}\hat{U}_{(n)}(\boldsymbol{\theta})\hat{\rho}^{[n]}\right){\rm Tr}\left(i\frac{\partial\hat{U}_{(n)}^\dagger(\boldsymbol{\theta})}{\partial\theta_k}\hat{U}_{(n)}(\boldsymbol{\theta})\hat{\rho}^{[n]}\right)\right]\\
&+4{\rm Re}\sum_{n\neq m}\left[{\rm Tr}\left\lbrace\left(\frac{\partial\hat{U}_{(n)}^{\dagger}(\boldsymbol{\theta})}{\partial\theta_j}\hat{U}_{(n)}(\boldsymbol{\theta})\otimes\hat{U}_{(m)}^\dagger(\boldsymbol{\theta})\frac{\partial\hat{U}_{(m)}(\boldsymbol{\theta})}{\partial\theta_k}\right)\hat{\rho}^{[n,m]}\right\rbrace-{\rm Tr}\left(i\frac{\partial\hat{U}_{(n)}^\dagger(\boldsymbol{\theta})}{\partial\theta_j}\hat{U}_{(n)}(\boldsymbol{\theta})\hat{\rho}^{[n]}\right){\rm Tr}\left(i\frac{\partial\hat{U}_{(m)}^\dagger(\boldsymbol{\theta})}{\partial\theta_k}\hat{U}_{(m)}(\boldsymbol{\theta})\hat{\rho}^{[m]}\right)\right]\\
&=J_Q^{jk,[n]}+J_Q^{jk,[n,m]}.
\end{split}
\end{equation*}\normalsize\vspace*{-3mm}
\end{widetext}

Now, note that the first term $J_Q^{jk,[n]}$ is of O($N$) and the second term $J_Q^{jk,[n,m]}$ is of O($N^2$), as they involve $N$ and $N(N-1)/2$ terms, respectively. Then, the term $J_Q^{jk,[n,m]}$ should be non-zero, implying that quantum correlations amongst the particles play a role in attaining the Heisenberg scaling of $1/N$. As observed earlier, if the probe state is a product state, i.e.~$\hat{\rho}=\bigotimes_{n=1}^N\hat{\rho}^{[n]}$, then we have $\hat{\rho}^{[n,m]}=\hat{\rho}^{[n]}\otimes\hat{\rho}^{[m]}$, and consequently $J_Q^{jk,[n,m]}=0$, such that the Heisenberg scaling is lost and the covariance scales as $1/\sqrt{N}$ at best. Also, if both $\hat{\rho}^{[n]}$ and $\hat{\rho}^{[n,m]}$ are maximally mixed, the Heisenberg scaling is lost again and the best scaling for the covariance is $1/\sqrt{N}$, implying that too much quantum correlations harms the quantum advantage with $N$ parallel resources. Classical correlations in the initial probe state cannot be converted into quantum correlations by a unitary channel and cannot allow for an advantage over the scaling $1/\sqrt{N}$. Thus, any quantum enhancement to the estimation precision is provided by the two-particle reduced density matrices of the probe state. Notice that the saturability condition (\ref{eq:qcrb_saturate}) here yields:

\begin{widetext}
\small
\begin{equation}
\begin{split}
&4{\rm Im}\sum_n{\rm Tr}\left(\frac{\partial\hat{U}_{(n)}^{\dagger}(\boldsymbol{\theta})}{\partial\theta_j}\frac{\partial\hat{U}_{(n)}(\boldsymbol{\theta})}{\partial\theta_k}\hat{\rho}^{[n]}\right)+4{\rm Im}\sum_{n\neq m}{\rm Tr}\left\lbrace\left(\frac{\partial\hat{U}_{(n)}^{\dagger}(\boldsymbol{\theta})}{\partial\theta_j}\hat{U}_{(n)}(\boldsymbol{\theta})\otimes\hat{U}_{(m)}^\dagger(\boldsymbol{\theta})\frac{\partial\hat{U}_{(m)}(\boldsymbol{\theta})}{\partial\theta_k}\right)\hat{\rho}^{[n,m]}\right\rbrace=0\\
\Rightarrow &4{\rm Im}\sum_n{\rm Tr}\left(\frac{\partial\hat{U}_{(n)}^{\dagger}(\boldsymbol{\theta})}{\partial\theta_j}\frac{\partial\hat{U}_{(n)}(\boldsymbol{\theta})}{\partial\theta_k}\hat{\rho}^{[n]}\right)=0,
\end{split}
\end{equation}\normalsize\vspace*{-3mm}
\end{widetext}
since $\hat{U}_{(n/m)}^{\dagger}(\boldsymbol{\theta})\hat{U}_{(n/m)}(\boldsymbol{\theta})=\mathbb{1}_2 \, \forall n,m$. Clearly, the attainability of the quantum enhancement to the estimation precision is determined solely by the one-particle reduced density matrices of the initial mixed probe state.

Next, consider the case of a mixed initial probe state, comprising $N$ particles, evolving through a noisy quantum channel, and that the $N$ particles of the initial probe state undergo $N$ independent $\boldsymbol{\theta}$-dependent evolutions, i.e.~the Kraus operator of the noisy quantum channel is a product of $N$ independent Kraus operators $\hat{\Pi}_l(\boldsymbol{\theta})=\bigotimes_{n=1}^N\hat{\Pi}_{l_n}^{(n)}(\boldsymbol{\theta})$, where we have $l=(l_1,l_2,\ldots,l_N)$.

Then, (\ref{eq:noisy_qfim2}) takes the form:
\begin{widetext}\vspace*{-3mm}
\small
\begin{equation*}
\begin{split}
C_Q^{jk}&=4{\rm Re}\sum_n\left[{\rm Tr}\left(\sum_{l_n}\frac{\partial\hat{\Pi}_{l_n}^{(n)\dagger}(\boldsymbol{\theta})}{\partial\theta_j}\frac{\partial\hat{\Pi}_{l_n}^{(n)}(\boldsymbol{\theta})}{\partial\theta_k}\hat{\rho}^{[n]}\right)-{\rm Tr}\left(i\sum_{l_{n_1}}\frac{\partial\hat{\Pi}_{l_{n_1}}^{(n)\dagger}(\boldsymbol{\theta})}{\partial\theta_j}\hat{\Pi}_{l_{n_1}}(\boldsymbol{\theta})\hat{\rho}^{[n]}\right){\rm Tr}\left(i\sum_{l_{n_2}}\frac{\partial\hat{\Pi}_{l_{n_2}}^{(n)\dagger}(\boldsymbol{\theta})}{\partial\theta_k}\hat{\Pi}_{l_{n_2}}^{(n)}(\boldsymbol{\theta})\hat{\rho}^{[n]}\right)\right]\\
&+4{\rm Re}\sum_{n\neq m}\sum_{l_n,l_m}\left[{\rm Tr}\left\lbrace\left(\frac{\partial\hat{\Pi}_{l_n}^{(n)\dagger}(\boldsymbol{\theta})}{\partial\theta_j}\hat{\Pi}_{l_n}^{(n)}(\boldsymbol{\theta})\otimes\hat{\Pi}_{l_m}^{(m)\dagger}(\boldsymbol{\theta})\frac{\partial\hat{\Pi}_{l_m}^{(m)}(\boldsymbol{\theta})}{\partial\theta_k}\right)\hat{\rho}^{[n,m]}\right\rbrace\right.\\
&\left.-{\rm Tr}\left(i\frac{\partial\hat{\Pi}_{l_n}^{(n)\dagger}(\boldsymbol{\theta})}{\partial\theta_j}\hat{\Pi}_{l_n}^{(n)}(\boldsymbol{\theta})\hat{\rho}^{[n]}\right){\rm Tr}\left(i\frac{\partial\hat{\Pi}_{l_m}^{(m)\dagger}(\boldsymbol{\theta})}{\partial\theta_k}\hat{\Pi}_{l_m}^{(m)}(\boldsymbol{\theta})\hat{\rho}^{[m]}\right)\right]\\
&=C_Q^{jk,[n]}+C_Q^{jk,[n,m]}.
\end{split}
\end{equation*}\normalsize
\end{widetext}

Again, note that the first term $C_Q^{jk,[n]}$ is of O($N$) and the second term $C_Q^{jk,[n,m]}$ is of O($N^2$), as they involve $N$ and $N(N-1)/2$ terms, respectively. Then, the term $C_Q^{jk,[n,m]}$ should be non-zero, implying that quantum correlations amongst the particles play a role in attaining the Heisenberg scaling of $1/N$ or better. Now, if the initial probe state is separable but not a product state, then that leads to $C_Q^{jk,[n,m]}\neq 0$. This is because, as noted earlier, although noise is widely known to reduce quantum correlations in a system in most cases \cite{HHHH,NC}, noise can also introduce or increase quantum correlations in a system \cite{DB,BFP,SKB,OCMBRM}, that may then be activated into entanglement \cite{MCWV,PGACHW}. Even without quantum correlations between the particles of the initial probe state, an estimation precision scaling of $1/N$ or better can be achieved, when the initial probe state has classical correlations, that can be converted into quantum correlations by non-unital noise in the channel, unlike in cases of mixed state evolving unitarily or unitally considered earlier. Thus, noise in the quantum channel can act as a feature rather than a bug, since we see that the estimation precision that can be achieved with a noisy channel in some situations is impossible with a noiseless channel. However, if both $\hat{\rho}^{[n]}$ and $\hat{\rho}^{[n,m]}$ are maximally mixed, we get $C_Q^{jk,[n,m]}=0$, so a best precision scaling of $1/\sqrt{N}$ can be achieved.

Moreover, if there exists some Kraus representation $\hat{\Pi}_l(\boldsymbol{\theta})$ of the quantum channel which renders $C_Q^{jk,[n,m]}=0$, then the covariance scales as $1/\sqrt{N}$ at best, even when the particles of the initial probe state are entangled. Extending the argument from Ref.~\cite{EFD} to the multiparameter case, the covariance also scales as $1/\sqrt{N}$ at most, even in the presence of feedback control. Thus, any quantum enhancement to the estimation precision is provided by the two-particle reduced density matrices of the initial probe state. The saturability condition (\ref{eq:qfim_bound_saturate}) here becomes:
\begin{widetext}
\small
\begin{equation}
\begin{split}
&4{\rm Im}\sum_n{\rm Tr}\left(\sum_{l_n}\frac{\partial\hat{\Pi}_{l_n}^{(n)\dagger}(\boldsymbol{\theta})}{\partial\theta_j}\frac{\partial\hat{\Pi}_{l_n}^{(n)}(\boldsymbol{\theta})}{\partial\theta_k}\hat{\rho}^{[n]}\right)+4{\rm Im}\sum_{n\neq m}\sum_{l_n,l_m}{\rm Tr}\left\lbrace\left(\frac{\partial\hat{\Pi}_{l_n}^{(n)\dagger}(\boldsymbol{\theta})}{\partial\theta_j}\hat{\Pi}_{l_n}^{(n)}(\boldsymbol{\theta})\otimes\hat{\Pi}_{l_m}^{(m)\dagger}(\boldsymbol{\theta})\frac{\partial\hat{\Pi}_{l_m}^{(m)}(\boldsymbol{\theta})}{\partial\theta_k}\right)\hat{\rho}^{[n,m]}\right\rbrace=0\\
\Rightarrow &4{\rm Im}\sum_n{\rm Tr}\left(\sum_{l_n}\frac{\partial\hat{\Pi}_{l_n}^{(n)\dagger}(\boldsymbol{\theta})}{\partial\theta_j}\frac{\partial\hat{\Pi}_{l_n}^{(n)}(\boldsymbol{\theta})}{\partial\theta_k}\hat{\rho}^{[n]}\right)=0,
\end{split}
\end{equation}\normalsize\vspace*{-3mm}
\end{widetext}
since $\sum_{l_n/l_m}\hat{\Pi}_{l_n/l_m}^{(n/m)\dagger}(\boldsymbol{\theta})\hat{\Pi}_{l_n/l_m}^{(n/m)}(\boldsymbol{\theta})=\mathbb{1}_2 \, \forall n,m$. Clearly, the attainability of the quantum enhancement to the estimation precision is determined solely by the one-particle reduced density matrices of the probe state.

Now, in terms of the evolved probe state $\hat{\rho}(\boldsymbol{\theta})$, (\ref{eq:noisy_qfim2}) takes the following form. We get the below $C_Q$ from $J_Q$ defined in the $S+B$ space by tracing out the bath $B$, and this is equivalent to $C_Q$ in terms of the initial state.
\begin{widetext}\vspace*{-3mm}
\small
\begin{equation*}
\begin{split}
C_Q^{jk}&=4{\rm Re}\left[{\rm Tr}\left(\sum_l\hat{\Pi}_l(\boldsymbol{\theta})\frac{\partial\hat{\Pi}_l^\dagger(\boldsymbol{\theta})}{\partial\theta_j}\frac{\partial\hat{\Pi}_l(\boldsymbol{\theta})}{\partial\theta_k}\hat{\Pi}_l^\dagger(\boldsymbol{\theta})\hat{\rho}(\boldsymbol{\theta})\right)-{\rm Tr}\left(i\sum_p\hat{\Pi}_p(\boldsymbol{\theta})\frac{\partial\hat{\Pi}_p^\dagger(\boldsymbol{\theta})}{\partial\theta_j}\hat{\rho}(\boldsymbol{\theta})\right){\rm Tr}\left(i\sum_r\hat{\Pi}_r(\boldsymbol{\theta})\frac{\partial\hat{\Pi}_r^\dagger(\boldsymbol{\theta})}{\partial\theta_k}\hat{\rho}(\boldsymbol{\theta})\right)\right]\\
&=4{\rm Re}\sum_n\left[{\rm Tr}\left(\sum_{l_n}\hat{\Pi}_{l_n}^{(n)}(\boldsymbol{\theta})\frac{\partial\hat{\Pi}_{l_n}^{(n)\dagger}(\boldsymbol{\theta})}{\partial\theta_j}\frac{\partial\hat{\Pi}_{l_n}^{(n)}(\boldsymbol{\theta})}{\partial\theta_k}\hat{\Pi}_{l_n}^{(n)\dagger}(\boldsymbol{\theta})\hat{\rho}^{[n]}(\boldsymbol{\theta})\right)\right.\\
&\left.-{\rm Tr}\left(i\sum_{l_{n_1}}\hat{\Pi}_{l_{n_1}}^{(n)}(\boldsymbol{\theta})\frac{\partial\hat{\Pi}_{l_{n_1}}^{(n)\dagger}(\boldsymbol{\theta})}{\partial\theta_j}\hat{\rho}^{[n]}(\boldsymbol{\theta})\right){\rm Tr}\left(i\sum_{l_{n_2}}\hat{\Pi}_{l_{n_2}}^{(n)}(\boldsymbol{\theta})\frac{\partial\hat{\Pi}_{l_{n_2}}^{(n)\dagger}(\boldsymbol{\theta})}{\partial\theta_k}\hat{\rho}^{[n]}(\boldsymbol{\theta})\right)\right]\\
&+4{\rm Re}\sum_{n\neq m}\sum_{l_n,l_m}\left[{\rm Tr}\left\lbrace\left(\hat{\Pi}_{l_n}^{(n)}(\boldsymbol{\theta})\frac{\partial\hat{\Pi}_{l_n}^{(n)\dagger}(\boldsymbol{\theta})}{\partial\theta_j}\otimes\frac{\partial\hat{\Pi}_{l_m}^{(m)}(\boldsymbol{\theta})}{\partial\theta_k}\hat{\Pi}_{l_m}^{(m)\dagger}(\boldsymbol{\theta})\right)\hat{\rho}^{[n,m]}(\boldsymbol{\theta})\right\rbrace\right.\\
&\left.-{\rm Tr}\left(i\hat{\Pi}_{l_n}^{(n)}(\boldsymbol{\theta})\frac{\partial\hat{\Pi}_{l_n}^{(n)\dagger}(\boldsymbol{\theta})}{\partial\theta_j}\hat{\rho}^{[n]}(\boldsymbol{\theta})\right){\rm Tr}\left(i\hat{\Pi}_{l_m}^{(m)}(\boldsymbol{\theta})\frac{\partial\hat{\Pi}_{l_m}^{(m)\dagger}(\boldsymbol{\theta})}{\partial\theta_k}\hat{\rho}^{[m]}(\boldsymbol{\theta})\right)\right]\\
&=C_Q^{jk,[n]}+C_Q^{jk,[n,m]}.
\end{split}
\end{equation*}\normalsize
\end{widetext}

Clearly, if the final probe state is a product state, we get $C_Q^{jk,[n,m]}=0$, such that a scaling of $1/\sqrt{N}$ can be attained at best. This implies that noise in the channel should introduce quantum correlations between the particles of the probe state, in order to provide quantum advantage in achieving an estimation precision scaling of $1/N$ or better. Also, if both $\hat{\rho}^{[n]}(\boldsymbol{\theta})$ and $\hat{\rho}^{[n,m]}(\boldsymbol{\theta})$ are maximally mixed, we get $C_Q^{jk,[n,m]}=0$. This implies that a lot of noise in the channel can introduce too much quantum correlations between the particles of the probe state, such that a best precision scaling of $1/\sqrt{N}$ can be achieved. 

Thus, some amount of noise in the quantum channel can act as a feature rather than a bug by introducing quantum correlations into the system, but excessive noise destroys the achievable quantum advantage with $N$ parallel resources.

\section{Beating the Heisenberg Limit}\label{sec:heisenberg}
We show in Appendix \ref{sec:app11} that unless the following condition is also satisfied by the channel Kraus operators:
\begin{widetext}
\small
\begin{equation}\label{eq:saturate_cond}
\begin{split}
\sum_l\frac{\partial\hat{\Pi}_l(\boldsymbol{\theta})}{\partial\theta_k}\hat{\rho}\hat{\Pi}_l^\dagger(\boldsymbol{\theta})=\sum_l\frac{\partial\hat{\Pi}_l(\boldsymbol{\theta})}{\partial\theta_k}\hat{\rho}&\Rightarrow\sum_l{\rm Tr}\left[\frac{\partial\hat{\Pi}_l^\dagger(\boldsymbol{\theta})}{\partial\theta_j}\hat{\Pi}_l(\boldsymbol{\theta})\hat{\Pi}_l^\dagger(\boldsymbol{\theta})\frac{\partial\hat{\Pi}_l(\boldsymbol{\theta})}{\partial\theta_k}\hat{\rho}\right]=\sum_l{\rm Tr}\left[\frac{\partial\hat{\Pi}_l^\dagger(\boldsymbol{\theta})}{\partial\theta_j}\frac{\partial\hat{\Pi}_l(\boldsymbol{\theta})}{\partial\theta_k}\hat{\rho}\right]\\
&\Rightarrow\sum_l\hat{\Pi}_l(\boldsymbol{\theta})\hat{\Pi}_l^\dagger(\boldsymbol{\theta})=\mathbb{1},
\end{split}
\end{equation}\normalsize
\end{widetext}
i.e.~unless the channel is unital, any noise in the channel may beat the Heisenberg limit, when (\ref{eq:qfim_bound_saturate}) is satisfied.

However, since the Heisenberg limit is not ultimate, e.g.~see Refs.~\cite{AL,BL,RB,RL,BFCG,WMC,ARCPHLD,BDDFSC,BDFSBC,CS,NKDBSM,JPJMNS,KR,TACLA}, although this has sparked some controversy \cite{ZPK,LP,HWX,GM,GLM,HBZW,RSDHZ}, now the question is what is the fundamental ultimate quantum limit to the achievable estimation precision in the presence of optimal amount of noise in a non-unitary quantum channel. In other words, what should the quantity $C_Q$ look like when the precision achievable is maximum in a non-unitary channel. It is fairly easy to see that for optimal quantity of noise in the channel, the two-particle reduced density operators of the evolved probe state should be a maximally entangled mixed state (MEMS) and the one-particle reduced density operators of the evolved probe state should be a maximally mixed state \cite{AIDS,VAM,LZFFL}. Therefore, we must have the reduced density operators of the evolved probe state as follows: $\hat{\rho}^{[n]}(\boldsymbol{\theta})=\mathbb{1}_2/2$ and $\hat{\rho}^{[n,m]}_{MEMS}(\boldsymbol{\theta})\neq \hat{\rho}^{[n]}(\boldsymbol{\theta})\otimes\hat{\rho}^{[m]}(\boldsymbol{\theta})$.

Then, the fundamental quantum limit to the achievable estimation precision in a noisy channel is given by the following:

\begin{widetext}
\small
\begin{equation*}
\begin{split}
J_{SH}^{jk}&={\rm Re}\sum_n\left[2{\rm Tr}\left(\sum_{l_n}\hat{\Pi}_{l_n}^{(n)}(\boldsymbol{\theta})\frac{\partial\hat{\Pi}_{l_n}^{(n)\dagger}(\boldsymbol{\theta})}{\partial\theta_j}\frac{\partial\hat{\Pi}_{l_n}^{(n)}(\boldsymbol{\theta})}{\partial\theta_k}\hat{\Pi}_{l_n}^{(n)\dagger}(\boldsymbol{\theta})\right)\right.\\
&\left.-{\rm Tr}\left(i\sum_{l_{n_1}}\hat{\Pi}_{l_{n_1}}^{(n)}(\boldsymbol{\theta})\frac{\partial\hat{\Pi}_{l_{n_1}}^{(n)\dagger}(\boldsymbol{\theta})}{\partial\theta_j}\right){\rm Tr}\left(i\sum_{l_{n_2}}\hat{\Pi}_{l_{n_2}}^{(n)}(\boldsymbol{\theta})\frac{\partial\hat{\Pi}_{l_{n_2}}^{(n)\dagger}(\boldsymbol{\theta})}{\partial\theta_k}\right)\right]\\
&+{\rm Re}\sum_{n\neq m}\sum_{l_n,l_m}\left[4{\rm Tr}\left\lbrace\left(\hat{\Pi}_{l_n}^{(n)}(\boldsymbol{\theta})\frac{\partial\hat{\Pi}_{l_n}^{(n)\dagger}(\boldsymbol{\theta})}{\partial\theta_j}\otimes\frac{\partial\hat{\Pi}_{l_m}^{(m)}(\boldsymbol{\theta})}{\partial\theta_k}\hat{\Pi}_{l_m}^{(m)\dagger}(\boldsymbol{\theta})\right)\hat{\rho}^{[n,m]}_{MEMS}(\boldsymbol{\theta})\right\rbrace\right.\\
&\left.-{\rm Tr}\left(i\hat{\Pi}_{l_n}^{(n)}(\boldsymbol{\theta})\frac{\partial\hat{\Pi}_{l_n}^{(n)\dagger}(\boldsymbol{\theta})}{\partial\theta_j}\right){\rm Tr}\left(i\hat{\Pi}_{l_m}^{(m)}(\boldsymbol{\theta})\frac{\partial\hat{\Pi}_{l_m}^{(m)\dagger}(\boldsymbol{\theta})}{\partial\theta_k}\right)\right]=J_{SH}^{jk,[n]}+J_{SH}^{jk,[n,m]},
\end{split}
\end{equation*}\normalsize
\end{widetext}
where we have used the subscript ``$SH$" to denote `super-Heisenberg' \cite{NKDBSM} fundamental quantum estimation precision limit. The set of POVMs from (\ref{eq:qfim_bound_measure1}) then saturates this ultimate limit. Note that a maximally discordant mixed state (MDMS) need not be maximally entangled \cite{GGZ}. In fact, it can be not entangled at all, but then it can be at best as nonclassical as (and not more nonclassical than) a maximally entangled pure state \cite{PGACHW}, and therefore, cannot allow to beat the Heisenberg limit.

Note, however, that in order for entanglement to be activated from the quantum correlations in the probe state, multi-particle unitary maps (such as CNOT gates) are required \cite{MCWV,PGACHW}, if there was no entanglement in the initial probe state already or any entanglement in the initial probe state vanishes even if leaving the probe state maximally discordant. The Kraus representation of the channel is non-unique and is invariant under arbitrary unitary maps and so the above equations are invariant under addition of such unitary maps. But unless the quantum correlations are activated into entanglement, the above best estimation precision cannot be achieved. Thus, the active ancilla-assisted scheme from Ref.~\cite{DDM} can be strictly better than the passive ancilla-assisted scheme, since mixed entangled states can be more nonclassical than mixed separable states \cite{PGACHW}. Note that a unitary operator is also a Kraus operator, and an identity operator is trivially unitary.

Now, without the additional unitary maps, that can activate entanglement from quantum correlations in the probe state, the best estimation precision limit is determined by the two-particle reduced density matrices of the evolved probe state being separable and maximally discordant (MDMS) \cite{GGZ}, i.e.~the two-particle reduced density matrices having maximal dissonance \cite{MPSVW}. Therefore, we must have $\hat{\rho}^{[n,m]}_{MDMS}(\boldsymbol{\theta})\neq \hat{\rho}^{[n]}(\boldsymbol{\theta})\otimes\hat{\rho}^{[m]}(\boldsymbol{\theta})$, and then the fundamental limit is given by:
\begin{widetext}\vspace*{-4mm}
\small
\begin{equation*}
\begin{split}
J_Q^{jk}&=4{\rm Re}\sum_n\left[{\rm Tr}\left(\sum_{l_n}\hat{\Pi}_{l_n}^{(n)}(\boldsymbol{\theta})\frac{\partial\hat{\Pi}_{l_n}^{(n)\dagger}(\boldsymbol{\theta})}{\partial\theta_j}\frac{\partial\hat{\Pi}_{l_n}^{(n)}(\boldsymbol{\theta})}{\partial\theta_k}\hat{\Pi}_{l_n}^{(n)\dagger}(\boldsymbol{\theta})\hat{\rho}^{[n]}(\boldsymbol{\theta})\right)\right.\\
&\left.-{\rm Tr}\left(i\sum_{l_{n_1}}\hat{\Pi}_{l_{n_1}}^{(n)}(\boldsymbol{\theta})\frac{\partial\hat{\Pi}_{l_{n_1}}^{(n)\dagger}(\boldsymbol{\theta})}{\partial\theta_j}\hat{\rho}^{[n]}(\boldsymbol{\theta})\right){\rm Tr}\left(i\sum_{l_{n_2}}\hat{\Pi}_{l_{n_2}}^{(n)}(\boldsymbol{\theta})\frac{\partial\hat{\Pi}_{l_{n_2}}^{(n)\dagger}(\boldsymbol{\theta})}{\partial\theta_k}\hat{\rho}^{[n]}(\boldsymbol{\theta})\right)\right]\\
&+4{\rm Re}\sum_{n\neq m}\sum_{l_n,l_m}\left[{\rm Tr}\left\lbrace\left(\hat{\Pi}_{l_n}^{(n)}(\boldsymbol{\theta})\frac{\partial\hat{\Pi}_{l_n}^{(n)\dagger}(\boldsymbol{\theta})}{\partial\theta_j}\otimes\frac{\partial\hat{\Pi}_{l_m}^{(m)}(\boldsymbol{\theta})}{\partial\theta_k}\hat{\Pi}_{l_m}^{(m)\dagger}(\boldsymbol{\theta})\right)\hat{\rho}^{[n,m]}_{MDMS}(\boldsymbol{\theta})\right\rbrace\right.\\
&\left.-{\rm Tr}\left(i\hat{\Pi}_{l_n}^{(n)}(\boldsymbol{\theta})\frac{\partial\hat{\Pi}_{l_n}^{(n)\dagger}(\boldsymbol{\theta})}{\partial\theta_j}\hat{\rho}^{[n]}(\boldsymbol{\theta})\right){\rm Tr}\left(i\hat{\Pi}_{l_m}^{(m)}(\boldsymbol{\theta})\frac{\partial\hat{\Pi}_{l_m}^{(m)\dagger}(\boldsymbol{\theta})}{\partial\theta_k}\hat{\rho}^{[m]}(\boldsymbol{\theta})\right)\right]=J_Q^{jk,[n]}+J_Q^{jk,[n,m]},
\end{split}
\end{equation*}\normalsize\vspace*{-4mm}
\end{widetext}
which corresponds to a precision scaling of $1/N$ for maximal pairwise quantum correlations, without entanglement, amongst the final probe particles \cite{MCWV}, since mixed separable states can be as nonclassical as entangled pure states \cite{PGACHW}. Since the best estimation precision achievable with quantum correlations without entanglement coincides with and does not beat the Heisenberg limit, we used the subscript ``$Q$" above.

Next, with the additional unitary maps and entanglement activated from the quantum correlations in the probe state, since the super-Heisenberg limit is obtained for the two-particle reduced density operators of the evolved probe state being maximally entangled and the one-particle reduced density operators being maximally mixed, the super-Heisenberg limit corresponds to a precision scaling of $1/N^2$ for maximal pairwise quantum correlations including entanglement amongst the final probe particles \cite{BFCG}. This is because mixed entangled bipartite states can be twice as nonclassical as maximally entangled bipartite pure states \cite{PGACHW}.

Note that the precision scaling that could be achieved, e.g.~in Ref.~\cite{BFCG}, using two-particle Hamiltonians for a unitary channel, is achieved using one-particle Kraus operators for a noisy channel here, i.e.~local noise inducing quantum correlations including entanglement amongst the two particles \cite{SKB,OCMBRM}. Notice that we did not get precision scaling better than $1/N$ when we studied the unitary channel case in this paper, since we considered only one-particle Hamiltonians. If we further considered $\gamma$-particle (instead of one-particle) Kraus operators for the noisy channel case here, with $\gamma>1$, each set of Kraus operators can generate quantum correlations including entanglement induced by a common bath amongst the $\gamma$ particles \cite{DB,BFP}. Then, the best super-Heisenberg precision scaling of $1/N^{2\gamma}$ may be attained, that is known to be only attainable using $2\gamma$-particle Hamiltonians for a unitary channel. For example, using three-particle Kraus operators for a noisy channel, the best precision scaling of $1/N^6$ can be achieved, that is otherwise known to be possible with six-particle Hamiltonians for a unitary channel. This is again because mixed entangled states can be twice as nonclassical as pure entangled states \cite{PGACHW}.

Considering again one-particle Kraus operators for the noisy channel, although the quantum Cram\'{e}r-Rao bound (QCRB) can be beaten in the system space, the QCRB for the enlarged system plus bath space, for which the evolution is unitary, is not beaten. This also holds for multi-particle Kraus operators for the channel, where entanglement is induced by common baths. This implies that the estimation in the system space alone is not unbiased, when the QCRB, and therefore, the Heisenberg limit are beaten \cite{WM,LP}. However, when the estimation involving measurements beats the QCRB, and therefore, the Heisenberg limit, it does not violate Robertson's generalized formulation of Heisenberg's uncertainty relation \cite{HPR,NC,WM}, that does not include the measurement process. Note that the QCRB can be derived from the general Heisenberg's uncertainty relation, upon considering that the estimator is unbiased \cite{WM}. Thus, beating the QCRB implies that the estimator bias is no longer zero (also see Appendix \ref{sec:app1}), but does not violate the general Heisenberg's uncertainty principle. Nonetheless, without including measurements, it is noteworthy that entanglement amongst the particles of a state allows for lower bounds for the dispersions of non-commuting observables than that furnished by the traditional Heisenberg's uncertainty relation, originally derived for one particle \cite{GR}.

Finally, note that the super-Heisenberg limit will not necessarily be strictly less than the Heisenberg limit, such as when there are quantum correlations without entanglement in the evolved probe state. Moreover, if the two-particle reduced density matrices of the initial probe state are already maximally entangled, the super-Heisenberg limit will equal the Heisenberg limit. This is because it is only entanglement generated in the channel, i.e.~in the evolution stage, that can contribute to a precision scaling better than the Heisenberg limit, and entanglement in the preparation and measurement stages are inessential \cite{RB}. Furthermore, the Heisenberg limit is not beaten, when the Kraus operators of the channel satisfy the condition (\ref{eq:saturate_cond}). When the QCRB and the Heisenberg limit are not beaten, the estimator in the $S$ space alone will be unbiased. Otherwise, when they are beaten, the estimator in the $S$ space alone will be biased and may be of limited interest in practice.

The upper bound (\ref{eq:noisy_qfim2}) to the QFIM reduces to the following actual QFIM, when (\ref{eq:saturate_cond}) is satisfied:
\begin{widetext}\vspace*{-3mm}
\small
\begin{equation*}
\begin{split}
J_Q^{jk}&=4{\rm Re}\sum_n\left[{\rm Tr}\left(\sum_{l_n}\frac{\partial\hat{\Pi}_{l_n}^{(n)\dagger}(\boldsymbol{\theta})}{\partial\theta_j}\frac{\partial\hat{\Pi}_{l_n}^{(n)}(\boldsymbol{\theta})}{\partial\theta_k}\hat{\rho}^{[n]}\right)-{\rm Tr}\left(\sum_{l_{n_1}}\frac{\partial\hat{\Pi}_{l_{n_1}}^{(n)\dagger}(\boldsymbol{\theta})}{\partial\theta_j}\hat{\Pi}_{l_{n_1}}^{(n)}(\boldsymbol{\theta})\hat{\rho}^{[n]}\right){\rm Tr}\left(\sum_{l_{n_2}}\hat{\Pi}_{l_{n_2}}^{(n)\dagger}(\boldsymbol{\theta})\frac{\partial\hat{\Pi}_{l_{n_2}}^{(n)}(\boldsymbol{\theta})}{\partial\theta_k}\hat{\rho}^{[n]}\right)\right]\\
&+4{\rm Re}\sum_{n\neq m}\sum_{l_n,l_m}\left[{\rm Tr}\left\lbrace\left(\frac{\partial\hat{\Pi}_{l_n}^{(n)\dagger}(\boldsymbol{\theta})}{\partial\theta_j}\hat{\Pi}_{l_n}^{(n)}(\boldsymbol{\theta})\otimes\hat{\Pi}_{l_m}^{(m)\dagger}(\boldsymbol{\theta})\frac{\partial\hat{\Pi}_{l_m}^{(m)}(\boldsymbol{\theta})}{\partial\theta_k}\right)\hat{\rho}^{[n,m]}\right\rbrace\right.\\
&\left.-{\rm Tr}\left(\frac{\partial\hat{\Pi}_{l_n}^{(n)\dagger}(\boldsymbol{\theta})}{\partial\theta_j}\hat{\Pi}_{l_n}^{(n)}(\boldsymbol{\theta})\hat{\rho}^{[n]}\right){\rm Tr}\left(\hat{\Pi}_{l_m}^{(m)\dagger}(\boldsymbol{\theta})\frac{\partial\hat{\Pi}_{l_m}^{(m)}(\boldsymbol{\theta})}{\partial\theta_k}\hat{\rho}^{[m]}\right)\right]=J_Q^{jk,[n]}+J_Q^{jk,[n,m]}.
\end{split}
\end{equation*}\normalsize\vspace*{-3mm}
\end{widetext}

This was the case for unital channel of the form in Section \ref{sec:qfim_bound}. Notice that if the initial probe state is maximally mixed, i.e.~$\hat{\rho}=\mathbb{1}_{2^N}/2^N$, we get $\hat{\rho}(\boldsymbol{\theta})=\mathbb{1}_{2^N}/2^N$ too in that section. This is why quantum correlations are reduced, and cannot be created from any classical correlation in the probe state by the noise in a unital channel \cite{SKB}, and so the QCRB and the Heisenberg limit are not beaten and the estimator remains unbiased. When there are no correlations or too much quantum correlations in the two-particle reduced density matrix of the initial probe state, the best achievable precision scaling is $1/\sqrt{N}$ with a unital channel, like the unitary channel case. Thus, as long as (\ref{eq:saturate_cond}) is satisfied, a noisy channel can at best attain the Heisenberg limit but not beat it, so that the estimator remains unbiased. However, (\ref{eq:saturate_cond}) will not be satisfied by non-unital channels, such as local dissipation of the form in Section \ref{sec:mag_fld}, so that quantum correlations can be created from classical correlations in the probe state by noise in the channel. Notice that in this case, if the initial probe state is maximally mixed, the evolved state will not be maximally mixed. Thus, it may be possible to beat the Heisenberg limit with non-unital channels, and the estimator would be biased when the Heisenberg limit is beaten.

Moreover, the fact that dissonance is more robust to decoherence than entanglement \cite{WSFVB} suggests that it is more probable to attain the Heisenberg limit with a mixed state input than a pure entangled state input to a unital channel. In fact, it may not be possible at all to attain the Heisenberg limit with an input pure entangled state because of entanglement sudden death \cite{AMHMSWSRD,YE}. Furthermore, since dissonance can grow and give rise to entanglement in the presence of dissipation, it is more probable to attain or surpass the Heisenberg limit with a mixed state input than a pure entangled state input to a non-unital channel. In fact, it is never possible to attain or surpass the Heisenberg limit with an input pure entangled state because of no initial classical correlations and entanglement sudden death. On the other hand, the fact that entanglement is the intrinsic and minimal discord capturing nonlocal quantum correlations, as opposed to dissonance, which is the extrinsic discord capturing local quantum correlations that cannot be shared \cite{SL,SZ}, is the reason why the Heisenberg limit can be surpassed only when entanglement and not just dissonance is generated in a non-unital channel fed with a mixed state.

In summary, it may appear that noisy quantum states or channels may require the same or less resources to achieve as much as noiseless quantum states or channels, by exploiting additional resources from the environment. That is why, the overall resources required by the noisy cases in the enlarged noiseless system plus bath space are the same as those known to be required by the noiseless cases in the system space alone. However, any channel can be expressed by Kraus operators, which has the same effect as performing a measurement and discarding the result. To have a measurement on a pure state that is the same as the measurement of the pure state after noise, one would just need to have a POVM that combines the POVM elements used for the mixed state with the Kraus operators of the channel, without requiring any extra resource. Thus, a precision scaling of $1/N^{2\gamma}$ can, in principle, be achieved with a pure initial probe state evolving through a unitary channel, described by $\gamma$-particle Hamiltonians, by using a POVM, that combines the POVM elements used here with the $\gamma$-particle Kraus operators of the noisy channel and the Kraus operators used to prepare the initial mixed probe state considered here. Thus, entangling measurements \cite{RGMSSGB} may also contribute to a precision scaling surpassing the Heisenberg limit, unlike as noted earlier. Similarly, a precision scaling of $1/N^{2\gamma}$ can, in principle, be also achieved with a mixed initial probe state evolving through a unitary channel, described by $\gamma$-particle Hamiltonians, by using a POVM, that combines the POVM elements used here with the $\gamma$-particle Kraus operators of the noisy channel considered here. But using entangling measurements with our noisy channel, it is possible to obtain even better precision scaling, so the noisy case is still superior.

Nonetheless, although it may likewise seem that it should be possible too to achieve a precision scaling of $1/N^{2\gamma}$ with a pure initial probe state evolving through the noisy channel, described by $\gamma$-particle Kraus operators, by using a POVM, obtained by combining the POVM elements used here with the Kraus operators used to prepare the initial mixed probe state from the pure state, that is not true even if the initial pure probe state is maximally entangled and/or if the channel is non-unital. This is because of no initial classical or local quantum correlations in the probe state and sudden death of any entanglement in the probe state caused by the noise in the channel, as discussed earlier. This is the distinct important advantage, unique to mixed state metrology \cite{MCWV}.

\section{Conclusion}\label{sec:conc}
We studied fundamental quantum limits in noisy quantum multiparameter estimation using a quantum Fisher information matrix (QFIM) defined in terms of anti-symmetric logarithmic derivatives (ALDs), that lend a convenient way to study noisy metrology. We presented a QFIM for multiparameter estimation using a mixed probe state evolving unitarily. We then considered a mixed state evolving via a noisy channel, and presented an upper bound to the QFIM for this general-most case.

We found that the bounds are such that the quantum enhancement in the estimation precision is provided by the two-particle reduced density matrices and the attainability of the quantum enhancement is solely determined by the one-particle reduced density matrices of the initial probe state, when the channel is described by one-particle evolution operators. We showed conditions and accordingly measurements to saturate these explicitly computable bounds (e.g.~in terms of the Kraus operators of the channel), not known to exist with conventional symmetric logarithmic derivatives (SLDs) for these general-most cases. We saw that the Heisenberg limit can be achieved even in these most general noisy cases.

Moreover, for the most part of the past century since the inception of quantum physics, weird quantum phenomena, such as superposition and entanglement, were perceived as bugs, until the $80$s when the scientists started to exploit them as features \cite{CK}. Today, the biggest hurdle to quantum technologies, e.g.~in building a scalable quantum computer, is noise. The results here suggest that some noise in the initial probe state or the quantum channel can actually serve as a feature rather than a bug, because we saw that the achievable estimation precision scaling in the presence of noise is not possible in the absence of any noise in the initial probe state or the quantum channel. Noise in the initial probe state or the channel provides with a quantum advantage by introducing quantum correlations into the system. However, too much noise in the initial probe state or the channel is detrimental, since it introduces too much quantum correlations into the system, and, in turn, harms the quantum advantage achievable with $N$ parallel resources.

Furthermore, we found that it is possible to beat the Heisenberg limit by exploiting the noise in the quantum channel. The fundamental super-Heisenberg precision limit for non-unitary channel is then determined by two-particle reduced density operators of the evolved probe state being maximally entangled and one-particle reduced density operators being maximally mixed, and corresponds to a precision scaling of $1/N^2$, achieved with one-particle Kraus operators. Further, using $\gamma$-particle (instead of one-particle) Kraus operators for a noisy channel, where $\gamma>1$, the best scaling of $1/N^{2\gamma}$ can be attained, that is known to be only possible with $2\gamma$-particle Hamiltonians for a noiseless channel. Such a precision scaling can be achieved with an initial pure or mixed probe state evolving through a unitary channel without requiring additional resources, but not with an initial pure probe state evolving through a noisy channel.

These results may be experimentally demonstrated, as part of future work, with more practically implementable measurements that may exist than those presented here.

\begin{acknowledgments}
This work was partially supported by the UK National Quantum Technologies Programme (EP/M01326X/1, EP/M013243/1). The author thanks Christos Gagatsos, Dominic Branford, Animesh Datta, Ranjith Nair, Mankei Tsang, Andy Chia, Pieter Kok, Rafał Demkowicz-Dobrza\'{n}ski, Dominic Berry, Pragya Shukla, Rahul Gupta, Anindya Banerji, Sai Vinjanampathy and Jamie Friel for stimulating discussions in relation to this work.
\end{acknowledgments}

\bibliography{noisy_qcrb_biblio}

\begin{thebibliography}{110}%
\makeatletter
\providecommand \@ifxundefined [1]{%
 \@ifx{#1\undefined}
}%
\providecommand \@ifnum [1]{%
 \ifnum #1\expandafter \@firstoftwo
 \else \expandafter \@secondoftwo
 \fi
}%
\providecommand \@ifx [1]{%
 \ifx #1\expandafter \@firstoftwo
 \else \expandafter \@secondoftwo
 \fi
}%
\providecommand \natexlab [1]{#1}%
\providecommand \enquote  [1]{``#1''}%
\providecommand \bibnamefont  [1]{#1}%
\providecommand \bibfnamefont [1]{#1}%
\providecommand \citenamefont [1]{#1}%
\providecommand \href@noop [0]{\@secondoftwo}%
\providecommand \href [0]{\begingroup \@sanitize@url \@href}%
\providecommand \@href[1]{\@@startlink{#1}\@@href}%
\providecommand \@@href[1]{\endgroup#1\@@endlink}%
\providecommand \@sanitize@url [0]{\catcode `\\12\catcode `\$12\catcode
  `\&12\catcode `\#12\catcode `\^12\catcode `\_12\catcode `\%12\relax}%
\providecommand \@@startlink[1]{}%
\providecommand \@@endlink[0]{}%
\providecommand \url  [0]{\begingroup\@sanitize@url \@url }%
\providecommand \@url [1]{\endgroup\@href {#1}{\urlprefix }}%
\providecommand \urlprefix  [0]{URL }%
\providecommand \Eprint [0]{\href }%
\providecommand \doibase [0]{http://dx.doi.org/}%
\providecommand \selectlanguage [0]{\@gobble}%
\providecommand \bibinfo  [0]{\@secondoftwo}%
\providecommand \bibfield  [0]{\@secondoftwo}%
\providecommand \translation [1]{[#1]}%
\providecommand \BibitemOpen [0]{}%
\providecommand \bibitemStop [0]{}%
\providecommand \bibitemNoStop [0]{.\EOS\space}%
\providecommand \EOS [0]{\spacefactor3000\relax}%
\providecommand \BibitemShut  [1]{\csname bibitem#1\endcsname}%
\let\auto@bib@innerbib\@empty
\bibitem [{\citenamefont {Genoni}\ \emph {et~al.}(2013)\citenamefont {Genoni},
  \citenamefont {Paris}, \citenamefont {Adesso}, \citenamefont {Nha},
  \citenamefont {Knight},\ and\ \citenamefont {Kim}}]{GPANKK}%
  \BibitemOpen
  \bibfield  {author} {\bibinfo {author} {\bibfnamefont {M.~G.}\ \bibnamefont
  {Genoni}}, \bibinfo {author} {\bibfnamefont {M.~G.~A.}\ \bibnamefont
  {Paris}}, \bibinfo {author} {\bibfnamefont {G.}~\bibnamefont {Adesso}},
  \bibinfo {author} {\bibfnamefont {H.}~\bibnamefont {Nha}}, \bibinfo {author}
  {\bibfnamefont {P.~L.}\ \bibnamefont {Knight}}, \ and\ \bibinfo {author}
  {\bibfnamefont {M.~S.}\ \bibnamefont {Kim}},\ }\bibfield  {title} {\enquote
  {\bibinfo {title} {Optimal estimation of joint parameters in phase space},}\
  }\href@noop {} {\bibfield  {journal} {\bibinfo  {journal} {Physical Review
  A}\ }\textbf {\bibinfo {volume} {87}},\ \bibinfo {pages} {012107} (\bibinfo
  {year} {2013})}\BibitemShut {NoStop}%
\bibitem [{\citenamefont {Vidrighin}\ \emph {et~al.}(2014)\citenamefont
  {Vidrighin}, \citenamefont {Donati}, \citenamefont {Genoni}, \citenamefont
  {Jin}, \citenamefont {Kolthammer}, \citenamefont {Kim}, \citenamefont
  {Datta}, \citenamefont {Barbieri},\ and\ \citenamefont
  {Walmsley}}]{VDGJKKDBW}%
  \BibitemOpen
  \bibfield  {author} {\bibinfo {author} {\bibfnamefont {M.~D.}\ \bibnamefont
  {Vidrighin}}, \bibinfo {author} {\bibfnamefont {G.}~\bibnamefont {Donati}},
  \bibinfo {author} {\bibfnamefont {M~G.}\ \bibnamefont {Genoni}}, \bibinfo
  {author} {\bibfnamefont {X.}~\bibnamefont {Jin}}, \bibinfo {author}
  {\bibfnamefont {W.~S.}\ \bibnamefont {Kolthammer}}, \bibinfo {author}
  {\bibfnamefont {M.~S.}\ \bibnamefont {Kim}}, \bibinfo {author} {\bibfnamefont
  {A.}~\bibnamefont {Datta}}, \bibinfo {author} {\bibfnamefont
  {M.}~\bibnamefont {Barbieri}}, \ and\ \bibinfo {author} {\bibfnamefont
  {I.~A.}\ \bibnamefont {Walmsley}},\ }\bibfield  {title} {\enquote {\bibinfo
  {title} {Joint estimation of phase and phase diffusion for quantum
  metrology},}\ }\href@noop {} {\bibfield  {journal} {\bibinfo  {journal}
  {Nature Communications}\ }\textbf {\bibinfo {volume} {5}},\ \bibinfo {pages}
  {3532} (\bibinfo {year} {2014})}\BibitemShut {NoStop}%
\bibitem [{\citenamefont {Crowley}\ \emph {et~al.}(2014)\citenamefont
  {Crowley}, \citenamefont {Datta}, \citenamefont {Barbieri},\ and\
  \citenamefont {Walmsley}}]{CDBW}%
  \BibitemOpen
  \bibfield  {author} {\bibinfo {author} {\bibfnamefont {P.~J.~D.}\
  \bibnamefont {Crowley}}, \bibinfo {author} {\bibfnamefont {A.}~\bibnamefont
  {Datta}}, \bibinfo {author} {\bibfnamefont {M.}~\bibnamefont {Barbieri}}, \
  and\ \bibinfo {author} {\bibfnamefont {I.~A.}\ \bibnamefont {Walmsley}},\
  }\bibfield  {title} {\enquote {\bibinfo {title} {Tradeoff in simultaneous
  quantum-limited phase and loss estimation in interferometry},}\ }\href@noop
  {} {\bibfield  {journal} {\bibinfo  {journal} {Physical Review A}\ }\textbf
  {\bibinfo {volume} {89}},\ \bibinfo {pages} {023845} (\bibinfo {year}
  {2014})}\BibitemShut {NoStop}%
\bibitem [{\citenamefont {Humphreys}\ \emph {et~al.}(2013)\citenamefont
  {Humphreys}, \citenamefont {Barbieri}, \citenamefont {Datta},\ and\
  \citenamefont {Walmsley}}]{HBDW}%
  \BibitemOpen
  \bibfield  {author} {\bibinfo {author} {\bibfnamefont {P.~C.}\ \bibnamefont
  {Humphreys}}, \bibinfo {author} {\bibfnamefont {M.}~\bibnamefont {Barbieri}},
  \bibinfo {author} {\bibfnamefont {A.}~\bibnamefont {Datta}}, \ and\ \bibinfo
  {author} {\bibfnamefont {I.~A.}\ \bibnamefont {Walmsley}},\ }\bibfield
  {title} {\enquote {\bibinfo {title} {Quantum enhanced multiple phase
  estimation},}\ }\href@noop {} {\bibfield  {journal} {\bibinfo  {journal}
  {Physical Review Letters}\ }\textbf {\bibinfo {volume} {111}},\ \bibinfo
  {pages} {070403} (\bibinfo {year} {2013})}\BibitemShut {NoStop}%
\bibitem [{\citenamefont {Baumgratz}\ and\ \citenamefont {Datta}(2016)}]{BD}%
  \BibitemOpen
  \bibfield  {author} {\bibinfo {author} {\bibfnamefont {T.}~\bibnamefont
  {Baumgratz}}\ and\ \bibinfo {author} {\bibfnamefont {A.}~\bibnamefont
  {Datta}},\ }\bibfield  {title} {\enquote {\bibinfo {title} {Quantum enhanced
  estimation of a multi-dimensional field},}\ }\href@noop {} {\bibfield
  {journal} {\bibinfo  {journal} {Physical Review Letters}\ }\textbf {\bibinfo
  {volume} {116}},\ \bibinfo {pages} {030801} (\bibinfo {year}
  {2016})}\BibitemShut {NoStop}%
\bibitem [{\citenamefont {Szczykulska}\ \emph {et~al.}(2016)\citenamefont
  {Szczykulska}, \citenamefont {Baumgratz},\ and\ \citenamefont {Datta}}]{SBD}%
  \BibitemOpen
  \bibfield  {author} {\bibinfo {author} {\bibfnamefont {M.}~\bibnamefont
  {Szczykulska}}, \bibinfo {author} {\bibfnamefont {T.}~\bibnamefont
  {Baumgratz}}, \ and\ \bibinfo {author} {\bibfnamefont {A.}~\bibnamefont
  {Datta}},\ }\bibfield  {title} {\enquote {\bibinfo {title} {Multi-parameter
  quantum metrology},}\ }\href@noop {} {\bibfield  {journal} {\bibinfo
  {journal} {Advances in Physics: X}\ }\textbf {\bibinfo {volume} {1}},\
  \bibinfo {pages} {621--639} (\bibinfo {year} {2016})}\BibitemShut {NoStop}%
\bibitem [{\citenamefont {Pezz\`{e}}\ \emph {et~al.}(2017)\citenamefont
  {Pezz\`{e}}, \citenamefont {Ciampini}, \citenamefont {Spagnolo},
  \citenamefont {Humphreys}, \citenamefont {Datta}, \citenamefont {Walmsley},
  \citenamefont {Barbieri}, \citenamefont {Sciarrino},\ and\ \citenamefont
  {Smerzi}}]{PCSHDWBSS}%
  \BibitemOpen
  \bibfield  {author} {\bibinfo {author} {\bibfnamefont {L.}~\bibnamefont
  {Pezz\`{e}}}, \bibinfo {author} {\bibfnamefont {M.~A.}\ \bibnamefont
  {Ciampini}}, \bibinfo {author} {\bibfnamefont {N.}~\bibnamefont {Spagnolo}},
  \bibinfo {author} {\bibfnamefont {P.~C.}\ \bibnamefont {Humphreys}}, \bibinfo
  {author} {\bibfnamefont {A.}~\bibnamefont {Datta}}, \bibinfo {author}
  {\bibfnamefont {I.~A.}\ \bibnamefont {Walmsley}}, \bibinfo {author}
  {\bibfnamefont {M.}~\bibnamefont {Barbieri}}, \bibinfo {author}
  {\bibfnamefont {F.}~\bibnamefont {Sciarrino}}, \ and\ \bibinfo {author}
  {\bibfnamefont {A.}~\bibnamefont {Smerzi}},\ }\bibfield  {title} {\enquote
  {\bibinfo {title} {Optimal measurements for simultaneous quantum estimation
  of multiple phases},}\ }\href@noop {} {\bibfield  {journal} {\bibinfo
  {journal} {Physical Review Letters}\ }\textbf {\bibinfo {volume} {119}},\
  \bibinfo {pages} {130504} (\bibinfo {year} {2017})}\BibitemShut {NoStop}%
\bibitem [{\citenamefont {Gagatsos}\ \emph {et~al.}(2016)\citenamefont
  {Gagatsos}, \citenamefont {Branford},\ and\ \citenamefont {Datta}}]{GBD}%
  \BibitemOpen
  \bibfield  {author} {\bibinfo {author} {\bibfnamefont {C.~N.}\ \bibnamefont
  {Gagatsos}}, \bibinfo {author} {\bibfnamefont {D.}~\bibnamefont {Branford}},
  \ and\ \bibinfo {author} {\bibfnamefont {A.}~\bibnamefont {Datta}},\
  }\bibfield  {title} {\enquote {\bibinfo {title} {Gaussian systems for
  quantum-enhanced multiple phase estimation},}\ }\href@noop {} {\bibfield
  {journal} {\bibinfo  {journal} {Physical Review A}\ }\textbf {\bibinfo
  {volume} {94}},\ \bibinfo {pages} {042342} (\bibinfo {year}
  {2016})}\BibitemShut {NoStop}%
\bibitem [{\citenamefont {Nichols}\ \emph {et~al.}(2018)\citenamefont
  {Nichols}, \citenamefont {Liuzzo-Scorpo}, \citenamefont {Knott},\ and\
  \citenamefont {Adesso}}]{NLKA}%
  \BibitemOpen
  \bibfield  {author} {\bibinfo {author} {\bibfnamefont {R.}~\bibnamefont
  {Nichols}}, \bibinfo {author} {\bibfnamefont {P.}~\bibnamefont
  {Liuzzo-Scorpo}}, \bibinfo {author} {\bibfnamefont {P.~A.}\ \bibnamefont
  {Knott}}, \ and\ \bibinfo {author} {\bibfnamefont {G.}~\bibnamefont
  {Adesso}},\ }\bibfield  {title} {\enquote {\bibinfo {title} {Multiparameter
  {G}aussian quantum metrology},}\ }\href@noop {} {\bibfield  {journal}
  {\bibinfo  {journal} {Physical Review A}\ }\textbf {\bibinfo {volume} {98}},\
  \bibinfo {pages} {012114} (\bibinfo {year} {2018})}\BibitemShut {NoStop}%
\bibitem [{\citenamefont {Liu}\ and\ \citenamefont {Yuan}(2017)}]{LY}%
  \BibitemOpen
  \bibfield  {author} {\bibinfo {author} {\bibfnamefont {J.}~\bibnamefont
  {Liu}}\ and\ \bibinfo {author} {\bibfnamefont {H.}~\bibnamefont {Yuan}},\
  }\bibfield  {title} {\enquote {\bibinfo {title} {Control-enhanced
  multiparameter quantum estimation},}\ }\href@noop {} {\bibfield  {journal}
  {\bibinfo  {journal} {Physical Review A}\ }\textbf {\bibinfo {volume} {96}},\
  \bibinfo {pages} {042114} (\bibinfo {year} {2017})}\BibitemShut {NoStop}%
\bibitem [{\citenamefont {Roccia}\ \emph {et~al.}(2017)\citenamefont {Roccia},
  \citenamefont {Gianani}, \citenamefont {Mancino}, \citenamefont {Sbroscia},
  \citenamefont {Somma1}, \citenamefont {Genoni},\ and\ \citenamefont
  {Barbieri}}]{RGMSSGB}%
  \BibitemOpen
  \bibfield  {author} {\bibinfo {author} {\bibfnamefont {E.}~\bibnamefont
  {Roccia}}, \bibinfo {author} {\bibfnamefont {I.}~\bibnamefont {Gianani}},
  \bibinfo {author} {\bibfnamefont {L.}~\bibnamefont {Mancino}}, \bibinfo
  {author} {\bibfnamefont {M.}~\bibnamefont {Sbroscia}}, \bibinfo {author}
  {\bibfnamefont {F.}~\bibnamefont {Somma1}}, \bibinfo {author} {\bibfnamefont
  {M.~G.}\ \bibnamefont {Genoni}}, \ and\ \bibinfo {author} {\bibfnamefont
  {M.}~\bibnamefont {Barbieri}},\ }\bibfield  {title} {\enquote {\bibinfo
  {title} {Entangling measurements for multiparameter estimation with two
  qubits},}\ }\href@noop {} {\bibfield  {journal} {\bibinfo  {journal} {Quantum
  Science and Technology}\ }\textbf {\bibinfo {volume} {3}},\ \bibinfo {pages}
  {01LT01} (\bibinfo {year} {2017})}\BibitemShut {NoStop}%
\bibitem [{\citenamefont {Ciampini}\ \emph {et~al.}(2016)\citenamefont
  {Ciampini}, \citenamefont {Spagnolo}, \citenamefont {Vitelli}, \citenamefont
  {Pezz\`{e}}, \citenamefont {Smerzi},\ and\ \citenamefont
  {Sciarrino}}]{CSVPSS}%
  \BibitemOpen
  \bibfield  {author} {\bibinfo {author} {\bibfnamefont {M.~A.}\ \bibnamefont
  {Ciampini}}, \bibinfo {author} {\bibfnamefont {N.}~\bibnamefont {Spagnolo}},
  \bibinfo {author} {\bibfnamefont {C.}~\bibnamefont {Vitelli}}, \bibinfo
  {author} {\bibfnamefont {L.}~\bibnamefont {Pezz\`{e}}}, \bibinfo {author}
  {\bibfnamefont {A.}~\bibnamefont {Smerzi}}, \ and\ \bibinfo {author}
  {\bibfnamefont {F.}~\bibnamefont {Sciarrino}},\ }\bibfield  {title} {\enquote
  {\bibinfo {title} {Quantum-enhanced multiparameter estimation in multiarm
  interferometers},}\ }\href@noop {} {\bibfield  {journal} {\bibinfo  {journal}
  {Scientific Reports}\ }\textbf {\bibinfo {volume} {6}},\ \bibinfo {pages}
  {28881} (\bibinfo {year} {2016})}\BibitemShut {NoStop}%
\bibitem [{\citenamefont {Proctor}\ \emph {et~al.}(2018)\citenamefont
  {Proctor}, \citenamefont {Knott},\ and\ \citenamefont {Dunningham}}]{PKD}%
  \BibitemOpen
  \bibfield  {author} {\bibinfo {author} {\bibfnamefont {T.~J.}\ \bibnamefont
  {Proctor}}, \bibinfo {author} {\bibfnamefont {P.~A.}\ \bibnamefont {Knott}},
  \ and\ \bibinfo {author} {\bibfnamefont {J.~A.}\ \bibnamefont {Dunningham}},\
  }\bibfield  {title} {\enquote {\bibinfo {title} {Multiparameter estimation in
  networked quantum sensors},}\ }\href@noop {} {\bibfield  {journal} {\bibinfo
  {journal} {Physical Review Letters}\ }\textbf {\bibinfo {volume} {120}},\
  \bibinfo {pages} {080501} (\bibinfo {year} {2018})}\BibitemShut {NoStop}%
\bibitem [{\citenamefont {Gessner}\ \emph {et~al.}(2018)\citenamefont
  {Gessner}, \citenamefont {Pezz\`{e}},\ and\ \citenamefont {Smerzi}}]{GPS}%
  \BibitemOpen
  \bibfield  {author} {\bibinfo {author} {\bibfnamefont {M.}~\bibnamefont
  {Gessner}}, \bibinfo {author} {\bibfnamefont {L.}~\bibnamefont {Pezz\`{e}}},
  \ and\ \bibinfo {author} {\bibfnamefont {A.}~\bibnamefont {Smerzi}},\
  }\bibfield  {title} {\enquote {\bibinfo {title} {Sensitivity bounds for
  multiparameter quantum metrology},}\ }\href@noop {} {\bibfield  {journal}
  {\bibinfo  {journal} {Physical Review Letters}\ }\textbf {\bibinfo {volume}
  {121}},\ \bibinfo {pages} {130503} (\bibinfo {year} {2018})}\BibitemShut
  {NoStop}%
\bibitem [{\citenamefont {Giovannetti}\ \emph {et~al.}(2011)\citenamefont
  {Giovannetti}, \citenamefont {Lloyd},\ and\ \citenamefont {Maccone}}]{GLM1}%
  \BibitemOpen
  \bibfield  {author} {\bibinfo {author} {\bibfnamefont {V.}~\bibnamefont
  {Giovannetti}}, \bibinfo {author} {\bibfnamefont {S.}~\bibnamefont {Lloyd}},
  \ and\ \bibinfo {author} {\bibfnamefont {L.}~\bibnamefont {Maccone}},\
  }\bibfield  {title} {\enquote {\bibinfo {title} {Advances in quantum
  metrology},}\ }\href@noop {} {\bibfield  {journal} {\bibinfo  {journal}
  {Nature Photonics}\ }\textbf {\bibinfo {volume} {5}},\ \bibinfo {pages}
  {222--229} (\bibinfo {year} {2011})}\BibitemShut {NoStop}%
\bibitem [{\citenamefont {Giovannetti}\ \emph {et~al.}(2006)\citenamefont
  {Giovannetti}, \citenamefont {Lloyd},\ and\ \citenamefont {Maccone}}]{GLM2}%
  \BibitemOpen
  \bibfield  {author} {\bibinfo {author} {\bibfnamefont {V.}~\bibnamefont
  {Giovannetti}}, \bibinfo {author} {\bibfnamefont {S.}~\bibnamefont {Lloyd}},
  \ and\ \bibinfo {author} {\bibfnamefont {L.}~\bibnamefont {Maccone}},\
  }\bibfield  {title} {\enquote {\bibinfo {title} {Quantum metrology},}\
  }\href@noop {} {\bibfield  {journal} {\bibinfo  {journal} {Physical Review
  Letters}\ }\textbf {\bibinfo {volume} {96}},\ \bibinfo {pages} {010401}
  (\bibinfo {year} {2006})}\BibitemShut {NoStop}%
\bibitem [{\citenamefont {Friis}\ \emph {et~al.}(2017)\citenamefont {Friis},
  \citenamefont {Orsucci}, \citenamefont {Skotiniotis}, \citenamefont
  {P.Sekatski}, \citenamefont {Dunjko}, \citenamefont {Briegel},\ and\
  \citenamefont {D\"{u}r}}]{FOSSDBD}%
  \BibitemOpen
  \bibfield  {author} {\bibinfo {author} {\bibfnamefont {N.}~\bibnamefont
  {Friis}}, \bibinfo {author} {\bibfnamefont {D.}~\bibnamefont {Orsucci}},
  \bibinfo {author} {\bibfnamefont {M.}~\bibnamefont {Skotiniotis}}, \bibinfo
  {author} {\bibnamefont {P.Sekatski}}, \bibinfo {author} {\bibfnamefont
  {V.}~\bibnamefont {Dunjko}}, \bibinfo {author} {\bibfnamefont {H.~J.}\
  \bibnamefont {Briegel}}, \ and\ \bibinfo {author} {\bibfnamefont
  {W.}~\bibnamefont {D\"{u}r}},\ }\bibfield  {title} {\enquote {\bibinfo
  {title} {Flexible resources for quantum metrology},}\ }\href@noop {}
  {\bibfield  {journal} {\bibinfo  {journal} {New Journal of Physics}\ }\textbf
  {\bibinfo {volume} {19}},\ \bibinfo {pages} {063044} (\bibinfo {year}
  {2017})}\BibitemShut {NoStop}%
\bibitem [{\citenamefont {Giovannetti}\ \emph {et~al.}(2004)\citenamefont
  {Giovannetti}, \citenamefont {Lloyd},\ and\ \citenamefont {Maccone}}]{GLM3}%
  \BibitemOpen
  \bibfield  {author} {\bibinfo {author} {\bibfnamefont {V.}~\bibnamefont
  {Giovannetti}}, \bibinfo {author} {\bibfnamefont {S.}~\bibnamefont {Lloyd}},
  \ and\ \bibinfo {author} {\bibfnamefont {L.}~\bibnamefont {Maccone}},\
  }\bibfield  {title} {\enquote {\bibinfo {title} {Quantum-enhanced
  measurements: Beating the standard quantum limit},}\ }\href@noop {}
  {\bibfield  {journal} {\bibinfo  {journal} {Science}\ }\textbf {\bibinfo
  {volume} {306}},\ \bibinfo {pages} {1330--1336} (\bibinfo {year}
  {2004})}\BibitemShut {NoStop}%
\bibitem [{\citenamefont {Helstrom}(1967)}]{CWH}%
  \BibitemOpen
  \bibfield  {author} {\bibinfo {author} {\bibfnamefont {C.~W.}\ \bibnamefont
  {Helstrom}},\ }\href@noop {} {\emph {\bibinfo {title} {Quantum detection and
  estimation theory, Mathematics in Science and Engineering}}}\ (\bibinfo
  {publisher} {Academic Press, Massachusetts},\ \bibinfo {year}
  {1967})\BibitemShut {NoStop}%
\bibitem [{\citenamefont {Paris}(2009)}]{MGAP}%
  \BibitemOpen
  \bibfield  {author} {\bibinfo {author} {\bibfnamefont {M.~G.~A.}\
  \bibnamefont {Paris}},\ }\bibfield  {title} {\enquote {\bibinfo {title}
  {Quantum estimation for quantum technology},}\ }\href@noop {} {\bibfield
  {journal} {\bibinfo  {journal} {International Journal of Quantum
  Information}\ }\textbf {\bibinfo {volume} {7}},\ \bibinfo {pages} {125}
  (\bibinfo {year} {2009})}\BibitemShut {NoStop}%
\bibitem [{\citenamefont {Liu}\ and\ \citenamefont {Cable}(2017)}]{LC}%
  \BibitemOpen
  \bibfield  {author} {\bibinfo {author} {\bibfnamefont {N.}~\bibnamefont
  {Liu}}\ and\ \bibinfo {author} {\bibfnamefont {H.}~\bibnamefont {Cable}},\
  }\bibfield  {title} {\enquote {\bibinfo {title} {Quantum-enhanced
  multi-parameter estimation for unitary photonic systems},}\ }\href@noop {}
  {\bibfield  {journal} {\bibinfo  {journal} {Quantum Science and Technology}\
  }\textbf {\bibinfo {volume} {2}},\ \bibinfo {pages} {025008} (\bibinfo {year}
  {2017})}\BibitemShut {NoStop}%
\bibitem [{\citenamefont {Yue}\ \emph {et~al.}(2014)\citenamefont {Yue},
  \citenamefont {Zhang},\ and\ \citenamefont {Fan}}]{YZF}%
  \BibitemOpen
  \bibfield  {author} {\bibinfo {author} {\bibfnamefont {J.-D.}\ \bibnamefont
  {Yue}}, \bibinfo {author} {\bibfnamefont {Y.-R.}\ \bibnamefont {Zhang}}, \
  and\ \bibinfo {author} {\bibfnamefont {H.}~\bibnamefont {Fan}},\ }\bibfield
  {title} {\enquote {\bibinfo {title} {Quantum-enhanced metrology for multiple
  phase estimation with noise},}\ }\href@noop {} {\bibfield  {journal}
  {\bibinfo  {journal} {Scientific Reports}\ }\textbf {\bibinfo {volume} {4}},\
  \bibinfo {pages} {5933} (\bibinfo {year} {2014})}\BibitemShut {NoStop}%
\bibitem [{\citenamefont {Wiseman}\ and\ \citenamefont {Milburn}(2010)}]{WM}%
  \BibitemOpen
  \bibfield  {author} {\bibinfo {author} {\bibfnamefont {H.~M.}\ \bibnamefont
  {Wiseman}}\ and\ \bibinfo {author} {\bibfnamefont {G.~J.}\ \bibnamefont
  {Milburn}},\ }\href@noop {} {\emph {\bibinfo {title} {Quantum Measurement and
  Control}}}\ (\bibinfo  {publisher} {Cambridge University Press},\ \bibinfo
  {year} {2010})\BibitemShut {NoStop}%
\bibitem [{\citenamefont {Nielsen}\ and\ \citenamefont {Chuang}(2002)}]{NC}%
  \BibitemOpen
  \bibfield  {author} {\bibinfo {author} {\bibfnamefont {M.~A.}\ \bibnamefont
  {Nielsen}}\ and\ \bibinfo {author} {\bibfnamefont {I.~L.}\ \bibnamefont
  {Chuang}},\ }\href@noop {} {\emph {\bibinfo {title} {Quantum Computation and
  Quantum Information}}}\ (\bibinfo  {publisher} {Cambridge University Press},\
  \bibinfo {year} {2002})\BibitemShut {NoStop}%
\bibitem [{\citenamefont {Braunstein}\ \emph {et~al.}(1996)\citenamefont
  {Braunstein}, \citenamefont {Caves},\ and\ \citenamefont {Milburn}}]{BCM}%
  \BibitemOpen
  \bibfield  {author} {\bibinfo {author} {\bibfnamefont {S.~L.}\ \bibnamefont
  {Braunstein}}, \bibinfo {author} {\bibfnamefont {C.~M.}\ \bibnamefont
  {Caves}}, \ and\ \bibinfo {author} {\bibfnamefont {G.~J.}\ \bibnamefont
  {Milburn}},\ }\bibfield  {title} {\enquote {\bibinfo {title} {Generalized
  uncertainty relations: Theory, examples, and {L}orentz invariance},}\
  }\href@noop {} {\bibfield  {journal} {\bibinfo  {journal} {Annals of
  Physics}\ }\textbf {\bibinfo {volume} {247}},\ \bibinfo {pages} {135--173}
  (\bibinfo {year} {1996})}\BibitemShut {NoStop}%
\bibitem [{\citenamefont {Escher}\ \emph {et~al.}(2011)\citenamefont {Escher},
  \citenamefont {de~Matos~Filho},\ and\ \citenamefont {Davidovich}}]{EFD}%
  \BibitemOpen
  \bibfield  {author} {\bibinfo {author} {\bibfnamefont {B.~M.}\ \bibnamefont
  {Escher}}, \bibinfo {author} {\bibfnamefont {R.~L.}\ \bibnamefont
  {de~Matos~Filho}}, \ and\ \bibinfo {author} {\bibfnamefont {L.}~\bibnamefont
  {Davidovich}},\ }\bibfield  {title} {\enquote {\bibinfo {title} {General
  framework for estimating the ultimate precision limit in noisy
  quantum-enhanced metrology},}\ }\href@noop {} {\bibfield  {journal} {\bibinfo
   {journal} {Nature Physics}\ }\textbf {\bibinfo {volume} {7}},\ \bibinfo
  {pages} {406--411} (\bibinfo {year} {2011})}\BibitemShut {NoStop}%
\bibitem [{\citenamefont {Szczykulska}\ \emph {et~al.}(2017)\citenamefont
  {Szczykulska}, \citenamefont {Baumgratz},\ and\ \citenamefont
  {Datta}}]{SBD1}%
  \BibitemOpen
  \bibfield  {author} {\bibinfo {author} {\bibfnamefont {M.}~\bibnamefont
  {Szczykulska}}, \bibinfo {author} {\bibfnamefont {T.}~\bibnamefont
  {Baumgratz}}, \ and\ \bibinfo {author} {\bibfnamefont {A.}~\bibnamefont
  {Datta}},\ }\bibfield  {title} {\enquote {\bibinfo {title} {Reaching for the
  quantum limits in the simultaneous estimation of phase and phase
  diffusion},}\ }\href@noop {} {\bibfield  {journal} {\bibinfo  {journal}
  {Quantum Science and Technology}\ }\textbf {\bibinfo {volume} {2}},\ \bibinfo
  {pages} {044004} (\bibinfo {year} {2017})}\BibitemShut {NoStop}%
\bibitem [{\citenamefont {Pasquale}\ \emph {et~al.}(2013)\citenamefont
  {Pasquale}, \citenamefont {Rossini}, \citenamefont {Facchi},\ and\
  \citenamefont {Giovannetti}}]{DRFG}%
  \BibitemOpen
  \bibfield  {author} {\bibinfo {author} {\bibfnamefont {A.~De}\ \bibnamefont
  {Pasquale}}, \bibinfo {author} {\bibfnamefont {D.}~\bibnamefont {Rossini}},
  \bibinfo {author} {\bibfnamefont {P.}~\bibnamefont {Facchi}}, \ and\ \bibinfo
  {author} {\bibfnamefont {V.}~\bibnamefont {Giovannetti}},\ }\bibfield
  {title} {\enquote {\bibinfo {title} {Quantum parameter estimation affected by
  unitary disturbance},}\ }\href@noop {} {\bibfield  {journal} {\bibinfo
  {journal} {Physical Review A}\ }\textbf {\bibinfo {volume} {88}},\ \bibinfo
  {pages} {052117} (\bibinfo {year} {2013})}\BibitemShut {NoStop}%
\bibitem [{\citenamefont {Kołodyński}\ and\ \citenamefont
  {Demkowicz-Dobrzański}(2013)}]{KDD}%
  \BibitemOpen
  \bibfield  {author} {\bibinfo {author} {\bibfnamefont {J.}~\bibnamefont
  {Kołodyński}}\ and\ \bibinfo {author} {\bibfnamefont {R.}~\bibnamefont
  {Demkowicz-Dobrzański}},\ }\bibfield  {title} {\enquote {\bibinfo {title}
  {Efficient tools for quantum metrology with uncorrelated noise},}\
  }\href@noop {} {\bibfield  {journal} {\bibinfo  {journal} {New Journal of
  Physics}\ }\textbf {\bibinfo {volume} {15}},\ \bibinfo {pages} {073043}
  (\bibinfo {year} {2013})}\BibitemShut {NoStop}%
\bibitem [{\citenamefont {Demkowicz-Dobrza\'{n}ski}\ \emph
  {et~al.}(2012)\citenamefont {Demkowicz-Dobrza\'{n}ski}, \citenamefont
  {Kołody\'{n}ski},\ and\ \citenamefont {Guţă}}]{DDKG}%
  \BibitemOpen
  \bibfield  {author} {\bibinfo {author} {\bibfnamefont {R.}~\bibnamefont
  {Demkowicz-Dobrza\'{n}ski}}, \bibinfo {author} {\bibfnamefont
  {J.}~\bibnamefont {Kołody\'{n}ski}}, \ and\ \bibinfo {author} {\bibfnamefont
  {M.}~\bibnamefont {Guţă}},\ }\bibfield  {title} {\enquote {\bibinfo {title}
  {The elusive {H}eisenberg limit in quantum-enhanced metrology},}\ }\href@noop
  {} {\bibfield  {journal} {\bibinfo  {journal} {Nature Communications}\
  }\textbf {\bibinfo {volume} {3}},\ \bibinfo {pages} {1063} (\bibinfo {year}
  {2012})}\BibitemShut {NoStop}%
\bibitem [{\citenamefont {Demkowicz-Dobrza\'{n}ski}\ and\ \citenamefont
  {Maccone}(2014)}]{DDM}%
  \BibitemOpen
  \bibfield  {author} {\bibinfo {author} {\bibfnamefont {R.}~\bibnamefont
  {Demkowicz-Dobrza\'{n}ski}}\ and\ \bibinfo {author} {\bibfnamefont
  {L.}~\bibnamefont {Maccone}},\ }\bibfield  {title} {\enquote {\bibinfo
  {title} {Using entanglement against noise in quantum metrology},}\
  }\href@noop {} {\bibfield  {journal} {\bibinfo  {journal} {Physical Review
  Letters}\ }\textbf {\bibinfo {volume} {113}},\ \bibinfo {pages} {250801}
  (\bibinfo {year} {2014})}\BibitemShut {NoStop}%
\bibitem [{\citenamefont {Demkowicz-Dobrzański}\ \emph
  {et~al.}(2017)\citenamefont {Demkowicz-Dobrzański}, \citenamefont
  {Czajkowski},\ and\ \citenamefont {Sekatski}}]{DDCS}%
  \BibitemOpen
  \bibfield  {author} {\bibinfo {author} {\bibfnamefont {R.}~\bibnamefont
  {Demkowicz-Dobrzański}}, \bibinfo {author} {\bibfnamefont {J.}~\bibnamefont
  {Czajkowski}}, \ and\ \bibinfo {author} {\bibfnamefont {P.}~\bibnamefont
  {Sekatski}},\ }\bibfield  {title} {\enquote {\bibinfo {title} {Adaptive
  quantum metrology under general {M}arkovian noise},}\ }\href@noop {}
  {\bibfield  {journal} {\bibinfo  {journal} {Physical Review X}\ }\textbf
  {\bibinfo {volume} {7}},\ \bibinfo {pages} {041009} (\bibinfo {year}
  {2017})}\BibitemShut {NoStop}%
\bibitem [{\citenamefont {Nichols}\ \emph {et~al.}(2016)\citenamefont
  {Nichols}, \citenamefont {Bromley}, \citenamefont {Correa},\ and\
  \citenamefont {Adesso}}]{NBCA}%
  \BibitemOpen
  \bibfield  {author} {\bibinfo {author} {\bibfnamefont {R.}~\bibnamefont
  {Nichols}}, \bibinfo {author} {\bibfnamefont {T.~R.}\ \bibnamefont
  {Bromley}}, \bibinfo {author} {\bibfnamefont {L.~A.}\ \bibnamefont {Correa}},
  \ and\ \bibinfo {author} {\bibfnamefont {G.}~\bibnamefont {Adesso}},\
  }\bibfield  {title} {\enquote {\bibinfo {title} {Practical quantum metrology
  in noisy environments},}\ }\href@noop {} {\bibfield  {journal} {\bibinfo
  {journal} {Physical Review A}\ }\textbf {\bibinfo {volume} {94}},\ \bibinfo
  {pages} {042101} (\bibinfo {year} {2016})}\BibitemShut {NoStop}%
\bibitem [{\citenamefont {Yao}\ \emph {et~al.}(2014)\citenamefont {Yao},
  \citenamefont {Ge}, \citenamefont {Xiao}, \citenamefont {Wang},\ and\
  \citenamefont {Sun}}]{YGXWS}%
  \BibitemOpen
  \bibfield  {author} {\bibinfo {author} {\bibfnamefont {Y.}~\bibnamefont
  {Yao}}, \bibinfo {author} {\bibfnamefont {L.}~\bibnamefont {Ge}}, \bibinfo
  {author} {\bibfnamefont {X.}~\bibnamefont {Xiao}}, \bibinfo {author}
  {\bibfnamefont {X.}~\bibnamefont {Wang}}, \ and\ \bibinfo {author}
  {\bibfnamefont {C.~P.}\ \bibnamefont {Sun}},\ }\bibfield  {title} {\enquote
  {\bibinfo {title} {Multiple phase estimation for arbitrary pure states under
  white noise},}\ }\href@noop {} {\bibfield  {journal} {\bibinfo  {journal}
  {Physical Review A}\ }\textbf {\bibinfo {volume} {90}},\ \bibinfo {pages}
  {062113} (\bibinfo {year} {2014})}\BibitemShut {NoStop}%
\bibitem [{\citenamefont {Roccia}\ \emph {et~al.}(2018)\citenamefont {Roccia},
  \citenamefont {Cimini}, \citenamefont {Sbroscia}, \citenamefont {Gianani},
  \citenamefont {Ruggiero}, \citenamefont {Mancino}, \citenamefont {Genoni},
  \citenamefont {Ricci},\ and\ \citenamefont {Barbieri}}]{RCSGRMGRB}%
  \BibitemOpen
  \bibfield  {author} {\bibinfo {author} {\bibfnamefont {E.}~\bibnamefont
  {Roccia}}, \bibinfo {author} {\bibfnamefont {V.}~\bibnamefont {Cimini}},
  \bibinfo {author} {\bibfnamefont {M.}~\bibnamefont {Sbroscia}}, \bibinfo
  {author} {\bibfnamefont {I.}~\bibnamefont {Gianani}}, \bibinfo {author}
  {\bibfnamefont {L.}~\bibnamefont {Ruggiero}}, \bibinfo {author}
  {\bibfnamefont {L.}~\bibnamefont {Mancino}}, \bibinfo {author} {\bibfnamefont
  {M.~G.}\ \bibnamefont {Genoni}}, \bibinfo {author} {\bibfnamefont {M.~A.}\
  \bibnamefont {Ricci}}, \ and\ \bibinfo {author} {\bibfnamefont
  {M.}~\bibnamefont {Barbieri}},\ }\bibfield  {title} {\enquote {\bibinfo
  {title} {Multiparameter quantum estimation of noisy phase shifts},}\
  }\href@noop {} {\bibfield  {journal} {\bibinfo  {journal} {Optica}\ }\textbf
  {\bibinfo {volume} {5}},\ \bibinfo {pages} {1171--1176} (\bibinfo {year}
  {2018})}\BibitemShut {NoStop}%
\bibitem [{\citenamefont {Jeske}\ \emph {et~al.}(2014)\citenamefont {Jeske},
  \citenamefont {Cole},\ and\ \citenamefont {Huelga}}]{JCH}%
  \BibitemOpen
  \bibfield  {author} {\bibinfo {author} {\bibfnamefont {J.}~\bibnamefont
  {Jeske}}, \bibinfo {author} {\bibfnamefont {J.~H.}\ \bibnamefont {Cole}}, \
  and\ \bibinfo {author} {\bibfnamefont {S.~F.}\ \bibnamefont {Huelga}},\
  }\bibfield  {title} {\enquote {\bibinfo {title} {Quantum metrology subject to
  spatially correlated {M}arkovian noise: restoring the {H}eisenberg limit},}\
  }\href@noop {} {\bibfield  {journal} {\bibinfo  {journal} {New Journal of
  Physics}\ }\textbf {\bibinfo {volume} {16}},\ \bibinfo {pages} {073039}
  (\bibinfo {year} {2014})}\BibitemShut {NoStop}%
\bibitem [{\citenamefont {Sekatski}\ \emph {et~al.}(2016)\citenamefont
  {Sekatski}, \citenamefont {Skotiniotis},\ and\ \citenamefont
  {D\"{u}r}}]{SSD}%
  \BibitemOpen
  \bibfield  {author} {\bibinfo {author} {\bibfnamefont {P.}~\bibnamefont
  {Sekatski}}, \bibinfo {author} {\bibfnamefont {M.}~\bibnamefont
  {Skotiniotis}}, \ and\ \bibinfo {author} {\bibfnamefont {W.}~\bibnamefont
  {D\"{u}r}},\ }\bibfield  {title} {\enquote {\bibinfo {title} {Dynamical
  decoupling leads to improved scaling in noisy quantum metrology},}\
  }\href@noop {} {\bibfield  {journal} {\bibinfo  {journal} {New Journal of
  Physics}\ }\textbf {\bibinfo {volume} {18}},\ \bibinfo {pages} {073034}
  (\bibinfo {year} {2016})}\BibitemShut {NoStop}%
\bibitem [{\citenamefont {Sekatski}\ \emph {et~al.}(2017)\citenamefont
  {Sekatski}, \citenamefont {Skotiniotis}, \citenamefont
  {Ko{\l{}}ody{\'{n}}ski},\ and\ \citenamefont {D{\"{u}}r}}]{SSKD}%
  \BibitemOpen
  \bibfield  {author} {\bibinfo {author} {\bibfnamefont {P.}~\bibnamefont
  {Sekatski}}, \bibinfo {author} {\bibfnamefont {M.}~\bibnamefont
  {Skotiniotis}}, \bibinfo {author} {\bibfnamefont {J.}~\bibnamefont
  {Ko{\l{}}ody{\'{n}}ski}}, \ and\ \bibinfo {author} {\bibfnamefont
  {W.}~\bibnamefont {D{\"{u}}r}},\ }\bibfield  {title} {\enquote {\bibinfo
  {title} {Quantum metrology with full and fast quantum control},}\ }\href@noop
  {} {\bibfield  {journal} {\bibinfo  {journal} {{Quantum}}\ }\textbf {\bibinfo
  {volume} {1}},\ \bibinfo {pages} {27} (\bibinfo {year} {2017})}\BibitemShut
  {NoStop}%
\bibitem [{\citenamefont {Falaye}\ \emph {et~al.}(2017)\citenamefont {Falaye},
  \citenamefont {Adepoju}, \citenamefont {Aliyu}, \citenamefont {Melchor},
  \citenamefont {Liman}, \citenamefont {Oluwadare}, \citenamefont
  {González-Ramírez},\ and\ \citenamefont {Oyewumi}}]{FAAMLOGO}%
  \BibitemOpen
  \bibfield  {author} {\bibinfo {author} {\bibfnamefont {B.~J.}\ \bibnamefont
  {Falaye}}, \bibinfo {author} {\bibfnamefont {A.~G.}\ \bibnamefont {Adepoju}},
  \bibinfo {author} {\bibfnamefont {A.~S.}\ \bibnamefont {Aliyu}}, \bibinfo
  {author} {\bibfnamefont {M.~M.}\ \bibnamefont {Melchor}}, \bibinfo {author}
  {\bibfnamefont {M.~S.}\ \bibnamefont {Liman}}, \bibinfo {author}
  {\bibfnamefont {O.~J.}\ \bibnamefont {Oluwadare}}, \bibinfo {author}
  {\bibfnamefont {M.~D.}\ \bibnamefont {González-Ramírez}}, \ and\ \bibinfo
  {author} {\bibfnamefont {K.~J.}\ \bibnamefont {Oyewumi}},\ }\bibfield
  {title} {\enquote {\bibinfo {title} {Investigating quantum metrology in noisy
  channels},}\ }\href@noop {} {\bibfield  {journal} {\bibinfo  {journal}
  {Scientific Reports}\ }\textbf {\bibinfo {volume} {7}},\ \bibinfo {pages}
  {16622} (\bibinfo {year} {2017})}\BibitemShut {NoStop}%
\bibitem [{\citenamefont {Huang}\ \emph {et~al.}(2018)\citenamefont {Huang},
  \citenamefont {Macchiavello},\ and\ \citenamefont {Maccone}}]{HMM}%
  \BibitemOpen
  \bibfield  {author} {\bibinfo {author} {\bibfnamefont {Z.}~\bibnamefont
  {Huang}}, \bibinfo {author} {\bibfnamefont {C.}~\bibnamefont {Macchiavello}},
  \ and\ \bibinfo {author} {\bibfnamefont {L.}~\bibnamefont {Maccone}},\
  }\bibfield  {title} {\enquote {\bibinfo {title} {Noise-dependent optimal
  strategies for quantum metrology},}\ }\href@noop {} {\bibfield  {journal}
  {\bibinfo  {journal} {Physical Review A}\ }\textbf {\bibinfo {volume} {97}},\
  \bibinfo {pages} {032333} (\bibinfo {year} {2018})}\BibitemShut {NoStop}%
\bibitem [{\citenamefont {Zhou}\ \emph {et~al.}(2018)\citenamefont {Zhou},
  \citenamefont {Zhang}, \citenamefont {Preskill},\ and\ \citenamefont
  {Jiang}}]{ZZPJ}%
  \BibitemOpen
  \bibfield  {author} {\bibinfo {author} {\bibfnamefont {S.}~\bibnamefont
  {Zhou}}, \bibinfo {author} {\bibfnamefont {M.}~\bibnamefont {Zhang}},
  \bibinfo {author} {\bibfnamefont {J.}~\bibnamefont {Preskill}}, \ and\
  \bibinfo {author} {\bibfnamefont {L.}~\bibnamefont {Jiang}},\ }\bibfield
  {title} {\enquote {\bibinfo {title} {Achieving the {H}eisenberg limit in
  quantum metrology using quantum error correction},}\ }\href@noop {}
  {\bibfield  {journal} {\bibinfo  {journal} {Nature Communications}\ }\textbf
  {\bibinfo {volume} {9}},\ \bibinfo {pages} {78} (\bibinfo {year}
  {2018})}\BibitemShut {NoStop}%
\bibitem [{\citenamefont {Chen}\ \emph {et~al.}(2018)\citenamefont {Chen},
  \citenamefont {Aharon}, \citenamefont {Sun}, \citenamefont {Zhang},
  \citenamefont {Zhang}, \citenamefont {He}, \citenamefont {Tang},
  \citenamefont {Xu}, \citenamefont {Kedem}, \citenamefont {Li},\ and\
  \citenamefont {Guo}}]{CASZZHTXKLG}%
  \BibitemOpen
  \bibfield  {author} {\bibinfo {author} {\bibfnamefont {G.}~\bibnamefont
  {Chen}}, \bibinfo {author} {\bibfnamefont {N.}~\bibnamefont {Aharon}},
  \bibinfo {author} {\bibfnamefont {Y-N.}\ \bibnamefont {Sun}}, \bibinfo
  {author} {\bibfnamefont {Z-H.}\ \bibnamefont {Zhang}}, \bibinfo {author}
  {\bibfnamefont {W-H.}\ \bibnamefont {Zhang}}, \bibinfo {author}
  {\bibfnamefont {D-Y.}\ \bibnamefont {He}}, \bibinfo {author} {\bibfnamefont
  {J-S.}\ \bibnamefont {Tang}}, \bibinfo {author} {\bibfnamefont {X-Y.}\
  \bibnamefont {Xu}}, \bibinfo {author} {\bibfnamefont {Y.}~\bibnamefont
  {Kedem}}, \bibinfo {author} {\bibfnamefont {C-F.}\ \bibnamefont {Li}}, \ and\
  \bibinfo {author} {\bibfnamefont {G-C.}\ \bibnamefont {Guo}},\ }\bibfield
  {title} {\enquote {\bibinfo {title} {{H}eisenberg-scaling measurement of the
  single-photon {K}err non-linearity using mixed states},}\ }\href@noop {}
  {\bibfield  {journal} {\bibinfo  {journal} {Nature Communications}\ }\textbf
  {\bibinfo {volume} {9}},\ \bibinfo {pages} {93} (\bibinfo {year}
  {2018})}\BibitemShut {NoStop}%
\bibitem [{\citenamefont {Demkowicz-Dobrzański}\ \emph
  {et~al.}(2015)\citenamefont {Demkowicz-Dobrzański}, \citenamefont
  {Jarzyna},\ and\ \citenamefont {Kołodyński}}]{DDJK}%
  \BibitemOpen
  \bibfield  {author} {\bibinfo {author} {\bibfnamefont {R.}~\bibnamefont
  {Demkowicz-Dobrzański}}, \bibinfo {author} {\bibfnamefont {M.}~\bibnamefont
  {Jarzyna}}, \ and\ \bibinfo {author} {\bibfnamefont {J.}~\bibnamefont
  {Kołodyński}},\ }\bibfield  {title} {\enquote {\bibinfo {title} {Quantum
  limits in optical interferometry},}\ }\href@noop {} {\bibfield  {journal}
  {\bibinfo  {journal} {Progress in Optics}\ }\textbf {\bibinfo {volume}
  {60}},\ \bibinfo {pages} {345--435} (\bibinfo {year} {2015})}\BibitemShut
  {NoStop}%
\bibitem [{\citenamefont {Benatti}\ \emph {et~al.}(2014)\citenamefont
  {Benatti}, \citenamefont {Alipour},\ and\ \citenamefont {Rezakhani}}]{BAR}%
  \BibitemOpen
  \bibfield  {author} {\bibinfo {author} {\bibfnamefont {F.}~\bibnamefont
  {Benatti}}, \bibinfo {author} {\bibfnamefont {S.}~\bibnamefont {Alipour}}, \
  and\ \bibinfo {author} {\bibfnamefont {A.~T.}\ \bibnamefont {Rezakhani}},\
  }\bibfield  {title} {\enquote {\bibinfo {title} {Dissipative quantum
  metrology in manybody systems of identical particles},}\ }\href@noop {}
  {\bibfield  {journal} {\bibinfo  {journal} {New Journal of Physics}\ }\textbf
  {\bibinfo {volume} {16}},\ \bibinfo {pages} {015023} (\bibinfo {year}
  {2014})}\BibitemShut {NoStop}%
\bibitem [{\citenamefont {Beau}\ and\ \citenamefont {del Campo}(2017)}]{BdC}%
  \BibitemOpen
  \bibfield  {author} {\bibinfo {author} {\bibfnamefont {M.}~\bibnamefont
  {Beau}}\ and\ \bibinfo {author} {\bibfnamefont {A.}~\bibnamefont {del
  Campo}},\ }\bibfield  {title} {\enquote {\bibinfo {title} {Nonlinear quantum
  metrology of many-body open systems},}\ }\href@noop {} {\bibfield  {journal}
  {\bibinfo  {journal} {Physical Review Letters}\ }\textbf {\bibinfo {volume}
  {119}},\ \bibinfo {pages} {010403} (\bibinfo {year} {2017})}\BibitemShut
  {NoStop}%
\bibitem [{\citenamefont {Alipour}\ \emph {et~al.}(2014)\citenamefont
  {Alipour}, \citenamefont {Mehboudi},\ and\ \citenamefont {Rezakhani}}]{AMR}%
  \BibitemOpen
  \bibfield  {author} {\bibinfo {author} {\bibfnamefont {S.}~\bibnamefont
  {Alipour}}, \bibinfo {author} {\bibfnamefont {M.}~\bibnamefont {Mehboudi}}, \
  and\ \bibinfo {author} {\bibfnamefont {A.~T.}\ \bibnamefont {Rezakhani}},\
  }\bibfield  {title} {\enquote {\bibinfo {title} {Quantum metrology in open
  systems: Dissipative {C}ram\'{e}r-{R}ao bound},}\ }\href@noop {} {\bibfield
  {journal} {\bibinfo  {journal} {Physical Review Letters}\ }\textbf {\bibinfo
  {volume} {112}},\ \bibinfo {pages} {120405} (\bibinfo {year}
  {2014})}\BibitemShut {NoStop}%
\bibitem [{\citenamefont {Yuan}\ and\ \citenamefont {Fung}(2017)}]{YF}%
  \BibitemOpen
  \bibfield  {author} {\bibinfo {author} {\bibfnamefont {H.}~\bibnamefont
  {Yuan}}\ and\ \bibinfo {author} {\bibfnamefont {C-H.~F.}\ \bibnamefont
  {Fung}},\ }\bibfield  {title} {\enquote {\bibinfo {title} {Quantum parameter
  estimation with general dynamics},}\ }\href@noop {} {\bibfield  {journal}
  {\bibinfo  {journal} {npj Quantum Information}\ }\textbf {\bibinfo {volume}
  {3}},\ \bibinfo {pages} {14} (\bibinfo {year} {2017})}\BibitemShut {NoStop}%
\bibitem [{\citenamefont {Tsang}(2013)}]{MT}%
  \BibitemOpen
  \bibfield  {author} {\bibinfo {author} {\bibfnamefont {M.}~\bibnamefont
  {Tsang}},\ }\bibfield  {title} {\enquote {\bibinfo {title} {Quantum metrology
  with open dynamical systems},}\ }\href@noop {} {\bibfield  {journal}
  {\bibinfo  {journal} {New Journal of Physics}\ }\textbf {\bibinfo {volume}
  {15}},\ \bibinfo {pages} {073005} (\bibinfo {year} {2013})}\BibitemShut
  {NoStop}%
\bibitem [{\citenamefont {Albarelli}\ \emph {et~al.}(2018)\citenamefont
  {Albarelli}, \citenamefont {Rossi}, \citenamefont {Tamascelli},\ and\
  \citenamefont {Genoni}}]{ARTG}%
  \BibitemOpen
  \bibfield  {author} {\bibinfo {author} {\bibfnamefont {F.}~\bibnamefont
  {Albarelli}}, \bibinfo {author} {\bibfnamefont {M.~A.~C.}\ \bibnamefont
  {Rossi}}, \bibinfo {author} {\bibfnamefont {D.}~\bibnamefont {Tamascelli}}, \
  and\ \bibinfo {author} {\bibfnamefont {M.~G.}\ \bibnamefont {Genoni}},\
  }\bibfield  {title} {\enquote {\bibinfo {title} {Restoring {H}eisenberg
  scaling in noisy quantum metrology by monitoring the environment},}\
  }\href@noop {} {\bibfield  {journal} {\bibinfo  {journal} {Quantum}\ }\textbf
  {\bibinfo {volume} {2}},\ \bibinfo {pages} {110} (\bibinfo {year}
  {2018})}\BibitemShut {NoStop}%
\bibitem [{\citenamefont {Haase}\ \emph {et~al.}(2018)\citenamefont {Haase},
  \citenamefont {Smirne}, \citenamefont {Kołodyński}, \citenamefont
  {Demkowicz-Dobrzański},\ and\ \citenamefont {Huelga}}]{HSKDDH}%
  \BibitemOpen
  \bibfield  {author} {\bibinfo {author} {\bibfnamefont {J.~F.}\ \bibnamefont
  {Haase}}, \bibinfo {author} {\bibfnamefont {A.}~\bibnamefont {Smirne}},
  \bibinfo {author} {\bibfnamefont {J.}~\bibnamefont {Kołodyński}}, \bibinfo
  {author} {\bibfnamefont {R.}~\bibnamefont {Demkowicz-Dobrzański}}, \ and\
  \bibinfo {author} {\bibfnamefont {S.~F.}\ \bibnamefont {Huelga}},\
  }\href@noop {} {\enquote {\bibinfo {title} {Precision limits in quantum
  metrology with open quantum systems},}\ } (\bibinfo {year} {2018}),\ \bibinfo
  {note} {arXiv:1807.11882}\BibitemShut {NoStop}%
\bibitem [{\citenamefont {Tsang}\ \emph {et~al.}(2011)\citenamefont {Tsang},
  \citenamefont {Wiseman},\ and\ \citenamefont {Caves}}]{TWC}%
  \BibitemOpen
  \bibfield  {author} {\bibinfo {author} {\bibfnamefont {M.}~\bibnamefont
  {Tsang}}, \bibinfo {author} {\bibfnamefont {H.~M.}\ \bibnamefont {Wiseman}},
  \ and\ \bibinfo {author} {\bibfnamefont {C.~M.}\ \bibnamefont {Caves}},\
  }\bibfield  {title} {\enquote {\bibinfo {title} {Fundamental quantum limit to
  waveform estimation},}\ }\href@noop {} {\bibfield  {journal} {\bibinfo
  {journal} {Physical Review Letters}\ }\textbf {\bibinfo {volume} {106}},\
  \bibinfo {pages} {090401} (\bibinfo {year} {2011})}\BibitemShut {NoStop}%
\bibitem [{\citenamefont {Fujiwara}\ and\ \citenamefont {Nagaoka}(1995)}]{FN}%
  \BibitemOpen
  \bibfield  {author} {\bibinfo {author} {\bibfnamefont {A.}~\bibnamefont
  {Fujiwara}}\ and\ \bibinfo {author} {\bibfnamefont {H.}~\bibnamefont
  {Nagaoka}},\ }\bibfield  {title} {\enquote {\bibinfo {title} {Quantum
  {F}isher metric and estimation for pure state models},}\ }\href@noop {}
  {\bibfield  {journal} {\bibinfo  {journal} {Physics Letters A}\ }\textbf
  {\bibinfo {volume} {201}},\ \bibinfo {pages} {119--124} (\bibinfo {year}
  {1995})}\BibitemShut {NoStop}%
\bibitem [{\citenamefont {Fujiwara}\ and\ \citenamefont {Imai}(2008)}]{FI}%
  \BibitemOpen
  \bibfield  {author} {\bibinfo {author} {\bibfnamefont {A.}~\bibnamefont
  {Fujiwara}}\ and\ \bibinfo {author} {\bibfnamefont {H.}~\bibnamefont
  {Imai}},\ }\bibfield  {title} {\enquote {\bibinfo {title} {A fibre bundle
  over manifolds of quantum channels and its application to quantum
  statistics},}\ }\href@noop {} {\bibfield  {journal} {\bibinfo  {journal}
  {Journal of Physics A: Mathematical and Theoretical}\ }\textbf {\bibinfo
  {volume} {41}},\ \bibinfo {pages} {255304} (\bibinfo {year}
  {2008})}\BibitemShut {NoStop}%
\bibitem [{\citenamefont {Braunstein}(1992)}]{SLB}%
  \BibitemOpen
  \bibfield  {author} {\bibinfo {author} {\bibfnamefont {S.~L.}\ \bibnamefont
  {Braunstein}},\ }\bibfield  {title} {\enquote {\bibinfo {title} {How large a
  sample is needed for the maximum likelihood estimator to be approximately
  {G}aussian?}}\ }\href@noop {} {\bibfield  {journal} {\bibinfo  {journal}
  {Journal of Physics A: Mathematical and General}\ }\textbf {\bibinfo {volume}
  {25}},\ \bibinfo {pages} {3813} (\bibinfo {year} {1992})}\BibitemShut
  {NoStop}%
\bibitem [{\citenamefont {Matsumoto}(2002)}]{KM}%
  \BibitemOpen
  \bibfield  {author} {\bibinfo {author} {\bibfnamefont {K.}~\bibnamefont
  {Matsumoto}},\ }\bibfield  {title} {\enquote {\bibinfo {title} {A new
  approach to the {C}ram\'{e}r-{R}ao-type bound of the pure-state model},}\
  }\href@noop {} {\bibfield  {journal} {\bibinfo  {journal} {Journal of Physics
  A: Mathematical and General}\ }\textbf {\bibinfo {volume} {35}},\ \bibinfo
  {pages} {3111} (\bibinfo {year} {2002})}\BibitemShut {NoStop}%
\bibitem [{\citenamefont {Ragy}\ \emph {et~al.}(2016)\citenamefont {Ragy},
  \citenamefont {Jarzyna},\ and\ \citenamefont
  {Demkowicz-Dobrza\'{n}ski}}]{RJD}%
  \BibitemOpen
  \bibfield  {author} {\bibinfo {author} {\bibfnamefont {S.}~\bibnamefont
  {Ragy}}, \bibinfo {author} {\bibfnamefont {M.}~\bibnamefont {Jarzyna}}, \
  and\ \bibinfo {author} {\bibfnamefont {R.}~\bibnamefont
  {Demkowicz-Dobrza\'{n}ski}},\ }\bibfield  {title} {\enquote {\bibinfo {title}
  {Compatibility in multiparameter quantum metrology},}\ }\href@noop {}
  {\bibfield  {journal} {\bibinfo  {journal} {Physical Review A}\ }\textbf
  {\bibinfo {volume} {94}},\ \bibinfo {pages} {052108} (\bibinfo {year}
  {2016})}\BibitemShut {NoStop}%
\bibitem [{\citenamefont {Wilcox}(1967)}]{RMW}%
  \BibitemOpen
  \bibfield  {author} {\bibinfo {author} {\bibfnamefont {R.~M.}\ \bibnamefont
  {Wilcox}},\ }\bibfield  {title} {\enquote {\bibinfo {title} {Exponential
  operators and parameter differentiation in quantum physics},}\ }\href@noop {}
  {\bibfield  {journal} {\bibinfo  {journal} {Journal of Mathematical Physics}\
  }\textbf {\bibinfo {volume} {8}},\ \bibinfo {pages} {962} (\bibinfo {year}
  {1967})}\BibitemShut {NoStop}%
\bibitem [{\citenamefont {Hyllus}\ \emph {et~al.}(2010)\citenamefont {Hyllus},
  \citenamefont {G\"{u}hne},\ and\ \citenamefont {Smerzi}}]{HGS}%
  \BibitemOpen
  \bibfield  {author} {\bibinfo {author} {\bibfnamefont {P.}~\bibnamefont
  {Hyllus}}, \bibinfo {author} {\bibfnamefont {O.}~\bibnamefont {G\"{u}hne}}, \
  and\ \bibinfo {author} {\bibfnamefont {A.}~\bibnamefont {Smerzi}},\
  }\bibfield  {title} {\enquote {\bibinfo {title} {Not all pure entangled
  states are useful for sub-shot-noise interferometry},}\ }\href@noop {}
  {\bibfield  {journal} {\bibinfo  {journal} {Physical Review A}\ }\textbf
  {\bibinfo {volume} {82}},\ \bibinfo {pages} {012337} (\bibinfo {year}
  {2010})}\BibitemShut {NoStop}%
\bibitem [{\citenamefont {Jarzyna}\ and\ \citenamefont
  {Demkowicz-Dobrza\'{n}ski}(2015)}]{JD}%
  \BibitemOpen
  \bibfield  {author} {\bibinfo {author} {\bibfnamefont {M.}~\bibnamefont
  {Jarzyna}}\ and\ \bibinfo {author} {\bibfnamefont {R.}~\bibnamefont
  {Demkowicz-Dobrza\'{n}ski}},\ }\bibfield  {title} {\enquote {\bibinfo {title}
  {True precision limits in quantum metrology},}\ }\href@noop {} {\bibfield
  {journal} {\bibinfo  {journal} {New Journal of Physics}\ }\textbf {\bibinfo
  {volume} {17}},\ \bibinfo {pages} {013010} (\bibinfo {year}
  {2015})}\BibitemShut {NoStop}%
\bibitem [{\citenamefont {Piani}\ \emph {et~al.}(2011)\citenamefont {Piani},
  \citenamefont {Gharibian}, \citenamefont {Adesso}, \citenamefont
  {Calsamiglia}, \citenamefont {Horodecki},\ and\ \citenamefont
  {Winter}}]{PGACHW}%
  \BibitemOpen
  \bibfield  {author} {\bibinfo {author} {\bibfnamefont {M.}~\bibnamefont
  {Piani}}, \bibinfo {author} {\bibfnamefont {S.}~\bibnamefont {Gharibian}},
  \bibinfo {author} {\bibfnamefont {G.}~\bibnamefont {Adesso}}, \bibinfo
  {author} {\bibfnamefont {J.}~\bibnamefont {Calsamiglia}}, \bibinfo {author}
  {\bibfnamefont {P.}~\bibnamefont {Horodecki}}, \ and\ \bibinfo {author}
  {\bibfnamefont {A.}~\bibnamefont {Winter}},\ }\bibfield  {title} {\enquote
  {\bibinfo {title} {All nonclassical correlations can be activated into
  distillable entanglement},}\ }\href@noop {} {\bibfield  {journal} {\bibinfo
  {journal} {Physical Review Letters}\ }\textbf {\bibinfo {volume} {106}},\
  \bibinfo {pages} {220403} (\bibinfo {year} {2011})}\BibitemShut {NoStop}%
\bibitem [{\citenamefont {Horodecki}\ \emph {et~al.}(2009)\citenamefont
  {Horodecki}, \citenamefont {Horodecki}, \citenamefont {Horodecki},\ and\
  \citenamefont {Horodecki}}]{HHHH}%
  \BibitemOpen
  \bibfield  {author} {\bibinfo {author} {\bibfnamefont {R.}~\bibnamefont
  {Horodecki}}, \bibinfo {author} {\bibfnamefont {P.}~\bibnamefont
  {Horodecki}}, \bibinfo {author} {\bibfnamefont {M.}~\bibnamefont
  {Horodecki}}, \ and\ \bibinfo {author} {\bibfnamefont {K.}~\bibnamefont
  {Horodecki}},\ }\bibfield  {title} {\enquote {\bibinfo {title} {Quantum
  entanglement},}\ }\href@noop {} {\bibfield  {journal} {\bibinfo  {journal}
  {Reviews of Modern Physics}\ }\textbf {\bibinfo {volume} {81}},\ \bibinfo
  {pages} {865} (\bibinfo {year} {2009})}\BibitemShut {NoStop}%
\bibitem [{\citenamefont {Braun}(2002)}]{DB}%
  \BibitemOpen
  \bibfield  {author} {\bibinfo {author} {\bibfnamefont {D.}~\bibnamefont
  {Braun}},\ }\bibfield  {title} {\enquote {\bibinfo {title} {Creation of
  entanglement by interaction with a common heat bath},}\ }\href@noop {}
  {\bibfield  {journal} {\bibinfo  {journal} {Physical Review Letters}\
  }\textbf {\bibinfo {volume} {89}},\ \bibinfo {pages} {277901} (\bibinfo
  {year} {2002})}\BibitemShut {NoStop}%
\bibitem [{\citenamefont {Benatti}\ \emph {et~al.}(2003)\citenamefont
  {Benatti}, \citenamefont {Floreanini},\ and\ \citenamefont {Piani}}]{BFP}%
  \BibitemOpen
  \bibfield  {author} {\bibinfo {author} {\bibfnamefont {F.}~\bibnamefont
  {Benatti}}, \bibinfo {author} {\bibfnamefont {R.}~\bibnamefont {Floreanini}},
  \ and\ \bibinfo {author} {\bibfnamefont {M.}~\bibnamefont {Piani}},\
  }\bibfield  {title} {\enquote {\bibinfo {title} {Environment induced
  entanglement in {M}arkovian dissipative dynamics},}\ }\href@noop {}
  {\bibfield  {journal} {\bibinfo  {journal} {Physical Review Letters}\
  }\textbf {\bibinfo {volume} {91}},\ \bibinfo {pages} {070402} (\bibinfo
  {year} {2003})}\BibitemShut {NoStop}%
\bibitem [{\citenamefont {Streltsov}\ \emph {et~al.}(2011)\citenamefont
  {Streltsov}, \citenamefont {Kampermann},\ and\ \citenamefont {Bruß}}]{SKB}%
  \BibitemOpen
  \bibfield  {author} {\bibinfo {author} {\bibfnamefont {A.}~\bibnamefont
  {Streltsov}}, \bibinfo {author} {\bibfnamefont {H.}~\bibnamefont
  {Kampermann}}, \ and\ \bibinfo {author} {\bibfnamefont {D.}~\bibnamefont
  {Bruß}},\ }\bibfield  {title} {\enquote {\bibinfo {title} {Behavior of
  quantum correlations under local noise},}\ }\href@noop {} {\bibfield
  {journal} {\bibinfo  {journal} {Physical Review Letters}\ }\textbf {\bibinfo
  {volume} {107}},\ \bibinfo {pages} {170502} (\bibinfo {year}
  {2011})}\BibitemShut {NoStop}%
\bibitem [{\citenamefont {Orieux}\ \emph {et~al.}(2015)\citenamefont {Orieux},
  \citenamefont {Ciampini}, \citenamefont {Mataloni}, \citenamefont {Bruß},
  \citenamefont {Rossi},\ and\ \citenamefont {Macchiavello}}]{OCMBRM}%
  \BibitemOpen
  \bibfield  {author} {\bibinfo {author} {\bibfnamefont {A.}~\bibnamefont
  {Orieux}}, \bibinfo {author} {\bibfnamefont {M.~A.}\ \bibnamefont
  {Ciampini}}, \bibinfo {author} {\bibfnamefont {P.}~\bibnamefont {Mataloni}},
  \bibinfo {author} {\bibfnamefont {D.}~\bibnamefont {Bruß}}, \bibinfo
  {author} {\bibfnamefont {M.}~\bibnamefont {Rossi}}, \ and\ \bibinfo {author}
  {\bibfnamefont {C.}~\bibnamefont {Macchiavello}},\ }\bibfield  {title}
  {\enquote {\bibinfo {title} {Experimental generation of robust entanglement
  from classical correlations via local dissipation},}\ }\href@noop {}
  {\bibfield  {journal} {\bibinfo  {journal} {Physical Review Letters}\
  }\textbf {\bibinfo {volume} {115}},\ \bibinfo {pages} {160503} (\bibinfo
  {year} {2015})}\BibitemShut {NoStop}%
\bibitem [{\citenamefont {Bollinger}\ \emph {et~al.}(1996)\citenamefont
  {Bollinger}, \citenamefont {Itano}, \citenamefont {Wineland},\ and\
  \citenamefont {Heinzen}}]{BIWH}%
  \BibitemOpen
  \bibfield  {author} {\bibinfo {author} {\bibfnamefont {J.~J.}\ \bibnamefont
  {Bollinger}}, \bibinfo {author} {\bibfnamefont {W.~M.}\ \bibnamefont
  {Itano}}, \bibinfo {author} {\bibfnamefont {D.~J.}\ \bibnamefont {Wineland}},
  \ and\ \bibinfo {author} {\bibfnamefont {D.~J.}\ \bibnamefont {Heinzen}},\
  }\bibfield  {title} {\enquote {\bibinfo {title} {Optimal frequency
  measurements with maximally correlated states},}\ }\href@noop {} {\bibfield
  {journal} {\bibinfo  {journal} {Physical Review A}\ }\textbf {\bibinfo
  {volume} {54}},\ \bibinfo {pages} {R4649} (\bibinfo {year}
  {1996})}\BibitemShut {NoStop}%
\bibitem [{\citenamefont {Shaji}\ and\ \citenamefont {Caves}(2007)}]{SC}%
  \BibitemOpen
  \bibfield  {author} {\bibinfo {author} {\bibfnamefont {A.}~\bibnamefont
  {Shaji}}\ and\ \bibinfo {author} {\bibfnamefont {C.~M.}\ \bibnamefont
  {Caves}},\ }\bibfield  {title} {\enquote {\bibinfo {title} {Qubit metrology
  and decoherence},}\ }\href@noop {} {\bibfield  {journal} {\bibinfo  {journal}
  {Physical Review A}\ }\textbf {\bibinfo {volume} {76}},\ \bibinfo {pages}
  {032111} (\bibinfo {year} {2007})}\BibitemShut {NoStop}%
\bibitem [{\citenamefont {Huver}\ \emph {et~al.}(2008)\citenamefont {Huver},
  \citenamefont {Wildfeuer},\ and\ \citenamefont {Dowling}}]{HWD}%
  \BibitemOpen
  \bibfield  {author} {\bibinfo {author} {\bibfnamefont {S.~D.}\ \bibnamefont
  {Huver}}, \bibinfo {author} {\bibfnamefont {C.~F.}\ \bibnamefont
  {Wildfeuer}}, \ and\ \bibinfo {author} {\bibfnamefont {J.~P.}\ \bibnamefont
  {Dowling}},\ }\bibfield  {title} {\enquote {\bibinfo {title} {Entangled
  {F}ock states for robust quantum optical metrology, imaging, and sensing},}\
  }\href@noop {} {\bibfield  {journal} {\bibinfo  {journal} {Physical Review
  A}\ }\textbf {\bibinfo {volume} {78}},\ \bibinfo {pages} {063828} (\bibinfo
  {year} {2008})}\BibitemShut {NoStop}%
\bibitem [{\citenamefont {Modi}\ \emph {et~al.}(2011)\citenamefont {Modi},
  \citenamefont {Cable}, \citenamefont {Williamson},\ and\ \citenamefont
  {Vedral}}]{MCWV}%
  \BibitemOpen
  \bibfield  {author} {\bibinfo {author} {\bibfnamefont {K.}~\bibnamefont
  {Modi}}, \bibinfo {author} {\bibfnamefont {H.}~\bibnamefont {Cable}},
  \bibinfo {author} {\bibfnamefont {M.}~\bibnamefont {Williamson}}, \ and\
  \bibinfo {author} {\bibfnamefont {V.}~\bibnamefont {Vedral}},\ }\bibfield
  {title} {\enquote {\bibinfo {title} {Quantum correlations in mixed-state
  metrology},}\ }\href@noop {} {\bibfield  {journal} {\bibinfo  {journal}
  {Physical Review X}\ }\textbf {\bibinfo {volume} {1}},\ \bibinfo {pages}
  {021022} (\bibinfo {year} {2011})}\BibitemShut {NoStop}%
\bibitem [{\citenamefont {Luis}(2007)}]{AL}%
  \BibitemOpen
  \bibfield  {author} {\bibinfo {author} {\bibfnamefont {A.}~\bibnamefont
  {Luis}},\ }\bibfield  {title} {\enquote {\bibinfo {title} {Quantum limits,
  nonseparable transformations, and nonlinear optics},}\ }\href@noop {}
  {\bibfield  {journal} {\bibinfo  {journal} {Physical Review A}\ }\textbf
  {\bibinfo {volume} {76}},\ \bibinfo {pages} {035801} (\bibinfo {year}
  {2007})}\BibitemShut {NoStop}%
\bibitem [{\citenamefont {Beltr\'{a}n}\ and\ \citenamefont {Luis}(2005)}]{BL}%
  \BibitemOpen
  \bibfield  {author} {\bibinfo {author} {\bibfnamefont {J.}~\bibnamefont
  {Beltr\'{a}n}}\ and\ \bibinfo {author} {\bibfnamefont {A.}~\bibnamefont
  {Luis}},\ }\bibfield  {title} {\enquote {\bibinfo {title} {Breaking the
  {H}eisenberg limit with inefficient detectors},}\ }\href@noop {} {\bibfield
  {journal} {\bibinfo  {journal} {Physical Review A}\ }\textbf {\bibinfo
  {volume} {72}},\ \bibinfo {pages} {045801} (\bibinfo {year}
  {2005})}\BibitemShut {NoStop}%
\bibitem [{\citenamefont {Roy}\ and\ \citenamefont {Braunstein}(2008)}]{RB}%
  \BibitemOpen
  \bibfield  {author} {\bibinfo {author} {\bibfnamefont {S.~M.}\ \bibnamefont
  {Roy}}\ and\ \bibinfo {author} {\bibfnamefont {S.~L.}\ \bibnamefont
  {Braunstein}},\ }\bibfield  {title} {\enquote {\bibinfo {title}
  {Exponentially enhanced quantum metrology},}\ }\href@noop {} {\bibfield
  {journal} {\bibinfo  {journal} {Physical Review Letters}\ }\textbf {\bibinfo
  {volume} {100}},\ \bibinfo {pages} {220501} (\bibinfo {year}
  {2008})}\BibitemShut {NoStop}%
\bibitem [{\citenamefont {Rivas}\ and\ \citenamefont {Luis}(2012)}]{RL}%
  \BibitemOpen
  \bibfield  {author} {\bibinfo {author} {\bibfnamefont {A.}~\bibnamefont
  {Rivas}}\ and\ \bibinfo {author} {\bibfnamefont {A.}~\bibnamefont {Luis}},\
  }\bibfield  {title} {\enquote {\bibinfo {title} {Sub-{H}eisenberg estimation
  of non-random phase shifts},}\ }\href@noop {} {\bibfield  {journal} {\bibinfo
   {journal} {New Journal of Physics}\ }\textbf {\bibinfo {volume} {14}},\
  \bibinfo {pages} {093052} (\bibinfo {year} {2012})}\BibitemShut {NoStop}%
\bibitem [{\citenamefont {Boixo}\ \emph {et~al.}(2007)\citenamefont {Boixo},
  \citenamefont {Flammia}, \citenamefont {Caves},\ and\ \citenamefont
  {Geremia}}]{BFCG}%
  \BibitemOpen
  \bibfield  {author} {\bibinfo {author} {\bibfnamefont {S.}~\bibnamefont
  {Boixo}}, \bibinfo {author} {\bibfnamefont {S.~T.}\ \bibnamefont {Flammia}},
  \bibinfo {author} {\bibfnamefont {C.~M.}\ \bibnamefont {Caves}}, \ and\
  \bibinfo {author} {\bibfnamefont {J.~M.}\ \bibnamefont {Geremia}},\
  }\bibfield  {title} {\enquote {\bibinfo {title} {Generalized limits for
  single-parameter quantum estimation},}\ }\href@noop {} {\bibfield  {journal}
  {\bibinfo  {journal} {Physical Review Letters}\ }\textbf {\bibinfo {volume}
  {98}},\ \bibinfo {pages} {090401} (\bibinfo {year} {2007})}\BibitemShut
  {NoStop}%
\bibitem [{\citenamefont {Woolley}\ \emph {et~al.}(2008)\citenamefont
  {Woolley}, \citenamefont {Milburn},\ and\ \citenamefont {Caves}}]{WMC}%
  \BibitemOpen
  \bibfield  {author} {\bibinfo {author} {\bibfnamefont {M.~J.}\ \bibnamefont
  {Woolley}}, \bibinfo {author} {\bibfnamefont {G.~J.}\ \bibnamefont
  {Milburn}}, \ and\ \bibinfo {author} {\bibfnamefont {C.~M.}\ \bibnamefont
  {Caves}},\ }\bibfield  {title} {\enquote {\bibinfo {title} {Nonlinear quantum
  metrology using coupled nanomechanical resonators},}\ }\href@noop {}
  {\bibfield  {journal} {\bibinfo  {journal} {New Journal of Physics}\ }\textbf
  {\bibinfo {volume} {10}},\ \bibinfo {pages} {125018} (\bibinfo {year}
  {2008})}\BibitemShut {NoStop}%
\bibitem [{\citenamefont {Anisimov}\ \emph {et~al.}(2010)\citenamefont
  {Anisimov}, \citenamefont {Raterman}, \citenamefont {Chiruvelli},
  \citenamefont {Plick}, \citenamefont {Huver}, \citenamefont {Lee},\ and\
  \citenamefont {Dowling}}]{ARCPHLD}%
  \BibitemOpen
  \bibfield  {author} {\bibinfo {author} {\bibfnamefont {P.~M.}\ \bibnamefont
  {Anisimov}}, \bibinfo {author} {\bibfnamefont {G.~M.}\ \bibnamefont
  {Raterman}}, \bibinfo {author} {\bibfnamefont {A.}~\bibnamefont
  {Chiruvelli}}, \bibinfo {author} {\bibfnamefont {W.~N.}\ \bibnamefont
  {Plick}}, \bibinfo {author} {\bibfnamefont {S.~D.}\ \bibnamefont {Huver}},
  \bibinfo {author} {\bibfnamefont {H.}~\bibnamefont {Lee}}, \ and\ \bibinfo
  {author} {\bibfnamefont {J.~P.}\ \bibnamefont {Dowling}},\ }\bibfield
  {title} {\enquote {\bibinfo {title} {Quantum metrology with two-mode squeezed
  vacuum: Parity detection beats the {H}eisenberg limit},}\ }\href@noop {}
  {\bibfield  {journal} {\bibinfo  {journal} {Physical Review Letters}\
  }\textbf {\bibinfo {volume} {104}},\ \bibinfo {pages} {103602} (\bibinfo
  {year} {2010})}\BibitemShut {NoStop}%
\bibitem [{\citenamefont {Boixo}\ \emph
  {et~al.}(2008{\natexlab{a}})\citenamefont {Boixo}, \citenamefont {Datta},
  \citenamefont {Davis}, \citenamefont {Flammia}, \citenamefont {Shaji},\ and\
  \citenamefont {Caves}}]{BDDFSC}%
  \BibitemOpen
  \bibfield  {author} {\bibinfo {author} {\bibfnamefont {S.}~\bibnamefont
  {Boixo}}, \bibinfo {author} {\bibfnamefont {A.}~\bibnamefont {Datta}},
  \bibinfo {author} {\bibfnamefont {M.~J.}\ \bibnamefont {Davis}}, \bibinfo
  {author} {\bibfnamefont {S.~T.}\ \bibnamefont {Flammia}}, \bibinfo {author}
  {\bibfnamefont {A.}~\bibnamefont {Shaji}}, \ and\ \bibinfo {author}
  {\bibfnamefont {C.~M.}\ \bibnamefont {Caves}},\ }\bibfield  {title} {\enquote
  {\bibinfo {title} {Quantum metrology: Dynamics versus entanglement},}\
  }\href@noop {} {\bibfield  {journal} {\bibinfo  {journal} {Physical Review
  Letters}\ }\textbf {\bibinfo {volume} {101}},\ \bibinfo {pages} {040403}
  (\bibinfo {year} {2008}{\natexlab{a}})}\BibitemShut {NoStop}%
\bibitem [{\citenamefont {Boixo}\ \emph
  {et~al.}(2008{\natexlab{b}})\citenamefont {Boixo}, \citenamefont {Datta},
  \citenamefont {Flammia}, \citenamefont {Shaji}, \citenamefont {Bagan},\ and\
  \citenamefont {Caves}}]{BDFSBC}%
  \BibitemOpen
  \bibfield  {author} {\bibinfo {author} {\bibfnamefont {S.}~\bibnamefont
  {Boixo}}, \bibinfo {author} {\bibfnamefont {A.}~\bibnamefont {Datta}},
  \bibinfo {author} {\bibfnamefont {S.~T.}\ \bibnamefont {Flammia}}, \bibinfo
  {author} {\bibfnamefont {A.}~\bibnamefont {Shaji}}, \bibinfo {author}
  {\bibfnamefont {E.}~\bibnamefont {Bagan}}, \ and\ \bibinfo {author}
  {\bibfnamefont {C.~M.}\ \bibnamefont {Caves}},\ }\bibfield  {title} {\enquote
  {\bibinfo {title} {Quantum-limited metrology with product states},}\
  }\href@noop {} {\bibfield  {journal} {\bibinfo  {journal} {Physical Review
  A}\ }\textbf {\bibinfo {volume} {77}},\ \bibinfo {pages} {012317} (\bibinfo
  {year} {2008}{\natexlab{b}})}\BibitemShut {NoStop}%
\bibitem [{\citenamefont {Choi}\ and\ \citenamefont {Sundaram}(2008)}]{CS}%
  \BibitemOpen
  \bibfield  {author} {\bibinfo {author} {\bibfnamefont {S.}~\bibnamefont
  {Choi}}\ and\ \bibinfo {author} {\bibfnamefont {B.}~\bibnamefont
  {Sundaram}},\ }\bibfield  {title} {\enquote {\bibinfo {title}
  {Bose-{E}instein condensate as a nonlinear {R}amsey interferometer operating
  beyond the {H}eisenberg limit},}\ }\href@noop {} {\bibfield  {journal}
  {\bibinfo  {journal} {Physical Review A}\ }\textbf {\bibinfo {volume} {77}},\
  \bibinfo {pages} {053613} (\bibinfo {year} {2008})}\BibitemShut {NoStop}%
\bibitem [{\citenamefont {Napolitano}\ \emph {et~al.}(2011)\citenamefont
  {Napolitano}, \citenamefont {Koschorreck}, \citenamefont {Dubost},
  \citenamefont {Behbood}, \citenamefont {Sewell},\ and\ \citenamefont
  {Mitchell}}]{NKDBSM}%
  \BibitemOpen
  \bibfield  {author} {\bibinfo {author} {\bibfnamefont {M.}~\bibnamefont
  {Napolitano}}, \bibinfo {author} {\bibfnamefont {M.}~\bibnamefont
  {Koschorreck}}, \bibinfo {author} {\bibfnamefont {B.}~\bibnamefont {Dubost}},
  \bibinfo {author} {\bibfnamefont {N.}~\bibnamefont {Behbood}}, \bibinfo
  {author} {\bibfnamefont {R.~J.}\ \bibnamefont {Sewell}}, \ and\ \bibinfo
  {author} {\bibfnamefont {M.~W.}\ \bibnamefont {Mitchell}},\ }\bibfield
  {title} {\enquote {\bibinfo {title} {Interaction-based quantum metrology
  showing scaling beyond the {H}eisenberg limit},}\ }\href@noop {} {\bibfield
  {journal} {\bibinfo  {journal} {Nature (London)}\ }\textbf {\bibinfo {volume}
  {471}},\ \bibinfo {pages} {486--489} (\bibinfo {year} {2011})}\BibitemShut
  {NoStop}%
\bibitem [{\citenamefont {Joo}\ \emph {et~al.}(2012)\citenamefont {Joo},
  \citenamefont {Park}, \citenamefont {Jeong}, \citenamefont {Munro},
  \citenamefont {Nemoto},\ and\ \citenamefont {Spiller}}]{JPJMNS}%
  \BibitemOpen
  \bibfield  {author} {\bibinfo {author} {\bibfnamefont {J.}~\bibnamefont
  {Joo}}, \bibinfo {author} {\bibfnamefont {K.}~\bibnamefont {Park}}, \bibinfo
  {author} {\bibfnamefont {H.}~\bibnamefont {Jeong}}, \bibinfo {author}
  {\bibfnamefont {W.~J.}\ \bibnamefont {Munro}}, \bibinfo {author}
  {\bibfnamefont {K.}~\bibnamefont {Nemoto}}, \ and\ \bibinfo {author}
  {\bibfnamefont {T.~P.}\ \bibnamefont {Spiller}},\ }\bibfield  {title}
  {\enquote {\bibinfo {title} {Quantum metrology for nonlinear phase shifts
  with entangled coherent states},}\ }\href@noop {} {\bibfield  {journal}
  {\bibinfo  {journal} {Physical Review A}\ }\textbf {\bibinfo {volume} {86}},\
  \bibinfo {pages} {043828} (\bibinfo {year} {2012})}\BibitemShut {NoStop}%
\bibitem [{\citenamefont {Kish}\ and\ \citenamefont {Ralph}(2017)}]{KR}%
  \BibitemOpen
  \bibfield  {author} {\bibinfo {author} {\bibfnamefont {S.~P.}\ \bibnamefont
  {Kish}}\ and\ \bibinfo {author} {\bibfnamefont {T.~C.}\ \bibnamefont
  {Ralph}},\ }\bibfield  {title} {\enquote {\bibinfo {title} {Quantum-limited
  measurement of space-time curvature with scaling beyond the conventional
  {H}eisenberg limit},}\ }\href@noop {} {\bibfield  {journal} {\bibinfo
  {journal} {Physical Review A}\ }\textbf {\bibinfo {volume} {96}},\ \bibinfo
  {pages} {041801(R)} (\bibinfo {year} {2017})}\BibitemShut {NoStop}%
\bibitem [{\citenamefont {Tsarev}\ \emph {et~al.}(2018)\citenamefont {Tsarev},
  \citenamefont {Arakelian}, \citenamefont {Chuang}, \citenamefont {Lee},\ and\
  \citenamefont {Alodjants}}]{TACLA}%
  \BibitemOpen
  \bibfield  {author} {\bibinfo {author} {\bibfnamefont {D.~V.}\ \bibnamefont
  {Tsarev}}, \bibinfo {author} {\bibfnamefont {S.~M.}\ \bibnamefont
  {Arakelian}}, \bibinfo {author} {\bibfnamefont {Y.-L.}\ \bibnamefont
  {Chuang}}, \bibinfo {author} {\bibfnamefont {R.-K.}\ \bibnamefont {Lee}}, \
  and\ \bibinfo {author} {\bibfnamefont {A.~P.}\ \bibnamefont {Alodjants}},\
  }\bibfield  {title} {\enquote {\bibinfo {title} {Quantum metrology beyond
  {H}eisenberg limit with entangled matter wave solitons},}\ }\href@noop {}
  {\bibfield  {journal} {\bibinfo  {journal} {Optics Express}\ }\textbf
  {\bibinfo {volume} {26}},\ \bibinfo {pages} {19583--19595} (\bibinfo {year}
  {2018})}\BibitemShut {NoStop}%
\bibitem [{\citenamefont {Zwierz}\ \emph {et~al.}(2010)\citenamefont {Zwierz},
  \citenamefont {P\'{e}rez-Delgado},\ and\ \citenamefont {Kok}}]{ZPK}%
  \BibitemOpen
  \bibfield  {author} {\bibinfo {author} {\bibfnamefont {M.}~\bibnamefont
  {Zwierz}}, \bibinfo {author} {\bibfnamefont {C.~A.}\ \bibnamefont
  {P\'{e}rez-Delgado}}, \ and\ \bibinfo {author} {\bibfnamefont
  {P.}~\bibnamefont {Kok}},\ }\bibfield  {title} {\enquote {\bibinfo {title}
  {General optimality of the {H}eisenberg limit for quantum metrology},}\
  }\href@noop {} {\bibfield  {journal} {\bibinfo  {journal} {Physical Review
  Letters}\ }\textbf {\bibinfo {volume} {105}},\ \bibinfo {pages} {180402}
  (\bibinfo {year} {2010})}\BibitemShut {NoStop}%
\bibitem [{\citenamefont {Pezz\`{e}}(2013)}]{LP}%
  \BibitemOpen
  \bibfield  {author} {\bibinfo {author} {\bibfnamefont {L.}~\bibnamefont
  {Pezz\`{e}}},\ }\bibfield  {title} {\enquote {\bibinfo {title}
  {Sub-{H}eisenberg phase uncertainties},}\ }\href@noop {} {\bibfield
  {journal} {\bibinfo  {journal} {Physical Review A}\ }\textbf {\bibinfo
  {volume} {88}},\ \bibinfo {pages} {060101(R)} (\bibinfo {year}
  {2013})}\BibitemShut {NoStop}%
\bibitem [{\citenamefont {Hall}\ and\ \citenamefont {Wiseman}(2012)}]{HWX}%
  \BibitemOpen
  \bibfield  {author} {\bibinfo {author} {\bibfnamefont {M.~J.~W.}\
  \bibnamefont {Hall}}\ and\ \bibinfo {author} {\bibfnamefont {H.~M.}\
  \bibnamefont {Wiseman}},\ }\bibfield  {title} {\enquote {\bibinfo {title}
  {Does nonlinear metrology offer improved resolution? answers from quantum
  information theory},}\ }\href@noop {} {\bibfield  {journal} {\bibinfo
  {journal} {Physical Review X}\ }\textbf {\bibinfo {volume} {2}},\ \bibinfo
  {pages} {041006} (\bibinfo {year} {2012})}\BibitemShut {NoStop}%
\bibitem [{\citenamefont {Giovannetti}\ and\ \citenamefont
  {Maccone}(2012)}]{GM}%
  \BibitemOpen
  \bibfield  {author} {\bibinfo {author} {\bibfnamefont {V.}~\bibnamefont
  {Giovannetti}}\ and\ \bibinfo {author} {\bibfnamefont {L.}~\bibnamefont
  {Maccone}},\ }\bibfield  {title} {\enquote {\bibinfo {title}
  {Sub-{H}eisenberg estimation strategies are ineffective},}\ }\href@noop {}
  {\bibfield  {journal} {\bibinfo  {journal} {Physical Review Letters}\
  }\textbf {\bibinfo {volume} {108}},\ \bibinfo {pages} {210404} (\bibinfo
  {year} {2012})}\BibitemShut {NoStop}%
\bibitem [{\citenamefont {Giovannetti}\ \emph {et~al.}(2012)\citenamefont
  {Giovannetti}, \citenamefont {Lloyd},\ and\ \citenamefont {Maccone}}]{GLM}%
  \BibitemOpen
  \bibfield  {author} {\bibinfo {author} {\bibfnamefont {V.}~\bibnamefont
  {Giovannetti}}, \bibinfo {author} {\bibfnamefont {S.}~\bibnamefont {Lloyd}},
  \ and\ \bibinfo {author} {\bibfnamefont {L.}~\bibnamefont {Maccone}},\
  }\bibfield  {title} {\enquote {\bibinfo {title} {Quantum measurement bounds
  beyond the uncertainty relations},}\ }\href@noop {} {\bibfield  {journal}
  {\bibinfo  {journal} {Physical Review Letters}\ }\textbf {\bibinfo {volume}
  {108}},\ \bibinfo {pages} {260405} (\bibinfo {year} {2012})}\BibitemShut
  {NoStop}%
\bibitem [{\citenamefont {Hall}\ \emph {et~al.}(2012)\citenamefont {Hall},
  \citenamefont {Berry}, \citenamefont {Zwierz},\ and\ \citenamefont
  {Wiseman}}]{HBZW}%
  \BibitemOpen
  \bibfield  {author} {\bibinfo {author} {\bibfnamefont {M.~J.~W.}\
  \bibnamefont {Hall}}, \bibinfo {author} {\bibfnamefont {D.~W.}\ \bibnamefont
  {Berry}}, \bibinfo {author} {\bibfnamefont {M.}~\bibnamefont {Zwierz}}, \
  and\ \bibinfo {author} {\bibfnamefont {H.~M.}\ \bibnamefont {Wiseman}},\
  }\bibfield  {title} {\enquote {\bibinfo {title} {Universality of the
  {H}eisenberg limit for estimates of random phase shifts},}\ }\href@noop {}
  {\bibfield  {journal} {\bibinfo  {journal} {Physical Review A}\ }\textbf
  {\bibinfo {volume} {85}},\ \bibinfo {pages} {041802(R)} (\bibinfo {year}
  {2012})}\BibitemShut {NoStop}%
\bibitem [{\citenamefont {Rams}\ \emph {et~al.}(2018)\citenamefont {Rams},
  \citenamefont {Sierant}, \citenamefont {Dutta}, \citenamefont {Horodecki},\
  and\ \citenamefont {Zakrzewski}}]{RSDHZ}%
  \BibitemOpen
  \bibfield  {author} {\bibinfo {author} {\bibfnamefont {M.~M.}\ \bibnamefont
  {Rams}}, \bibinfo {author} {\bibfnamefont {P.}~\bibnamefont {Sierant}},
  \bibinfo {author} {\bibfnamefont {O.}~\bibnamefont {Dutta}}, \bibinfo
  {author} {\bibfnamefont {P.}~\bibnamefont {Horodecki}}, \ and\ \bibinfo
  {author} {\bibfnamefont {J.}~\bibnamefont {Zakrzewski}},\ }\bibfield  {title}
  {\enquote {\bibinfo {title} {At the limits of criticality-based quantum
  metrology: Apparent super-{H}eisenberg scaling revisited},}\ }\href@noop {}
  {\bibfield  {journal} {\bibinfo  {journal} {Physical Review X}\ }\textbf
  {\bibinfo {volume} {8}},\ \bibinfo {pages} {021022} (\bibinfo {year}
  {2018})}\BibitemShut {NoStop}%
\bibitem [{\citenamefont {Adesso}\ \emph {et~al.}(2003)\citenamefont {Adesso},
  \citenamefont {Illuminati},\ and\ \citenamefont {Siena}}]{AIDS}%
  \BibitemOpen
  \bibfield  {author} {\bibinfo {author} {\bibfnamefont {G.}~\bibnamefont
  {Adesso}}, \bibinfo {author} {\bibfnamefont {F.}~\bibnamefont {Illuminati}},
  \ and\ \bibinfo {author} {\bibfnamefont {S.~De}\ \bibnamefont {Siena}},\
  }\bibfield  {title} {\enquote {\bibinfo {title} {Characterizing entanglement
  with global and marginal entropic measures},}\ }\href@noop {} {\bibfield
  {journal} {\bibinfo  {journal} {Physical Review A}\ }\textbf {\bibinfo
  {volume} {68}},\ \bibinfo {pages} {062318} (\bibinfo {year}
  {2003})}\BibitemShut {NoStop}%
\bibitem [{\citenamefont {Verstraete}\ \emph {et~al.}(2001)\citenamefont
  {Verstraete}, \citenamefont {Audenaert},\ and\ \citenamefont {Moor}}]{VAM}%
  \BibitemOpen
  \bibfield  {author} {\bibinfo {author} {\bibfnamefont {F.}~\bibnamefont
  {Verstraete}}, \bibinfo {author} {\bibfnamefont {K.}~\bibnamefont
  {Audenaert}}, \ and\ \bibinfo {author} {\bibfnamefont {B.~D.}\ \bibnamefont
  {Moor}},\ }\bibfield  {title} {\enquote {\bibinfo {title} {Maximally
  entangled mixed states of two qubits},}\ }\href@noop {} {\bibfield  {journal}
  {\bibinfo  {journal} {Physical Review A}\ }\textbf {\bibinfo {volume} {64}},\
  \bibinfo {pages} {012316} (\bibinfo {year} {2001})}\BibitemShut {NoStop}%
\bibitem [{\citenamefont {Li}\ \emph {et~al.}(2012)\citenamefont {Li},
  \citenamefont {Zhao}, \citenamefont {Fei}, \citenamefont {Fan},\ and\
  \citenamefont {Liu}}]{LZFFL}%
  \BibitemOpen
  \bibfield  {author} {\bibinfo {author} {\bibfnamefont {Z.~G.}\ \bibnamefont
  {Li}}, \bibinfo {author} {\bibfnamefont {M.~J.}\ \bibnamefont {Zhao}},
  \bibinfo {author} {\bibfnamefont {S.~M.}\ \bibnamefont {Fei}}, \bibinfo
  {author} {\bibfnamefont {H.}~\bibnamefont {Fan}}, \ and\ \bibinfo {author}
  {\bibfnamefont {W.~M.}\ \bibnamefont {Liu}},\ }\bibfield  {title} {\enquote
  {\bibinfo {title} {Mixed maximally entangled states},}\ }\href@noop {}
  {\bibfield  {journal} {\bibinfo  {journal} {Quantum Information and
  Computation}\ }\textbf {\bibinfo {volume} {12}},\ \bibinfo {pages} {63--73}
  (\bibinfo {year} {2012})}\BibitemShut {NoStop}%
\bibitem [{\citenamefont {Galve}\ \emph {et~al.}(2011)\citenamefont {Galve},
  \citenamefont {Giorgi},\ and\ \citenamefont {Zambrini}}]{GGZ}%
  \BibitemOpen
  \bibfield  {author} {\bibinfo {author} {\bibfnamefont {F.}~\bibnamefont
  {Galve}}, \bibinfo {author} {\bibfnamefont {G.~L.}\ \bibnamefont {Giorgi}}, \
  and\ \bibinfo {author} {\bibfnamefont {R.}~\bibnamefont {Zambrini}},\
  }\bibfield  {title} {\enquote {\bibinfo {title} {Maximally discordant mixed
  states of two qubits},}\ }\href@noop {} {\bibfield  {journal} {\bibinfo
  {journal} {Physical Review A}\ }\textbf {\bibinfo {volume} {83}},\ \bibinfo
  {pages} {012102} (\bibinfo {year} {2011})}\BibitemShut {NoStop}%
\bibitem [{\citenamefont {Modi}\ \emph {et~al.}(2010)\citenamefont {Modi},
  \citenamefont {Paterek}, \citenamefont {Son}, \citenamefont {Vedral},\ and\
  \citenamefont {Williamson}}]{MPSVW}%
  \BibitemOpen
  \bibfield  {author} {\bibinfo {author} {\bibfnamefont {K.}~\bibnamefont
  {Modi}}, \bibinfo {author} {\bibfnamefont {T.}~\bibnamefont {Paterek}},
  \bibinfo {author} {\bibfnamefont {W.}~\bibnamefont {Son}}, \bibinfo {author}
  {\bibfnamefont {V.}~\bibnamefont {Vedral}}, \ and\ \bibinfo {author}
  {\bibfnamefont {M.}~\bibnamefont {Williamson}},\ }\bibfield  {title}
  {\enquote {\bibinfo {title} {Unified view of quantum and classical
  correlations},}\ }\href@noop {} {\bibfield  {journal} {\bibinfo  {journal}
  {Physical Review Letters}\ }\textbf {\bibinfo {volume} {104}},\ \bibinfo
  {pages} {080501} (\bibinfo {year} {2010})}\BibitemShut {NoStop}%
\bibitem [{\citenamefont {Robertson}(1929)}]{HPR}%
  \BibitemOpen
  \bibfield  {author} {\bibinfo {author} {\bibfnamefont {H.~P.}\ \bibnamefont
  {Robertson}},\ }\bibfield  {title} {\enquote {\bibinfo {title} {The
  uncertainty principle},}\ }\href@noop {} {\bibfield  {journal} {\bibinfo
  {journal} {Physical Review}\ }\textbf {\bibinfo {volume} {34}},\ \bibinfo
  {pages} {163} (\bibinfo {year} {1929})}\BibitemShut {NoStop}%
\bibitem [{\citenamefont {Rigolin}(2016)}]{GR}%
  \BibitemOpen
  \bibfield  {author} {\bibinfo {author} {\bibfnamefont {G.}~\bibnamefont
  {Rigolin}},\ }\bibfield  {title} {\enquote {\bibinfo {title} {Entanglement,
  identical particles and the uncertainty principle},}\ }\href@noop {}
  {\bibfield  {journal} {\bibinfo  {journal} {Communications in Theoretical
  Physics}\ }\textbf {\bibinfo {volume} {66}},\ \bibinfo {pages} {201--206}
  (\bibinfo {year} {2016})}\BibitemShut {NoStop}%
\bibitem [{\citenamefont {Werlang}\ \emph {et~al.}(2009)\citenamefont
  {Werlang}, \citenamefont {Souza}, \citenamefont {Fanchini},\ and\
  \citenamefont {Boas}}]{WSFVB}%
  \BibitemOpen
  \bibfield  {author} {\bibinfo {author} {\bibfnamefont {T.}~\bibnamefont
  {Werlang}}, \bibinfo {author} {\bibfnamefont {S.}~\bibnamefont {Souza}},
  \bibinfo {author} {\bibfnamefont {F.~F.}\ \bibnamefont {Fanchini}}, \ and\
  \bibinfo {author} {\bibfnamefont {C.~J.~Villas}\ \bibnamefont {Boas}},\
  }\bibfield  {title} {\enquote {\bibinfo {title} {Robustness of quantum
  discord to sudden death},}\ }\href@noop {} {\bibfield  {journal} {\bibinfo
  {journal} {Physical Review A}\ }\textbf {\bibinfo {volume} {80}},\ \bibinfo
  {pages} {024103} (\bibinfo {year} {2009})}\BibitemShut {NoStop}%
\bibitem [{\citenamefont {Almeida}\ \emph {et~al.}(2007)\citenamefont
  {Almeida}, \citenamefont {de~Melo}, \citenamefont {Hor-Meyll}, \citenamefont
  {Salles}, \citenamefont {Walborn}, \citenamefont {Ribeiro},\ and\
  \citenamefont {Davidovich}}]{AMHMSWSRD}%
  \BibitemOpen
  \bibfield  {author} {\bibinfo {author} {\bibfnamefont {M.~P.}\ \bibnamefont
  {Almeida}}, \bibinfo {author} {\bibfnamefont {F.}~\bibnamefont {de~Melo}},
  \bibinfo {author} {\bibfnamefont {M.}~\bibnamefont {Hor-Meyll}}, \bibinfo
  {author} {\bibfnamefont {A.}~\bibnamefont {Salles}}, \bibinfo {author}
  {\bibfnamefont {S.~P.}\ \bibnamefont {Walborn}}, \bibinfo {author}
  {\bibfnamefont {P.~H.~Souto}\ \bibnamefont {Ribeiro}}, \ and\ \bibinfo
  {author} {\bibfnamefont {L.}~\bibnamefont {Davidovich}},\ }\bibfield  {title}
  {\enquote {\bibinfo {title} {Environment-induced sudden death of
  entanglement},}\ }\href@noop {} {\bibfield  {journal} {\bibinfo  {journal}
  {Science}\ }\textbf {\bibinfo {volume} {316}},\ \bibinfo {pages} {579--582}
  (\bibinfo {year} {2007})}\BibitemShut {NoStop}%
\bibitem [{\citenamefont {Yu}\ and\ \citenamefont {Eberly}(2009)}]{YE}%
  \BibitemOpen
  \bibfield  {author} {\bibinfo {author} {\bibfnamefont {T.}~\bibnamefont
  {Yu}}\ and\ \bibinfo {author} {\bibfnamefont {J.~H.}\ \bibnamefont
  {Eberly}},\ }\bibfield  {title} {\enquote {\bibinfo {title} {Sudden death of
  entanglement},}\ }\href@noop {} {\bibfield  {journal} {\bibinfo  {journal}
  {Science}\ }\textbf {\bibinfo {volume} {323}},\ \bibinfo {pages} {598--601}
  (\bibinfo {year} {2009})}\BibitemShut {NoStop}%
\bibitem [{\citenamefont {Luo}(2016)}]{SL}%
  \BibitemOpen
  \bibfield  {author} {\bibinfo {author} {\bibfnamefont {S.}~\bibnamefont
  {Luo}},\ }\bibfield  {title} {\enquote {\bibinfo {title} {Entanglement as
  minimal discord over state extensions},}\ }\href@noop {} {\bibfield
  {journal} {\bibinfo  {journal} {Physical Review A}\ }\textbf {\bibinfo
  {volume} {94}},\ \bibinfo {pages} {032129} (\bibinfo {year}
  {2016})}\BibitemShut {NoStop}%
\bibitem [{\citenamefont {Streltsov}\ and\ \citenamefont {Zurek}(2013)}]{SZ}%
  \BibitemOpen
  \bibfield  {author} {\bibinfo {author} {\bibfnamefont {A.}~\bibnamefont
  {Streltsov}}\ and\ \bibinfo {author} {\bibfnamefont {W.~H.}\ \bibnamefont
  {Zurek}},\ }\bibfield  {title} {\enquote {\bibinfo {title} {Quantum discord
  cannot be shared},}\ }\href@noop {} {\bibfield  {journal} {\bibinfo
  {journal} {Physical Review Letters}\ }\textbf {\bibinfo {volume} {111}},\
  \bibinfo {pages} {040401} (\bibinfo {year} {2013})}\BibitemShut {NoStop}%
\bibitem [{\citenamefont {Coecke}\ and\ \citenamefont {Kissinger}(2017)}]{CK}%
  \BibitemOpen
  \bibfield  {author} {\bibinfo {author} {\bibfnamefont {B.}~\bibnamefont
  {Coecke}}\ and\ \bibinfo {author} {\bibfnamefont {A.}~\bibnamefont
  {Kissinger}},\ }\href@noop {} {\emph {\bibinfo {title} {Picturing Quantum
  Processes}}}\ (\bibinfo  {publisher} {Cambridge University Press},\ \bibinfo
  {year} {2017})\BibitemShut {NoStop}%
\bibitem [{\citenamefont {Nagaoka}(1989)}]{HNMH}%
  \BibitemOpen
  \bibfield  {author} {\bibinfo {author} {\bibfnamefont {H.}~\bibnamefont
  {Nagaoka}},\ }\bibfield  {title} {\enquote {\bibinfo {title} {A new approach
  to {C}ram\'{e}r-{R}ao bounds for quantum state estimation},}\ }in\ \href@noop
  {} {\emph {\bibinfo {booktitle} {Asymptotic Theory of Quantum Statistical
  Inference}}},\ Vol.~\bibinfo {volume} {1},\ \bibinfo {editor} {edited by\
  \bibinfo {editor} {\bibfnamefont {M.}~\bibnamefont {Hayashi}}}\ (\bibinfo
  {publisher} {World Scientific},\ \bibinfo {year} {1989})\ Chap.~\bibinfo
  {chapter} {8}\BibitemShut {NoStop}%
\bibitem [{\citenamefont {Braunstein}\ and\ \citenamefont {Caves}(1994)}]{BC}%
  \BibitemOpen
  \bibfield  {author} {\bibinfo {author} {\bibfnamefont {S.~L.}\ \bibnamefont
  {Braunstein}}\ and\ \bibinfo {author} {\bibfnamefont {C.~M.}\ \bibnamefont
  {Caves}},\ }\bibfield  {title} {\enquote {\bibinfo {title} {Statistical
  distance and the geometry of quantum states},}\ }\href@noop {} {\bibfield
  {journal} {\bibinfo  {journal} {Physical Review Letters}\ }\textbf {\bibinfo
  {volume} {72}},\ \bibinfo {pages} {3439} (\bibinfo {year}
  {1994})}\BibitemShut {NoStop}%
\bibitem [{\citenamefont {Hayashi}(2006)}]{MH}%
  \BibitemOpen
  \bibfield  {author} {\bibinfo {author} {\bibfnamefont {M.}~\bibnamefont
  {Hayashi}},\ }\href@noop {} {\emph {\bibinfo {title} {Quantum Information -
  An Introduction}}}\ (\bibinfo  {publisher} {Springer},\ \bibinfo {year}
  {2006})\BibitemShut {NoStop}%
\bibitem [{\citenamefont {Dittmann}(1999)}]{JDN}%
  \BibitemOpen
  \bibfield  {author} {\bibinfo {author} {\bibfnamefont {J.}~\bibnamefont
  {Dittmann}},\ }\bibfield  {title} {\enquote {\bibinfo {title} {Explicit
  formulae for the {B}ures metric},}\ }\href@noop {} {\bibfield  {journal}
  {\bibinfo  {journal} {Journal of Physics A: Mathematical and General}\
  }\textbf {\bibinfo {volume} {32}},\ \bibinfo {pages} {2663} (\bibinfo {year}
  {1999})}\BibitemShut {NoStop}%
\bibitem [{\citenamefont {Fawzi}\ and\ \citenamefont {Renner}(2015)}]{FR}%
  \BibitemOpen
  \bibfield  {author} {\bibinfo {author} {\bibfnamefont {O.}~\bibnamefont
  {Fawzi}}\ and\ \bibinfo {author} {\bibfnamefont {R.}~\bibnamefont {Renner}},\
  }\bibfield  {title} {\enquote {\bibinfo {title} {Quantum conditional mutual
  information and approximate {M}arkov chains},}\ }\href@noop {} {\bibfield
  {journal} {\bibinfo  {journal} {Communications in Mathematical Physics}\
  }\textbf {\bibinfo {volume} {340}},\ \bibinfo {pages} {575--611} (\bibinfo
  {year} {2015})}\BibitemShut {NoStop}%
\bibitem [{\citenamefont {Dodd}\ and\ \citenamefont {Nielsen}(2002)}]{DN}%
  \BibitemOpen
  \bibfield  {author} {\bibinfo {author} {\bibfnamefont {J.~L.}\ \bibnamefont
  {Dodd}}\ and\ \bibinfo {author} {\bibfnamefont {M.~A.}\ \bibnamefont
  {Nielsen}},\ }\bibfield  {title} {\enquote {\bibinfo {title} {Simple
  operational interpretation of the fidelity of mixed states},}\ }\href@noop {}
  {\bibfield  {journal} {\bibinfo  {journal} {Physical Review A}\ }\textbf
  {\bibinfo {volume} {66}},\ \bibinfo {pages} {044301} (\bibinfo {year}
  {2002})}\BibitemShut {NoStop}%
\bibitem [{\citenamefont {Petz}\ and\ \citenamefont {Ghinea}(2011)}]{PG}%
  \BibitemOpen
  \bibfield  {author} {\bibinfo {author} {\bibfnamefont {D.}~\bibnamefont
  {Petz}}\ and\ \bibinfo {author} {\bibfnamefont {C.}~\bibnamefont {Ghinea}},\
  }\enquote {\bibinfo {title} {Introduction to quantum {F}isher information},}\
  in\ \href@noop {} {\emph {\bibinfo {booktitle} {Quantum Probability and
  Related Topics}}}\ (\bibinfo  {publisher} {World Scientific},\ \bibinfo
  {year} {2011})\ pp.\ \bibinfo {pages} {261--281}\BibitemShut {NoStop}%
\end{thebibliography}%

\onecolumngrid
\appendix
\section{Proof for $\nu V\left[\boldsymbol{\tilde{\theta}}(m)\right]\geq \left[J_C(\boldsymbol{\theta})\right]^{-1}\geq \left[J_Q(\boldsymbol{\theta})\right]^{-1}$}\label{sec:app1}
Here, we prove the following quantum Cram\'{e}r-Rao inequality, as claimed in Section \ref{sec:qcrb}:
\begin{equation}\label{eq:supp_ald_qcrb}
\nu V\left[\boldsymbol{\tilde{\theta}}(m)\right]\geq \left[J_C(\boldsymbol{\theta})\right]^{-1}\geq \left[J_Q(\boldsymbol{\theta})\right]^{-1},
\end{equation}
where $\nu$ is the number of times the experiment is repeated,
\begin{equation}\label{eq:supp_est_err_cov}
V\left[\boldsymbol{\tilde{\theta}}(m)\right] = \sum_m p(m|\boldsymbol{\theta}) \left(\boldsymbol{\tilde{\theta}}(m)-\boldsymbol{\theta}\right)\left(\boldsymbol{\tilde{\theta}}(m)-\boldsymbol{\theta}\right)^T =: \Sigma
\end{equation}
is the estimation error covariance,
\begin{equation}\label{eq:supp_fim1}
J_C^{jk} = \sum_m\frac{1}{p(m|\boldsymbol{\theta})}\frac{\partial}{\partial\theta_j}p(m|\boldsymbol{\theta})\frac{\partial}{\partial\theta_k}p(m|\boldsymbol{\theta})
\end{equation}
is the classical Fisher information matrix (FIM), and
\begin{equation}\label{eq:supp_qfim1}
J_Q^{jk} = \frac{1}{2}{\rm Tr}\left[\left(\hat{L}_j^\dagger \hat{L}_k+\hat{L}_k^\dagger \hat{L}_j\right)\hat{\rho}(\boldsymbol{\theta})\right]
\end{equation}
is the quantum Fisher information matrix (QFIM), with the operators $\hat{L}_k$ satisfying
\begin{equation}\label{eq:supp_ald_diffeqn1}
\frac{1}{2}\left(\hat{L}_k\hat{\rho}(\boldsymbol{\theta})+\hat{\rho}(\boldsymbol{\theta})\hat{L}_k^\dagger\right)=\frac{\partial}{\partial\theta_k}\hat{\rho}(\boldsymbol{\theta}).
\end{equation}
The proof is adapted from Ref.~\cite{TWC} for frequentist multiparameter estimation problem here.

The estimates $\boldsymbol{\tilde{\theta}}(m) = \left[\begin{array}{cccc} \tilde{\theta}_1(m) & \tilde{\theta}_2(m) & \hdots & \tilde{\theta}_q(m) \end{array}\right]^T$ of the parameters $\boldsymbol{\theta} = \left[\begin{array}{cccc} \theta_1 & \theta_2 & \hdots & \theta_q \end{array}\right]^T$ are unbiased, if
\begin{equation}\label{eq:supp_unbiased_est}
\sum_m p(m|\boldsymbol{\theta})\tilde{\theta}_j(m) = \theta_j \qquad \forall j,
\end{equation}
where $p(m|\boldsymbol{\theta})={\rm Tr}\left(\hat{P}_m\hat{\rho}(\boldsymbol{\theta})\right)$ is the conditional probability to obtain the outcome $m$ from a measurement performed on the evolved probe state $\hat{\rho}(\boldsymbol{\theta})$ via a positive operator valued measure (POVM) $\{\hat{P}_m\}$, given that the parameters have the value $\boldsymbol{\theta}$. Differentiating (\ref{eq:supp_unbiased_est}) with respect to $\theta_k$, we get
\begin{equation}\label{eq:supp_deljk}
\delta_{jk}=\sum_m\left(\tilde{\theta}_j(m)-\theta_j\right)\frac{\partial p(m|\boldsymbol{\theta})}{\partial\theta_k}
={\rm Re}\sum_m\left(\tilde{\theta}_j(m)-\theta_j\right){\rm Tr}\left[\hat{P}_m\hat{L}_k\hat{\rho}(\boldsymbol{\theta})\right].
\end{equation}
Then, following Ref.~\cite{TWC}, since $\nu\geq 1$, we get:
\begin{equation}
\boldsymbol{v}^T\boldsymbol{u}=\sum_ju_jv_j\leq A^TB,\qquad
\boldsymbol{w}^T\boldsymbol{u}=\sum_ku_kw_k\leq{\rm Re}\left[{\rm Tr}\left(C^\dagger D\right)\right],
\end{equation}
where $\boldsymbol{u}$, $\boldsymbol{v}$, $\boldsymbol{w}$ are arbitrary real column vectors, and
\begin{equation}
\begin{split}
A^T&=\sum_kv_k\frac{\partial p(m|\boldsymbol{\theta})}{\partial\theta_k}\frac{1}{\sqrt{p(m|\boldsymbol{\theta})}},\qquad
B=\sum_ju_j\left(\tilde{\theta}_j(m)-\theta_j\right)\sqrt{\nu}\sqrt{p(m|\boldsymbol{\theta})},\\
C^\dagger&=\sum_lw_l\sqrt{\hat{P}_m}\hat{L}_l\sqrt{\hat{\rho}(\boldsymbol{\theta})},\qquad \quad \, \, \,
D=\sum_ju_j\left(\tilde{\theta}_j(m)-\theta_j\right)\sqrt{\nu}\sqrt{\hat{\rho}(\boldsymbol{\theta})}\sqrt{\hat{P}_m}.
\end{split}
\end{equation}
We assume that $\boldsymbol{v}^T\boldsymbol{u}$ and $\boldsymbol{w}^T\boldsymbol{u}$ are positive, which are valid assumptions given how we set these later. Then,
\begin{equation}\label{eq:supp_schwarz}
\begin{split}
\left(\boldsymbol{v}^T\boldsymbol{u}\right)^2&\leq\left(A^TB\right)^2\leq\left(A^TA\right)\left(B^TB\right),\\
\left(\boldsymbol{w}^T\boldsymbol{u}\right)^2&\leq\left|{\rm Tr}\left(C^\dagger D\right)\right|^2\leq{\rm Tr}\left(C^\dagger C\right){\rm Tr}\left(D^\dagger D\right),
\end{split}
\end{equation}
where the second inequalities in both lines are Schwarz inequalities.

Now, note that $A^TA=\boldsymbol{v}^TJ_C\boldsymbol{v}$, where $J_C$ is a real, symmetric and positive semidefinite classical Fisher information matrix (FIM) as defined in (\ref{eq:supp_fim1}), ${\rm Tr}\left(C^\dagger C\right)=\boldsymbol{w}^TJ_Q\boldsymbol{w}$, where $J_Q$ is a real, symmetric and positive semidefinite quantum Fisher information matrix (QFIM) as defined in (\ref{eq:supp_qfim1}), and $B^TB={\rm Tr}\left(D^\dagger D\right)=\boldsymbol{u}^T\nu\Sigma\boldsymbol{u}$, where $\nu\Sigma$ is the estimation error covariance matrix as defined in (\ref{eq:supp_est_err_cov}). Substituting these in (\ref{eq:supp_schwarz}), we find that
\begin{equation}
\left(\boldsymbol{v}^TJ_C\boldsymbol{v}\right)\left(\boldsymbol{u}^T\nu\Sigma\boldsymbol{u}\right) \geq \left(\boldsymbol{v}^T\boldsymbol{u}\right)\left(\boldsymbol{u}^T\boldsymbol{v}\right),\qquad
\left(\boldsymbol{w}^TJ_Q\boldsymbol{w}\right)\left(\boldsymbol{u}^T\nu\Sigma\boldsymbol{u}\right) \geq 
\left(\boldsymbol{w}^T\boldsymbol{u}\right)\left(\boldsymbol{u}^T\boldsymbol{w}\right).
\end{equation}
Setting $\boldsymbol{v}=J_C^{-1}\boldsymbol{u}$ implies that 
\begin{equation}
\boldsymbol{u}^T\left(\nu\Sigma-J_C^{-1}\right)\boldsymbol{u} \geq 0,
\end{equation}
for arbitrary real vectors $\boldsymbol{u}$. Since $\nu\Sigma-J_C^{-1}$ is real and symmetric, this implies that $\nu\Sigma-J_C^{-1}$ is positive semidefinite. Also, setting $\boldsymbol{w}=J_Q^{-1}\boldsymbol{u}$ implies that 
\begin{equation}
\boldsymbol{u}^T\left(\nu\Sigma-J_Q^{-1}\right)\boldsymbol{u} \geq 0.
\end{equation}
Since $\nu\Sigma-J_Q^{-1}$ is real and symmetric, this implies that $\nu\Sigma-J_Q^{-1}$ is positive semidefinite.

We now take $\boldsymbol{v}=\boldsymbol{w}$. Then, we have
\begin{equation}
\boldsymbol{v}^T\boldsymbol{u}\leq A^TB={\rm Re}\left[{\rm Tr}\left(C^\dagger D\right)\right]
\Rightarrow\left(\boldsymbol{v}^T\boldsymbol{u}\right)\left(\boldsymbol{u}^T\boldsymbol{v}\right)\leq\left|{\rm Tr}\left(C^\dagger D\right)\right|^2\leq{\rm Tr}\left(C^\dagger C\right){\rm Tr}\left(D^\dagger D\right)=\left(\boldsymbol{v}^TJ_Q\boldsymbol{v}\right)\left(\boldsymbol{u}^T\nu\Sigma\boldsymbol{u}\right).
\end{equation}
Then, again setting $\boldsymbol{v}=J_C^{-1}\boldsymbol{u}$ imply that
\begin{equation}
\left(\boldsymbol{u}^TJ_C^{-1}\boldsymbol{u}\right)\left(\boldsymbol{u}^TJ_C^{-1}\boldsymbol{u}\right)\leq\left(\boldsymbol{u}^TJ_C^{-1}J_QJ_C^{-1}\boldsymbol{u}\right)\left(\boldsymbol{u}^T\nu\Sigma\boldsymbol{u}\right).
\end{equation}
Now, since $\boldsymbol{u}^T\left(\nu\Sigma-J_C^{-1}\right)\boldsymbol{u} \geq 0$, we get from above
\begin{equation}
\boldsymbol{u}^TJ_C^{-1}\boldsymbol{u}\leq \boldsymbol{u}^TJ_C^{-1}J_QJ_C^{-1}\boldsymbol{u}
\Rightarrow J_C^{-1}\leq J_C^{-1}J_QJ_C^{-1}
\Rightarrow J_C^{-1}\geq J_Q^{-1}.
\end{equation}
Thus, we have (\ref{eq:supp_ald_qcrb}).

\pagebreak
\section{Saturability of ALD-based QCRB}\label{sec:app2}
Here, we prove that an ALD-based QCRB can be saturated when the expectation of the commutator of the ALDs vanishes, as claimed in Section \ref{sec:qcrb}:
\begin{equation}\label{eq:supp_qcrb_saturate}
{\rm Tr}\left[\left(\hat{L}_j^\dagger \hat{L}_k-\hat{L}_k^\dagger \hat{L}_j\right)\hat{\rho}(\boldsymbol{\theta})\right]={\rm Tr}\left(\left[\hat{L}_j,\hat{L}_k\right]\hat{\rho}(\boldsymbol{\theta})\right)= 0,
\end{equation}
where the operators $\hat{L}_k$ are anti-Hermitian. The proof presented here is directly adapted from Ref.~\cite{RJD} for ALDs, and relies on the fact that it is enough to show that the QFIM bound is equivalent to the Holevo bound when (\ref{eq:supp_qcrb_saturate}) is satisfied, because the Holevo bound is a tighter bound, known to be asymptotically saturable.

Given that the operators $\hat{L}_k$ are anti-Hermitian and satisfy
\begin{equation}
\frac{1}{2}\left(\hat{L}_k\hat{\rho}(\boldsymbol{\theta})-\hat{\rho}(\boldsymbol{\theta})\hat{L}_k\right)=\frac{\partial}{\partial\theta_k}\hat{\rho}(\boldsymbol{\theta}),
\end{equation}
and the QFIM $J_Q$ is given by (\ref{eq:supp_qfim1}), then (\ref{eq:supp_ald_qcrb}) implies that for a given cost matrix $G$, the estimation cost is bounded by
\begin{equation}\label{eq:supp_helstrom_bound}
{\rm tr}\left(G\nu V\left[\boldsymbol{\tilde{\theta}}(m)\right]\right) \geq {\rm tr}\left(GJ_Q^{-1}\right),
\end{equation}
where ${\rm tr}$ denotes the trace of a matrix in distinction from ${\rm Tr}$ for an operator. Then, the achievable estimation uncertainty is lower-bounded by the Holevo Cram\'{e}r-Rao bound \cite{RJD,HNMH}:
\begin{equation}\label{eq:supp_holevo_bound}
{\rm tr}\left(G\nu V\left[\boldsymbol{\tilde{\theta}}(m)\right]\right) \geq \min_{\{\hat{X}_j\}}\left\lbrace{\rm tr}\left(G{\rm Re}W\right)+||G{\rm Im}W||_1\right\rbrace,
\end{equation}
where $||\cdot||_1$ is the operator trace norm, the elements of the matrix $W$ are \cite{HNMH}
\begin{equation}
W_{jk}={\rm Tr}\left(\hat{X}_j^\dagger\hat{X}_k\hat{\rho}(\boldsymbol{\theta})\right),
\end{equation}
and the minimization is performed over the operators $\hat{X}_j$ satisfying
\begin{equation}
\frac{1}{2}{\rm Tr}\left[\left(\hat{X}_j^\dagger\hat{L}_k+\hat{L}_k^\dagger\hat{X}_j\right)\hat{\rho}(\boldsymbol{\theta})\right]=\delta_{jk}.
\end{equation}
In our case, the operators $\hat{X}_j$ are also anti-Hermitian. The bound (\ref{eq:supp_holevo_bound}) is stronger than the bound (\ref{eq:supp_helstrom_bound}), the right hand side of which can be rewritten in the form \cite{RJD}:
\begin{equation}\label{eq:supp_helstrom_qcrb}
{\rm tr}\left(GJ_Q^{-1}\right)=\min_{\{\hat{X}_j\}}{\rm tr}\left(G{\rm Re}W\right).
\end{equation}
Then, the solution to the minimization problem in (\ref{eq:supp_helstrom_qcrb}) is \cite{RJD}
\begin{equation}
\hat{X}_j = \sum_k \left(G^{-1}\Lambda\right)_{jk}\hat{L}_k = \sum_k \left(J_Q^{-1}\right)_{jk}\hat{L}_k,
\end{equation}
where $\Lambda$ is a matrix of Lagrange multipliers, chosen so that $G^{-1}\Lambda J_Q=\mathbb{1}$.

Now, the cost matrix $G$ and the QFIM $J_Q$ are assumed to be strictly positive. Firstly, we assume that (\ref{eq:supp_qcrb_saturate}) holds for all $j$, $k$. We saw that the optimal $\hat{X}_j=\sum_k\left(J_Q^{-1}\right)_{jk}\hat{L}_k$ are linear combinations of $\hat{L}_j$. This implies that ${\rm Tr}\left(\left[\hat{X}_j,\hat{X}_k\right]\hat{\rho}(\boldsymbol{\theta})\right)=0$ for all $j$, $k$. Hence, the same set of $\hat{X}_j$ minimizes the Holevo bound, since it makes the second term in (\ref{eq:supp_holevo_bound}) to equal zero. Thus, (\ref{eq:supp_qcrb_saturate}) is a sufficient condition for saturating the ALD-based QCRB corresponding to the QFIM (\ref{eq:supp_qfim1}).

Secondly, we assume that the Holevo bound coincides with the QFIM bound, and so for the $\hat{X}_j$ that minimize both (\ref{eq:supp_helstrom_bound}) and (\ref{eq:supp_holevo_bound}), the second term in (\ref{eq:supp_holevo_bound}) must equal zero. Since $G$ is strictly positive, the matrix ${\rm Im}W$ must be zero and hence ${\rm Tr}\left(\left[\hat{X}_j,\hat{X}_k\right]\hat{\rho}(\boldsymbol{\theta})\right)=0$ for all $j$, $k$. However, the $\hat{X}_j$ that minimizes (\ref{eq:supp_helstrom_bound}) is $\hat{X}_j=\sum_k\left(J_Q^{-1}\right)_{jk}\hat{L}_k$. Inverting this formula, we get $\hat{L}_j=\sum_k\left(\hat{J}_Q\right)_{jk}\hat{X}_k$. Hence, (\ref{eq:supp_qcrb_saturate}) holds for all $j$, $k$ and is also a necessary condition for saturating the ALD-based QCRB corresponding to the QFIM (\ref{eq:supp_qfim1}).

\pagebreak
\section{The states $\hat{\rho}^N$ are permutationally invariant}\label{sec:app3}
Here, we show that the first order and second order reduced density matrices are as claimed in Section \ref{sec:mag_fld} for the magnetic field example. 

First, considering the $N=2$ case:
\begin{equation}
\begin{split}
\hat{\rho}_k^{N=2} = \frac{1}{2}&\left[\hat{E}_0^{\otimes 2}|\phi_k^{+},\phi_k^{+}\rangle\langle\phi_k^{+},\phi_k^{+}|\hat{E}_0^{\otimes 2}+(\hat{E}_0\otimes\hat{E}_1)|\phi_k^{+},\phi_k^{+}\rangle\langle\phi_k^{+},\phi_k^{+}|(\hat{E}_0\otimes\hat{E}_1)\right.\\
&\left.+(\hat{E}_1\otimes\hat{E}_0)|\phi_k^{+},\phi_k^{+}\rangle\langle\phi_k^{+},\phi_k^{+}|(\hat{E}_1\otimes\hat{E}_0)+\hat{E}_1^{\otimes 2}|\phi_k^{+},\phi_k^{+}\rangle\langle\phi_k^{+},\phi_k^{+}|\hat{E}_1^{\otimes 2}\right.\\
&\left.+\hat{E}_0^{\otimes 2}|\phi_k^{+},\phi_k^{+}\rangle\langle\phi_k^{-},\phi_k^{-}|\hat{E}_0^{\otimes 2}+(\hat{E}_0\otimes\hat{E}_1)|\phi_k^{+},\phi_k^{+}\rangle\langle\phi_k^{-},\phi_k^{-}|(\hat{E}_0\otimes\hat{E}_1)\right.\\
&\left.+(\hat{E}_1\otimes\hat{E}_0)|\phi_k^{+},\phi_k^{+}\rangle\langle\phi_k^{-},\phi_k^{-}|(\hat{E}_1\otimes\hat{E}_0)+\hat{E}_1^{\otimes 2}|\phi_k^{+},\phi_k^{+}\rangle\langle\phi_k^{-},\phi_k^{-}|\hat{E}_1^{\otimes 2}\right.\\
&\left.+\hat{E}_0^{\otimes 2}|\phi_k^{-},\phi_k^{-}\rangle\langle\phi_k^{+},\phi_k^{+}|\hat{E}_0^{\otimes 2}+(\hat{E}_0\otimes\hat{E}_1)|\phi_k^{-},\phi_k^{-}\rangle\langle\phi_k^{+},\phi_k^{+}|(\hat{E}_0\otimes\hat{E}_1)\right.\\
&\left.+(\hat{E}_1\otimes\hat{E}_0)|\phi_k^{-},\phi_k^{-}\rangle\langle\phi_k^{+},\phi_k^{+}|(\hat{E}_1\otimes\hat{E}_0)+\hat{E}_1^{\otimes 2}|\phi_k^{-},\phi_k^{-}\rangle\langle\phi_k^{+},\phi_k^{+}|\hat{E}_1^{\otimes 2}\right.\\
&\left.+\hat{E}_0^{\otimes 2}|\phi_k^{-},\phi_k^{-}\rangle\langle\phi_k^{-},\phi_k^{-}|\hat{E}_0^{\otimes 2}+(\hat{E}_0\otimes\hat{E}_1)|\phi_k^{-},\phi_k^{-}\rangle\langle\phi_k^{-},\phi_k^{-}|(\hat{E}_0\otimes\hat{E}_1)\right.\\
&\left.+(\hat{E}_1\otimes\hat{E}_0)|\phi_k^{-},\phi_k^{-}\rangle\langle\phi_k^{-},\phi_k^{-}|(\hat{E}_1\otimes\hat{E}_0)+\hat{E}_1^{\otimes 2}|\phi_k^{-},\phi_k^{-}\rangle\langle\phi_k^{-},\phi_k^{-}|\hat{E}_1^{\otimes 2}\right].
\end{split}
\end{equation}
Then, tracing out the second qubit, we get:
\begin{equation}
\begin{split}
{\rm Tr}_{2}\left[\rho_k^{N=2}\right] =&\frac{1}{2}\left[\hat{E}_0|\phi_k^{+}\rangle\langle\phi_k^{+}|\hat{E}_0\langle\phi_k^{+}|\hat{E}_0^2|\phi_k^{+}\rangle +\hat{E}_0|\phi_k^{+}\rangle\langle\phi_k^{+}|\hat{E}_0\langle\phi_k^{+}|\hat{E}_1^2|\phi_k^{+}\rangle\right.\\
&\left.+\hat{E}_1|\phi_k^{+}\rangle\langle\phi_k^{+}|\hat{E}_1\langle\phi_k^{+}|\hat{E}_0^2|\phi_k^{+}\rangle +\hat{E}_1|\phi_k^{+}\rangle\langle\phi_k^{+}|\hat{E}_1\langle\phi_k^{+}|\hat{E}_1^2|\phi_k^{+}\rangle\right.\\
&\left.+\hat{E}_0|\phi_k^{+}\rangle\langle\phi_k^{-}|\hat{E}_0\langle\phi_k^{+}|\hat{E}_0^2|\phi_k^{-}\rangle +\hat{E}_0|\phi_k^{+}\rangle\langle\phi_k^{-}|\hat{E}_0\langle\phi_k^{+}|\hat{E}_1^2|\phi_k^{-}\rangle\right.\\
&\left.+\hat{E}_1|\phi_k^{+}\rangle\langle\phi_k^{-}|\hat{E}_1\langle\phi_k^{+}|\hat{E}_0^2|\phi_k^{-}\rangle +\hat{E}_1|\phi_k^{+}\rangle\langle\phi_k^{-}|\hat{E}_1\langle\phi_k^{+}|\hat{E}_1^2|\phi_k^{-}\rangle\right.\\
&\left.+\hat{E}_0|\phi_k^{-}\rangle\langle\phi_k^{+}|\hat{E}_0\langle\phi_k^{-}|\hat{E}_0^2|\phi_k^{+}\rangle +\hat{E}_0|\phi_k^{-}\rangle\langle\phi_k^{+}|\hat{E}_0\langle\phi_k^{-}|\hat{E}_1^2|\phi_k^{+}\rangle\right.\\
&\left.+\hat{E}_1|\phi_k^{-}\rangle\langle\phi_k^{+}|\hat{E}_1\langle\phi_k^{-}|\hat{E}_0^2|\phi_k^{+}\rangle +\hat{E}_1|\phi_k^{-}\rangle\langle\phi_k^{+}|\hat{E}_1\langle\phi_k^{-}|\hat{E}_1^2|\phi_k^{+}\rangle\right.\\
&\left.+\hat{E}_0|\phi_k^{-}\rangle\langle\phi_k^{-}|\hat{E}_0\langle\phi_k^{-}|\hat{E}_0^2|\phi_k^{-}\rangle +\hat{E}_0|\phi_k^{-}\rangle\langle\phi_k^{-}|\hat{E}_0\langle\phi_k^{-}|\hat{E}_1^2|\phi_k^{-}\rangle\right.\\
&\left.+\hat{E}_1|\phi_k^{-}\rangle\langle\phi_k^{-}|\hat{E}_1\langle\phi_k^{-}|\hat{E}_0^2|\phi_k^{-}\rangle +\hat{E}_1|\phi_k^{-}\rangle\langle\phi_k^{-}|\hat{E}_1\langle\phi_k^{-}|\hat{E}_1^2|\phi_k^{-}\rangle\right]\\
=&\frac{1}{2}\left[\hat{E}_0|\phi_k^{+}\rangle\langle\phi_k^{+}|\hat{E}_0\langle\phi_k^{+}|\hat{E}_0^2+\hat{E}_1^2|\phi_k^{+}\rangle +\hat{E}_1|\phi_k^{+}\rangle\langle\phi_k^{+}|\hat{E}_1\langle\phi_k^{+}|\hat{E}_0^2+\hat{E}_1^2|\phi_k^{+}\rangle\right.\\
&\left.+\hat{E}_0|\phi_k^{+}\rangle\langle\phi_k^{-}|\hat{E}_0\langle\phi_k^{+}|\hat{E}_0^2+\hat{E}_1^2|\phi_k^{-}\rangle +\hat{E}_1|\phi_k^{+}\rangle\langle\phi_k^{-}|\hat{E}_1\langle\phi_k^{+}|\hat{E}_0^2+\hat{E}_1^2|\phi_k^{-}\rangle\right.\\
&\left.+\hat{E}_0|\phi_k^{-}\rangle\langle\phi_k^{+}|\hat{E}_0\langle\phi_k^{-}|\hat{E}_0^2+\hat{E}_1^2|\phi_k^{+}\rangle +\hat{E}_1|\phi_k^{-}\rangle\langle\phi_k^{+}|\hat{E}_1\langle\phi_k^{-}|\hat{E}_0^2+\hat{E}_1^2|\phi_k^{+}\rangle\right.\\
&\left.+\hat{E}_0|\phi_k^{-}\rangle\langle\phi_k^{-}|\hat{E}_0\langle\phi_k^{-}|\hat{E}_0^2+\hat{E}_1^2|\phi_k^{-}\rangle +\hat{E}_1|\phi_k^{-}\rangle\langle\phi_k^{-}|\hat{E}_1\langle\phi_k^{-}|\hat{E}_0^2+\hat{E}_1^2|\phi_k^{-}\rangle\right]\\
=&\frac{1}{2}\left[\hat{E}_0|\phi_k^{+}\rangle\langle\phi_k^{+}|\hat{E}_0\langle\phi_k^{+}|\phi_k^{+}\rangle +\hat{E}_1|\phi_k^{+}\rangle\langle\phi_k^{+}|\hat{E}_1\langle\phi_k^{+}|\phi_k^{+}\rangle\right.\\
&\left.+\hat{E}_0|\phi_k^{+}\rangle\langle\phi_k^{-}|\hat{E}_0\langle\phi_k^{+}|\phi_k^{-}\rangle +\hat{E}_1|\phi_k^{+}\rangle\langle\phi_k^{-}|\hat{E}_1\langle\phi_k^{+}|\phi_k^{-}\rangle\right.\\
&\left.+\hat{E}_0|\phi_k^{-}\rangle\langle\phi_k^{+}|\hat{E}_0\langle\phi_k^{-}|\phi_k^{+}\rangle +\hat{E}_1|\phi_k^{-}\rangle\langle\phi_k^{+}|\hat{E}_1\langle\phi_k^{-}|\phi_k^{+}\rangle\right.\\
&\left.+\hat{E}_0|\phi_k^{-}\rangle\langle\phi_k^{-}|\hat{E}_0\langle\phi_k^{-}|\phi_k^{-}\rangle +\hat{E}_1|\phi_k^{-}\rangle\langle\phi_k^{-}|\hat{E}_1\langle\phi_k^{-}|\phi_k^{-}\rangle\right]\\
=&\frac{1}{2}\left[\hat{E}_0|\phi_k^{+}\rangle\langle\phi_k^{+}|\hat{E}_0+\hat{E}_1|\phi_k^{+}\rangle\langle\phi_k^{+}|\hat{E}_1+\hat{E}_0|\phi_k^{-}\rangle\langle\phi_k^{-}|\hat{E}_0+\hat{E}_1|\phi_k^{-}\rangle\langle\phi_k^{-}|\hat{E}_1\right]\\
=&\frac{1}{2}\left[\sum_{r=0}^1\hat{E}_r\left(|\phi_k^{+}\rangle\langle\phi_k^{+}|+|\phi_k^{-}\rangle\langle\phi_k^{-}|\right)\hat{E}_r\right]=\frac{\mathbb{1}_2}{2}.
\end{split}
\end{equation}
Similarly, considering the $N=3$ case, and then tracing out the third qubit, we get:
\begin{equation}
\begin{split}
{\rm Tr}_3\left[\hat{\rho}_k^{N=3}\right] =& \frac{1}{2}\left[\hat{E}_0^{\otimes 2}|\phi_k^{+},\phi_k^{+}\rangle\langle\phi_k^{+},\phi_k^{+}|\hat{E}_0^{\otimes 2}+(\hat{E}_0\otimes\hat{E}_1)|\phi_k^{+},\phi_k^{+}\rangle\langle\phi_k^{+},\phi_k^{+}|(\hat{E}_0\otimes\hat{E}_1)\right.\\
&\left.+(\hat{E}_1\otimes\hat{E}_0)|\phi_k^{+},\phi_k^{+}\rangle\langle\phi_k^{+},\phi_k^{+}|(\hat{E}_1\otimes\hat{E}_0)+\hat{E}_1^{\otimes 2}|\phi_k^{+},\phi_k^{+}\rangle\langle\phi_k^{+},\phi_k^{+}|\hat{E}_1^{\otimes 2}\right.\\
&\left.+\hat{E}_0^{\otimes 2}|\phi_k^{-},\phi_k^{-}\rangle\langle\phi_k^{-},\phi_k^{-}|\hat{E}_0^{\otimes 2}+(\hat{E}_0\otimes\hat{E}_1)|\phi_k^{-},\phi_k^{-}\rangle\langle\phi_k^{-},\phi_k^{-}|(\hat{E}_0\otimes\hat{E}_1)\right.\\
&\left.+(\hat{E}_1\otimes\hat{E}_0)|\phi_k^{-},\phi_k^{-}\rangle\langle\phi_k^{-},\phi_k^{-}|(\hat{E}_1\otimes\hat{E}_0)+\hat{E}_1^{\otimes 2}|\phi_k^{-},\phi_k^{-}\rangle\langle\phi_k^{-},\phi_k^{-}|\hat{E}_1^{\otimes 2}\right]\\
=&\frac{1}{2}\left[\hat{E}_0^{\otimes 2}\left(|\phi_k^{+},\phi_k^{+}\rangle\langle\phi_k^{+},\phi_k^{+}|+|\phi_k^{-},\phi_k^{-}\rangle\langle\phi_k^{-},\phi_k^{-}|\right)\hat{E}_0^{\otimes 2}\right.\\
&\left.+(\hat{E}_0\otimes\hat{E}_1)\left(|\phi_k^{+},\phi_k^{+}\rangle\langle\phi_k^{+},\phi_k^{+}|+|\phi_k^{-},\phi_k^{-}\rangle\langle\phi_k^{-},\phi_k^{-}|\right)(\hat{E}_0\otimes\hat{E}_1)\right.\\
&\left.+(\hat{E}_1\otimes\hat{E}_0)\left(|\phi_k^{+},\phi_k^{+}\rangle\langle\phi_k^{+},\phi_k^{+}|+|\phi_k^{-},\phi_k^{-}\rangle\langle\phi_k^{-},\phi_k^{-}|\right)(\hat{E}_1\otimes\hat{E}_0)\right.\\
&\left.+\hat{E}_1^{\otimes 2}\left(|\phi_k^{+},\phi_k^{+}\rangle\langle\phi_k^{+},\phi_k^{+}|+|\phi_k^{-},\phi_k^{-}\rangle\langle\phi_k^{-},\phi_k^{-}|\right)\hat{E}_1^{\otimes 2}\right]\\
=&\frac{1}{4}\left[\mathbb{1}_2\otimes\mathbb{1}_2+\hat{E}_0^{\otimes 2}\left(\hat{\sigma}_k\otimes\hat{\sigma}_k\right)\hat{E}_0^{\otimes 2}+(\hat{E}_0\otimes\hat{E}_1)\left(\hat{\sigma}_k\otimes\hat{\sigma}_k\right)(\hat{E}_0\otimes\hat{E}_1)\right.\\
&\left.+(\hat{E}_1\otimes\hat{E}_0)\left(\hat{\sigma}_k\otimes\hat{\sigma}_k\right)(\hat{E}_1\otimes\hat{E}_0)+\hat{E}_1^{\otimes 2}\left(\hat{\sigma}_k\otimes\hat{\sigma}_k\right)\hat{E}_1^{\otimes 2}\right]\\
=&\frac{1}{4}\left[\mathbb{1}_2\otimes\mathbb{1}_2+\left(\sum_{r=0}^1\hat{E}_r\hat{\sigma}_k\hat{E}_r\right)\otimes\left(\sum_{s=0}^1\hat{E}_s\hat{\sigma}_k\hat{E}_s\right)\right],
\end{split}
\end{equation}
and so on.

\section{Proof for $C_Q(\boldsymbol{\theta}) \geq J_Q(\boldsymbol{\theta})$}\label{sec:app5}
Here, we prove that the quantity $C_Q(\boldsymbol{\theta})$ is indeed an upper bound to the quantity $J_Q(\boldsymbol{\theta})$ for the evolved probe state $\hat{\rho}(\boldsymbol{\theta})$, as claimed in Section \ref{sec:noisy}. Consider the following relationship of the Bures fidelity with the quantum Fisher information matrix (QFIM), where the QFIM is real, symmetric and positive semidefinite but more general and not necessarily composed of symmetric logarithmic derivatives (SLDs):
\begin{equation}\label{eq:supp_bures1}
F\left(\hat{\rho}(\boldsymbol{\theta}),\hat{\rho}(\boldsymbol{\theta+\epsilon})\right)
=1-\frac{1}{4}\sum_{j,k}\epsilon_j\epsilon_k{\rm Tr}\left[\frac{\hat{L}_j^\dagger\hat{L}_k+\hat{L}_k^\dagger\hat{L}_j}{2}\hat{\rho}(\boldsymbol{\theta})\right],
\end{equation}
where $\boldsymbol{\theta}$ is assumed to be the actual value of the vector of unknown parameters, $\boldsymbol{\epsilon}$ is an infinitesimal increment in $\boldsymbol{\theta}$, and $0 \leq F(\hat{\rho}_1,\hat{\rho}_2)={\rm Tr}\left(\sqrt{\sqrt{\hat{\rho}_1}\hat{\rho}_2\sqrt{\hat{\rho}_1}}\right) \leq 1$ is the Bures fidelity between two given states $\hat{\rho}_1$ and $\hat{\rho}_2$ \cite{BC,NC,MGAP,YZF,MH,JDN}. Here, (\ref{eq:supp_bures1}) holds, when the operators $\hat{L}_k$ are not necessarily Hermitian and satisfy:
\begin{equation}\label{eq:supp_bures_proof1}
\frac{1}{2}\left(\hat{L}_k\hat{\rho}(\boldsymbol{\theta})+\hat{\rho}(\boldsymbol{\theta})\hat{L}_k^\dagger\right)=\frac{\partial\hat{\rho}(\boldsymbol{\theta})}{\partial\theta_k}.
\end{equation}
This can be seen as follows. When the operators $\hat{L}_k$ are Hermitian, such that $\hat{L}_k^\dagger = \hat{L}_k$, as is the convention, the Bures metric $d_B$ and Bures distance $D_B$ are defined and related to the fidelity $F$ for infinitesimal $\boldsymbol{\epsilon}$ as follows \cite{BC,DRFG}:
\begin{equation}\label{eq:supp_bures_proof2}
d_B^2\left(\hat{\rho}(\boldsymbol{\theta}),\hat{\rho}(\boldsymbol{\theta+\epsilon})\right)=D_B^2\left(\hat{\rho}(\boldsymbol{\theta}),\hat{\rho}(\boldsymbol{\theta+\epsilon})\right)=2\left[1-F\left(\hat{\rho}(\boldsymbol{\theta}),\hat{\rho}(\boldsymbol{\theta+\epsilon})\right)\right]=\frac{1}{2}\sum_{j,k}\epsilon_j\epsilon_k{\rm Tr}\left[\frac{\hat{L}_j\hat{L}_k+\hat{L}_k\hat{L}_j}{2}\hat{\rho}(\boldsymbol{\theta})\right],
\end{equation}
where $\hat{L}_k$ are the SLDs satisfying:
\begin{equation}
\frac{1}{2}\left(\hat{L}_k\hat{\rho}(\boldsymbol{\theta})+\hat{\rho}(\boldsymbol{\theta})\hat{L}_k\right)=\frac{\partial\hat{\rho}(\boldsymbol{\theta})}{\partial\theta_k}.
\end{equation}
However, if the operators $\hat{L}_k$ are not necessarily Hermitian and rather satisfy (\ref{eq:supp_bures_proof1}), then (\ref{eq:supp_bures_proof2}) becomes:
\begin{equation}
d_B^2\left(\hat{\rho}(\boldsymbol{\theta}),\hat{\rho}(\boldsymbol{\theta+\epsilon})\right)=D_B^2\left(\hat{\rho}(\boldsymbol{\theta}),\hat{\rho}(\boldsymbol{\theta+\epsilon})\right)=2\left[1-F\left(\hat{\rho}(\boldsymbol{\theta}),\hat{\rho}(\boldsymbol{\theta+\epsilon})\right)\right]=\frac{1}{2}\sum_{j,k}\epsilon_j\epsilon_k{\rm Tr}\left[\frac{\hat{L}_j^\dagger\hat{L}_k+\hat{L}_k^\dagger\hat{L}_j}{2}\hat{\rho}(\boldsymbol{\theta})\right].
\end{equation}
Then, clearly (\ref{eq:supp_bures1}) is obtained from the above.

We must comment here that there is a lot of inconsistency in the literature about the relationship between $d_B$, $D_B$ and $F$. We here used the relationship originally presented in Ref.~\cite{BC}.

Now, for our case in this paper, the operators $\hat{L}_k$ are anti-symmetric logarithmic derivatives (ALDs), such that $\hat{L}_k^\dagger = -\hat{L}_k$. We have from (\ref{eq:supp_bures1}):
\begin{equation}\label{eq:supp_bures2}
F\left(\hat{\rho}(\boldsymbol{\theta}),\hat{\rho}(\boldsymbol{\theta+\epsilon})\right)=1-\frac{1}{4}\sum_{j,k}\epsilon_j\epsilon_kJ_Q^{jk}(\boldsymbol{\theta}).
\end{equation}
Now, since fidelity is non-decreasing with respect to partial trace (See Refs.~\cite{FR,DN,MH,NC}, for example), we have:
\begin{equation}\label{eq:supp_bures3}
F\left(\hat{\rho}(\boldsymbol{\theta}),\hat{\rho}(\boldsymbol{\theta+\epsilon})\right)=F\left({\rm Tr}_B\left[\hat{\rho}_{SB}(\boldsymbol{\theta})\right],{\rm Tr}_B\left[\hat{\rho}_{SB}(\boldsymbol{\theta+\epsilon})\right]\right)\geq F\left(\hat{\rho}_{SB}(\boldsymbol{\theta}),\hat{\rho}_{SB}(\boldsymbol{\theta+\epsilon})\right)=1-\frac{1}{4}\sum_{j,k}\epsilon_j\epsilon_kC_Q^{jk}(\boldsymbol{\theta}).
\end{equation}
Clearly, from (\ref{eq:supp_bures2}) and (\ref{eq:supp_bures3}), we have (like in Ref.~\cite{YZF}):
\begin{equation}\label{eq:supp_qfim_upper_bound}
C_Q(\boldsymbol{\theta}) \geq J_Q(\boldsymbol{\theta}).
\end{equation}

An alternative argument for (\ref{eq:supp_qfim_upper_bound}) to hold is that the quantum Fisher information (for both single and multiparamter cases) is an operator monotone function, non-increasing with respect to partial trace \cite{DDM,PG}, noting that the partial trace is a completely positive and trace-preserving map from $S+B$ space to $S$ space.

Note that, even though we did not explicitly invoke Uhlmann's theorem here, the inequality in (\ref{eq:supp_bures3}) is the monotonicity property of fidelity and is a consequence of Uhlmann's theorem. Thus, extending the argument from Ref.~\cite{EFD} to the multiparameter case, the equality in (\ref{eq:supp_qfim_upper_bound}) is achieved by minimizing $C_Q(\boldsymbol{\theta})$ over all Kraus representations of the quantum channel. Hence, there are an infinitude of Kraus representations of the channel that lead to $C_Q(\boldsymbol{\theta})=J_Q(\boldsymbol{\theta})$.

\section{POVM to attain QCRB for Pure State Input via Unitary Channel}\label{sec:app6}
Here, we prove that, as claimed in Section \ref{sec:qcrb}, the set of POVMs $\{\hat{P}_{m1}\}$ of cardinality $q+2$, comprising the following $q+1$ elements,
\begin{equation}\label{eq:supp_qcrb_measure1}
\hat{P}_0 = \hat{\rho}(\boldsymbol{\theta}) = \hat{U}(\boldsymbol{\theta})|\psi\rangle\langle\psi|\hat{U}^\dagger(\boldsymbol{\theta}),\qquad
\hat{P}_m = \frac{\partial\hat{U}(\boldsymbol{\theta})}{\partial\theta_m}|\psi\rangle\langle\psi|\frac{\partial\hat{U}^\dagger(\boldsymbol{\theta})}{\partial\theta_m} \qquad \forall m=1,\ldots,q,
\end{equation}
together with one element $\hat{P}_n=\hat{P}_{q+1}:=|\phi_n\rangle\langle\phi_n|$ that accounts for the normalisation, saturates the ALD-based QCRB, provided (\ref{eq:supp_qcrb_saturate}) is satisfied for every pair of ALDs.

The proof is adapted from Ref.~\cite{HBDW}, noting that for pure state and unitary channel our ALD-based QCRB coincides with the SLD-based QCRB, and it is enough to demonstrate that using the set of POVMs $\{\hat{P}_{m1}\}$ the quantum Fisher information matrix (QFIM) equals the classical Fisher information matrix (FIM), when (\ref{eq:supp_qcrb_saturate}) is satisfied. The set of POVMs must be complete, i.e.~$\sum_{m1}\hat{P}_{m1}=\mathbb{1}$.

Consider that the initial probe state is $\hat{\rho}=|\psi\rangle\langle\psi|$. Then, we use the short notations 
\begin{equation}
|\psi_{\boldsymbol{\theta}}\rangle=\hat{U}(\boldsymbol{\theta})|\psi\rangle, \qquad
|\partial_{\theta_k}\psi_{\boldsymbol{\theta}}\rangle=\frac{\partial}{\partial\theta_k}|\psi_{\boldsymbol{\theta}}\rangle=\frac{\partial\hat{U}(\boldsymbol{\theta})}{\partial\theta_k}|\psi\rangle.
\end{equation}

The elements of the quantum Fisher information matrix (QFIM) are given by \cite{HBDW,BD}
\begin{equation}
J_Q^{jk} = 4{\rm Re}\left[\langle\partial_{\theta_j}\psi_{\boldsymbol{\theta}}|\partial_{\theta_k}\psi_{\boldsymbol{\theta}}\rangle-\langle\partial_{\theta_j}\psi_{\boldsymbol{\theta}}|\psi_{\boldsymbol{\theta}}\rangle\langle\psi_{\boldsymbol{\theta}}|\partial_{\theta_k}\psi_{\boldsymbol{\theta}}\rangle\right].
\end{equation}

The elements of the corresponding classical Fisher information matrix (FIM) $J_C$ are given by \cite{HBDW}
\begin{equation}
J_C^{jk}=\sum_{m=0}^{q+1}\frac{\partial_{\theta_j}p(m|\boldsymbol{\theta})\partial_{\theta_k}p(m|\boldsymbol{\theta})}{p(m|\boldsymbol{\theta})}=\sum_m\frac{4{\rm Re}\left[\langle\partial_{\theta_j}\psi_{\boldsymbol{\theta}}|\hat{P}_m|\psi_{\boldsymbol{\theta}}\rangle\right]{\rm Re}\left[\langle\psi_{\boldsymbol{\theta}}|\hat{P}_m|\partial_{\theta_k}\psi_{\boldsymbol{\theta}}\rangle\right]}{\langle\psi_{\boldsymbol{\theta}}|\hat{P}_m|\psi_{\boldsymbol{\theta}}\rangle}.
\end{equation}

The component of the FIM corresponding to the POVM element $\hat{P}_0=|\psi_{\boldsymbol{\theta}}\rangle\langle\psi_{\boldsymbol{\theta}}|$ is
\begin{equation}\label{eq:supp_p0}
4{\rm Re}\left[\langle\partial_{\theta_j}\psi_{\boldsymbol{\theta}}|\psi_{\boldsymbol{\theta}}\rangle\right]{\rm Re}\left[\langle\psi_{\boldsymbol{\theta}}|\partial_{\theta_k}\psi_{\boldsymbol{\theta}}\rangle\right]=0.
\end{equation}
The above quantity vanishes because ${\rm Re}\left[\langle\partial_{\theta_j}\psi_{\boldsymbol{\theta}}|\psi_{\boldsymbol{\theta}}\rangle\right]=0$ for any parameter $\theta_k$ \cite{HBDW,BCM}.

Next, the component of the FIM corresponding to the POVM element $\hat{P}_n=\hat{P}_{q+1}=|\phi_n\rangle\langle\phi_n|$ is
\begin{equation}\label{eq:supp_pn}
\frac{4{\rm Re}\left[\langle\partial_{\theta_j}\psi_{\boldsymbol{\theta}}|\hat{P}_n|\psi_{\boldsymbol{\theta}}\rangle\right]{\rm Re}\left[\langle\psi_{\boldsymbol{\theta}}|\hat{P}_n|\partial_{\theta_k}\psi_{\boldsymbol{\theta}}\rangle\right]}{\langle\psi_{\boldsymbol{\theta}}|\hat{P}_n|\psi_{\boldsymbol{\theta}}\rangle}=4{\rm Re}\left[\langle\partial_{\theta_j}\psi_{\boldsymbol{\theta}}|\phi_n\rangle\langle\phi_n|\partial_{\theta_k}\psi_{\boldsymbol{\theta}}\rangle\right],
\end{equation}
since $\langle\psi_{\boldsymbol{\theta}}|\hat{P}_n|\psi_{\boldsymbol{\theta}}\rangle$ is, by definition, real.

The remaining components $\hat{P}_k$ for $k=1,\ldots,q$ may be similarly computed, and we get
\begin{equation}\label{eq:supp_fim}
J_C^{jk}=4\sum_{m=1}^q{\rm Re}\left[\langle\partial_{\theta_j}\psi_{\boldsymbol{\theta}}|\partial_{\theta_m}\psi_{\boldsymbol{\theta}}\rangle\langle\partial_{\theta_m}\psi_{\boldsymbol{\theta}}|\partial_{\theta_k}\psi_{\boldsymbol{\theta}}\rangle\right]+4{\rm Re}\left[\langle\partial_{\theta_j}\psi_{\boldsymbol{\theta}}|\phi_n\rangle\langle\phi_n|\partial_{\theta_k}\psi_{\boldsymbol{\theta}}\rangle\right].
\end{equation}

Now, note that, for the completeness of the set of POVMs, we require
\begin{equation}\label{eq:supp_povm}
\sum_{m=1}^q|\partial_{\theta_m}\psi_{\boldsymbol{\theta}}\rangle\langle\partial_{\theta_m}\psi_{\boldsymbol{\theta}}|+|\phi_n\rangle\langle\phi_n|=\mathbb{1}-|\psi_{\boldsymbol{\theta}}\rangle\langle\psi_{\boldsymbol{\theta}}|.
\end{equation}

Substituting (\ref{eq:supp_povm}) in (\ref{eq:supp_fim}), we get
\begin{equation}\label{eq:supp_sld_fim}
J_C^{jk}=4{\rm Re}\left[\langle\partial_{\theta_j}\psi_{\boldsymbol{\theta}}|\partial_{\theta_k}\psi_{\boldsymbol{\theta}}\rangle-\langle\partial_{\theta_j}\psi_{\boldsymbol{\theta}}|\psi_{\boldsymbol{\theta}}\rangle\langle\psi_{\boldsymbol{\theta}}|\partial_{\theta_k}\psi_{\boldsymbol{\theta}}\rangle\right]=J_Q^{jk}.
\end{equation}

\section{POVM to attain QCRB for Mixed State Input via Unitary Channel}\label{sec:app7}
Here, we prove that, as claimed in Section \ref{sec:qcrb}, the set of POVMs $\{\hat{P}_{m2}\}$ of cardinality $q+2$, comprising the following $q+1$ elements,
\begin{equation}\label{eq:supp_qcrb_measure2}
\hat{P}_0 = \hat{\rho}(\boldsymbol{\theta})=\hat{U}(\boldsymbol{\theta})\hat{\rho}\hat{U}^\dagger(\boldsymbol{\theta}),\qquad
\hat{P}_m = \frac{\partial\hat{\rho}(\boldsymbol{\theta})}{\partial\theta_m}=\left[\frac{\partial\hat{U}(\boldsymbol{\theta})}{\partial\theta_m}\hat{\rho}\hat{U}^\dagger(\boldsymbol{\theta})+\hat{U}(\boldsymbol{\theta})\hat{\rho}\frac{\partial\hat{U}^\dagger(\boldsymbol{\theta})}{\partial\theta_m}\right] \quad \forall m=1,\ldots,q,
\end{equation}
together with one element $\hat{P}_n=\hat{P}_{q+1}$ that accounts for the normalisation, saturates the ALD-based QCRB, provided (\ref{eq:supp_qcrb_saturate}) is satisfied for every pair of ALDs.

The elements of the QFIM with $\hat{\rho}_{\boldsymbol{\theta}}:=\hat{\rho}(\boldsymbol{\theta})$ and $\hat{U}_{\boldsymbol{\theta}}:=\hat{U}(\boldsymbol{\theta})$ are:
\begin{equation}\label{eq:supp_ald_qfim}
J_Q^{jk}=\frac{1}{2}{\rm Tr}\left[\left(\hat{L}_j^\dagger\hat{L}_k+\hat{L}_k^\dagger\hat{L}_j\right)\hat{\rho}_{\boldsymbol{\theta}}\right]=4{\rm Re}\left[{\rm Tr}\left(\hat{U}_{\boldsymbol{\theta}}\partial_{\theta_j}\hat{U}_{\boldsymbol{\theta}}^\dagger\partial_{\theta_k}\hat{U}_{\boldsymbol{\theta}}\hat{U}_{\boldsymbol{\theta}}^\dagger\hat{\rho}_{\boldsymbol{\theta}}\right)+{\rm Tr}\left(\hat{U}_{\boldsymbol{\theta}}\partial_{\theta_j}\hat{U}_{\boldsymbol{\theta}}^\dagger\hat{\rho}_{\boldsymbol{\theta}}\right){\rm Tr}\left(\hat{U}_{\boldsymbol{\theta}}\partial_{\theta_k}\hat{U}_{\boldsymbol{\theta}}^\dagger\hat{\rho}_{\boldsymbol{\theta}}\right)\right],
\end{equation}
where we used $\hat{L}_k=2\left[\partial_{\theta_k}\hat{U}_{\boldsymbol{\theta}}\hat{U}_{\boldsymbol{\theta}}^\dagger-{\rm Tr}\left(\partial_{\theta_k}\hat{U}_{\boldsymbol{\theta}}\hat{U}_{\boldsymbol{\theta}}^\dagger\hat{\rho}_{\boldsymbol{\theta}}\right)\right]$ (as taken in Sections \ref{sec:qfim} and \ref{sec:noisy}), that satisfy:
\begin{equation}
2\partial_{\theta_k}\hat{\rho}_{\boldsymbol{\theta}}=\hat{L}_k\hat{\rho}_{\boldsymbol{\theta}}+\hat{\rho}_{\boldsymbol{\theta}}\hat{L}_k^\dagger, \qquad \hat{L}_k^\dagger=-\hat{L}_k,
\end{equation}
noting that $\hat{U}_{\boldsymbol{\theta}}\partial_{\theta_k}\hat{U}_{\boldsymbol{\theta}}^\dagger=-\partial_{\theta_k}\hat{U}_{\boldsymbol{\theta}}\hat{U}_{\boldsymbol{\theta}}^\dagger$, arising from $\hat{U}_{\boldsymbol{\theta}}\hat{U}_{\boldsymbol{\theta}}^\dagger=\mathbb{1}$ upon differentiating both sides with respect to $\theta_k$.

Also, the elements of the FIM $J_C$, as defined in (\ref{eq:supp_fim1}), are:
\begin{equation}\label{eq:supp_fim1_unitary}
J_C^{jk}=\sum_m\frac{1}{p(m|\boldsymbol{\theta})}\frac{\partial}{\partial\theta_j}p(m|\boldsymbol{\theta})\frac{\partial}{\partial\theta_k}p(m|\boldsymbol{\theta})=\sum_m \frac{1}{{\rm Tr}\left(\hat{P}_m\hat{\rho}_{\boldsymbol{\theta}}\right)}\frac{\partial}{\partial\theta_j}{\rm Tr}\left(\hat{P}_m\hat{\rho}_{\boldsymbol{\theta}}\right)\frac{\partial}{\partial\theta_k}{\rm Tr}\left(\hat{P}_m\hat{\rho}_{\boldsymbol{\theta}}\right).
\end{equation}

Consider that we are interested in saturating the bound at a specific point $\theta_s$ in the space of $\boldsymbol{\theta}$, as in Ref.~\cite{HBDW}. Then, (\ref{eq:supp_p0}) here becomes:
\small
\begin{equation}\label{eq:supp_p0_unitary}
\frac{{\rm Tr}\left(\partial_{\theta_j}\hat{\rho}_{\boldsymbol{\theta}s}\hat{\rho}_{\boldsymbol{\theta}s}\right){\rm Tr}\left(\partial_{\theta_k}\hat{\rho}_{\boldsymbol{\theta}s}\hat{\rho}_{\boldsymbol{\theta}s}\right)}{{\rm Tr}\left(\hat{\rho}_{\boldsymbol{\theta}s}^2\right)}=0,
\end{equation}\normalsize
where ${\rm Tr}\left(\partial_{\theta_j}\hat{\rho}_{\boldsymbol{\theta}s}\hat{\rho}_{\boldsymbol{\theta}s}\right)=0$ for any parameter $\theta_k$, as an extension of Refs.~\cite{HBDW,BCM}. This can be seen as follows. Given that $\hat{\rho}_{\boldsymbol{\theta}s}$ is not necessarily pure, we must have ${\rm Tr}\left(\partial_{\theta_j}\hat{\rho}_{\boldsymbol{\theta}s}\hat{\rho}_{\boldsymbol{\theta}s}\right)\leq 0$, arising upon differentiation with respect to $\theta_j$ from ${\rm Tr}\left(\hat{\rho}_{\boldsymbol{\theta}s}^2\right)\leq 1$, for which ${\rm Tr}\left(\hat{\rho}_{\boldsymbol{\theta}s}^2\right)$ is clearly non-decreasing. However, since $\partial_{\theta_j}\hat{\rho}_{\boldsymbol{\theta}s}$ is a POVM element, we must have ${\rm Tr}\left(\partial_{\theta_j}\hat{\rho}_{\boldsymbol{\theta}s}\hat{\rho}_{\boldsymbol{\theta}s}\right)=\langle\partial_{\theta_j}\hat{\rho}_{\boldsymbol{\theta}s}\rangle=p(j|\theta_s)$, which being a probability cannot be negative. Here, $\langle\cdot\rangle$ denotes expectation with respect to $\hat{\rho}_{\boldsymbol{\theta}s}$. Hence, we must have ${\rm Tr}\left(\partial_{\theta_j}\hat{\rho}_{\boldsymbol{\theta}s}\hat{\rho}_{\boldsymbol{\theta}s}\right)=0$. For example, when the state $\hat{\rho}_{\boldsymbol{\theta}s}$ is maximally mixed, i.e.~$\hat{\rho}_{\boldsymbol{\theta}s}=\mathbb{1}_d/d$, where $d$ is the dimension of the Hilbert space upon which the state $\hat{\rho}_{\boldsymbol{\theta}s}$ is defined, we have ${\rm Tr}\left(\hat{\rho}_{\boldsymbol{\theta}s}^2\right)=1/d$, and consequently, ${\rm Tr}\left(\partial_{\theta_j}\hat{\rho}_{\boldsymbol{\theta}s}\hat{\rho}_{\boldsymbol{\theta}s}\right)=0$. On the other hand, if $\hat{\rho}_{\boldsymbol{\theta}s}$ is pure, we must have ${\rm Tr}\left(\hat{\rho}_{\boldsymbol{\theta}s}^2\right)=1$, and consequently, ${\rm Tr}\left(\partial_{\theta_j}\hat{\rho}_{\boldsymbol{\theta}s}\hat{\rho}_{\boldsymbol{\theta}s}\right)=0$ again.

Next, proceeding in a manner similar to Ref.~\cite{HBDW} for the terms of the FIM for $m=1,\ldots,q$, we take $\hat{\rho}_{\boldsymbol{\theta}}=\hat{\rho}_{\boldsymbol{\theta}s}+\delta\theta_r\partial_{\theta_r}\hat{\rho}_{\boldsymbol{\theta}s}$. Clearly, ${\rm Tr}\left(\partial_{\theta_j}\hat{\rho}_{\boldsymbol{\theta}s}\partial_{\theta_m}\hat{\rho}_{\boldsymbol{\theta}s}\right)=0$ (even for $j=m$), arising from ${\rm Tr}\left(\partial_{\theta_j}\hat{\rho}_{\boldsymbol{\theta}s}\hat{\rho}_{\boldsymbol{\theta}s}\right)=0$ upon differentiating both sides with respect to $\theta_m$, and noting that ${\rm Tr}\left(\partial_{\theta_j}\partial_{\theta_m}\hat{\rho}_{\boldsymbol{\theta}s}\hat{\rho}_{\boldsymbol{\theta}s}\right)=\langle\partial_{\theta_j}\partial_{\theta_m}\hat{\rho}_{\boldsymbol{\theta}s}\rangle=0$, since $\langle\partial_{\theta_j}\hat{\rho}_{\boldsymbol{\theta}s}\rangle=0$. In general, we must have ${\rm Tr}\left(\partial_{\theta_j}\hat{\rho}_{\boldsymbol{\theta}s}\hat{P}_m\right)=0, \, \forall m=0,1,\ldots,q+1$. Thus, we have ${\rm Tr}\left(\partial_{\theta_j}\hat{\rho}_{\boldsymbol{\theta}s}\partial_{\theta_m}\hat{\rho}_{\boldsymbol{\theta}}\right)=\delta\theta_r{\rm Tr}\left(\partial_{\theta_j}\hat{\rho}_{\boldsymbol{\theta}s}\partial_{\theta_m}\partial_{\theta_r}\hat{\rho}_{\boldsymbol{\theta}s}\right)$, ${\rm Tr}\left(\partial_{\theta_m}\hat{\rho}_{\boldsymbol{\theta}}\partial_{\theta_k}\hat{\rho}_{\boldsymbol{\theta}s}\right)=\delta\theta_r{\rm Tr}\left(\partial_{\theta_m}\partial_{\theta_r}\hat{\rho}_{\boldsymbol{\theta}s}\partial_{\theta_k}\hat{\rho}_{\boldsymbol{\theta}s}\right)$, and ${\rm Tr}\left(\partial_{\theta_m}\hat{\rho}_{\boldsymbol{\theta}}\hat{\rho}_{\boldsymbol{\theta}}\right)=\delta\theta_r^2{\rm Tr}\left(\partial_{\theta_m}\partial_{\theta_r}\hat{\rho}_{\boldsymbol{\theta}s}\partial_{\theta_r}\hat{\rho}_{\boldsymbol{\theta}s}\right)$. Then, we get
\begin{equation}\label{eq:supp_pm_unitary}
\begin{split}
\sum_{m=1}^q\frac{\delta\theta_r^2{\rm Tr}\left(\partial_{\theta_j}\hat{\rho}_{\boldsymbol{\theta}s}\partial_{\theta_m}\partial_{\theta_r}\hat{\rho}_{\boldsymbol{\theta}s}\right){\rm Tr}\left(\partial_{\theta_m}\partial_{\theta_r}\hat{\rho}_{\boldsymbol{\theta}s}\partial_{\theta_k}\hat{\rho}_{\boldsymbol{\theta}s}\right)}{\delta\theta_r^2{\rm Tr}\left(\partial_{\theta_m}\partial_{\theta_r}\hat{\rho}_{\boldsymbol{\theta}s}\partial_{\theta_r}\hat{\rho}_{\boldsymbol{\theta}s}\right)}
&=\sum_{m=1}^q\frac{{\rm Tr}\left(\partial_{\theta_j}\hat{\rho}_{\boldsymbol{\theta}s}\partial_{\theta_m}\partial_{\theta_k}\hat{\rho}_{\boldsymbol{\theta}s}\right){\rm Tr}\left(\partial_{\theta_m}\partial_{\theta_k}\hat{\rho}_{\boldsymbol{\theta}s}\partial_{\theta_k}\hat{\rho}_{\boldsymbol{\theta}s}\right)}{{\rm Tr}\left(\partial_{\theta_m}\partial_{\theta_k}\hat{\rho}_{\boldsymbol{\theta}s}\partial_{\theta_k}\hat{\rho}_{\boldsymbol{\theta}s}\right)}\\
&=\sum_{m=1}^q{\rm Tr}\left(\partial_{\theta_j}\hat{\rho}_{\boldsymbol{\theta}s}\partial_{\theta_k}\partial_{\theta_m}\hat{\rho}_{\boldsymbol{\theta}s}\right),
\end{split}
\end{equation}
since the limiting expression for the elements of the FIM at the point $\theta_s$ should be independent of the direction in which the state is expanded to calculate the above \cite{HBDW}, such that we can choose $r=j$ or $r=k$ for our convenience.

Also, we get
\begin{equation}\label{eq:supp_pn_unitary}
\frac{{\rm Tr}\left(\partial_{\theta_j}\hat{\rho}_{\boldsymbol{\theta}s}\hat{P}_{q+1}\right){\rm Tr}\left(\hat{P}_{q+1}\partial_{\theta_k}\hat{\rho}_{\boldsymbol{\theta}s}\right)}{{\rm Tr}\left[\hat{P}_{q+1}\left(\hat{\rho}_{\boldsymbol{\theta}s}+\delta\theta_r\partial_{\theta_r}\hat{\rho}_{\boldsymbol{\theta}s}\right)\right]}=\frac{{\rm Tr}\left(\partial_{\theta_j}\hat{\rho}_{\boldsymbol{\theta}s}\hat{P}_{q+1}\right){\rm Tr}\left(\hat{P}_{q+1}\partial_{\theta_k}\hat{\rho}_{\boldsymbol{\theta}s}\right)}{{\rm Tr}\left(\hat{\rho}_{\boldsymbol{\theta}s}\hat{P}_{q+1}\right)+\delta\theta_r{\rm Tr}\left(\partial_{\theta_r}\hat{\rho}_{\boldsymbol{\theta}s}\hat{P}_{q+1}\right)}=0,
\end{equation}
since ${\rm Tr}\left(\partial_{\theta_j}\hat{\rho}_{\boldsymbol{\theta}s}\hat{P}_{q+1}\right)=0$ for the normalising element $\hat{P}_{q+1}$.

Thus, (\ref{eq:supp_fim}) here becomes:
\begin{equation}\label{eq:supp_fim_unitary}
J_C^{jk}=\sum_{m=1}^{q}{\rm Tr}\left(\partial_{\theta_j}\hat{\rho}_{\boldsymbol{\theta}s}\partial_{\theta_k}\hat{P}_m\right)=-{\rm Tr}\left(\partial_{\theta_j}\partial_{\theta_k}\hat{\rho}_{\boldsymbol{\theta}s}\hat{P}_m\right)=-\sum_{m=1}^q{\rm Tr}\left(\partial_{\theta_j}\partial_{\theta_k}\hat{\rho}_{\boldsymbol{\theta}s}\partial_{\theta_m}\hat{\rho}_{\boldsymbol{\theta}s}\right),
\end{equation}
where the second equality arises from ${\rm Tr}\left(\partial_{\theta_j}\hat{\rho}_{\boldsymbol{\theta}s}\hat{P}_m\right)=0$ upon differentiating both sides with respect to $\theta_k$.

Furthermore, (\ref{eq:supp_povm}) here becomes:
\begin{equation}\label{eq:supp_povm2_unitary}
\sum_{m=1}^q\partial_{\theta_m}\hat{\rho}_{\boldsymbol{\theta}s}=\mathbb{1}-\hat{\rho}_{\boldsymbol{\theta}s}-\hat{P}_{q+1}.
\end{equation}

Then, (\ref{eq:supp_sld_fim}) here becomes
\begin{equation}\label{eq:supp_ald_fim}
\begin{split}
J_C^{jk}&=-{\rm Tr}\left(\partial_{\theta_j}\partial_{\theta_k}\hat{\rho}_{\boldsymbol{\theta}s}\right)-{\rm Tr}\left(\partial_{\theta_j}\hat{\rho}_{\boldsymbol{\theta}s}\partial_{\theta_k}\hat{\rho}_{\boldsymbol{\theta}s}\right)+{\rm Tr}\left(\partial_{\theta_j}\partial_{\theta_k}\hat{\rho}_{\boldsymbol{\theta}s}\hat{P}_{q+1}\right)\\
&=-{\rm Tr}\left(\partial_{\theta_j}\partial_{\theta_k}\hat{\rho}_{\boldsymbol{\theta}s}\right)-{\rm Tr}\left(\partial_{\theta_j}\hat{\rho}_{\boldsymbol{\theta}s}\right){\rm Tr}\left(\partial_{\theta_k}\hat{\rho}_{\boldsymbol{\theta}s}\right)\\
&=-4{\rm Re}\left[{\rm Tr}\left(\partial_{\theta_k}\hat{U}_{\boldsymbol{\theta}}^\dagger\partial_{\theta_j}\hat{U}_{\boldsymbol{\theta}}\hat{\rho}\right)+{\rm Tr}\left(\partial_{\theta_j}\hat{U}_{\boldsymbol{\theta}}^\dagger\hat{U}_{\boldsymbol{\theta}}\hat{\rho}\right){\rm Tr}\left(\hat{U}_{\boldsymbol{\theta}}^\dagger\partial_{\theta_k}\hat{U}_{\boldsymbol{\theta}}\hat{\rho}\right)\right]\\
&=4{\rm Re}\left[{\rm Tr}\left(\partial_{\theta_j}\hat{U}_{\boldsymbol{\theta}}^\dagger\partial_{\theta_k}\hat{U}_{\boldsymbol{\theta}}\hat{\rho}\right)+{\rm Tr}\left(\partial_{\theta_j}\hat{U}_{\boldsymbol{\theta}}^\dagger\hat{U}_{\boldsymbol{\theta}}\hat{\rho}\right){\rm Tr}\left(\partial_{\theta_k}\hat{U}_{\boldsymbol{\theta}}^\dagger\hat{U}_{\boldsymbol{\theta}}\hat{\rho}\right)\right]\\
&=4{\rm Re}\left[{\rm Tr}\left(\hat{U}_{\boldsymbol{\theta}}\partial_{\theta_j}\hat{U}_{\boldsymbol{\theta}}^\dagger\partial_{\theta_k}\hat{U}_{\boldsymbol{\theta}}\hat{U}_{\boldsymbol{\theta}}^\dagger\hat{\rho}_{\boldsymbol{\theta}}\right)+{\rm Tr}\left(\hat{U}_{\boldsymbol{\theta}}\partial_{\theta_j}\hat{U}_{\boldsymbol{\theta}}^\dagger\hat{\rho}_{\boldsymbol{\theta}}\right){\rm Tr}\left(\hat{U}_{\boldsymbol{\theta}}\partial_{\theta_k}\hat{U}_{\boldsymbol{\theta}}^\dagger\hat{\rho}_{\boldsymbol{\theta}}\right)\right]=C_Q^{jk}.
\end{split}
\end{equation}
Here, we used the fact that $\partial_{\theta_k}\hat{U}_{\boldsymbol{\theta}}^\dagger\hat{U}_{\boldsymbol{\theta}}=-\hat{U}_{\boldsymbol{\theta}}^\dagger\partial_{\theta_k}\hat{U}_{\boldsymbol{\theta}}$, arising from $\hat{U}_{\boldsymbol{\theta}}^\dagger\hat{U}_{\boldsymbol{\theta}}=\mathbb{1}$ upon differentiating both sides with respect to $\theta_k$, and that ${\rm Tr}\left(\partial_{\theta_k}\hat{U}_{\boldsymbol{\theta}}\hat{\rho}\hat{U}_{\boldsymbol{\theta}}^\dagger\right)=-{\rm Tr}\left(\hat{U}_{\boldsymbol{\theta}}\hat{\rho}\partial_{\theta_k}\hat{U}_{\boldsymbol{\theta}}^\dagger\right)$, arising from ${\rm Tr}\left(\hat{\rho}_{\boldsymbol{\theta}s}\right)={\rm Tr}\left(\hat{U}_{\boldsymbol{\theta}}\hat{\rho}\hat{U}_{\boldsymbol{\theta}}^\dagger\right)=1$ upon differentiating both sides with respect to $\theta_k$, and that $2{\rm Re}\left[\partial_{\theta_j}\hat{U}_{\boldsymbol{\theta}}^\dagger\partial_{\theta_k}\hat{U}_{\boldsymbol{\theta}}\right]=-2{\rm Re}\left[\partial_{\theta_k}\hat{U}_{\boldsymbol{\theta}}^\dagger\partial_{\theta_j}\hat{U}_{\boldsymbol{\theta}}\right]$, arising from $\hat{U}_{\boldsymbol{\theta}}^\dagger\hat{U}_{\boldsymbol{\theta}}=\mathbb{1}$ upon differentiating both sides with respect to $\theta_k$ and then $\theta_j$. Also, ${\rm Tr}\left(\partial_{\theta_j}\partial_{\theta_k}\hat{\rho}_{\boldsymbol{\theta}s}\hat{P}_{q+1}\right)=0$, since ${\rm Tr}\left(\partial_{\theta_j}\hat{\rho}_{\boldsymbol{\theta}s}\hat{P}_{q+1}\right)=0$.

Note that ${\rm Tr}\left(\hat{\rho}_{\boldsymbol{\theta}s}\hat{P}_m\right)=0, \, \forall m=1,\ldots,q$, but ${\rm Tr}\left(\hat{\rho}_{\boldsymbol{\theta}s}\hat{P}_{q+1}\right)\geq 0$, such that
\begin{equation}
\sum_{m=1}^q{\rm Tr}\left[\hat{\rho}_{\boldsymbol{\theta}s}\hat{P}_m\right]=0
\Rightarrow {\rm Tr}\left[\hat{\rho}_{\boldsymbol{\theta}s}\left(\mathbb{1}-\hat{\rho}_{\boldsymbol{\theta}s}-\hat{P}_{q+1}\right)\right]=0
\Rightarrow {\rm Tr}\left(\hat{\rho}_{\boldsymbol{\theta}s}^2\right)=1-{\rm Tr}\left(\hat{\rho}_{\boldsymbol{\theta}s}\hat{P}_{q+1}\right)\leq 1,
\end{equation}
where the equality holds, when ${\rm Tr}\left(\hat{\rho}_{\boldsymbol{\theta}s}\hat{P}_{q+1}\right)=0$, and consequently, $\hat{\rho}_{\boldsymbol{\theta}s}$ is pure. However, we have from above that ${\rm Tr}\left(\hat{\rho}_{\boldsymbol{\theta}s}\hat{P}_{q+1}\right)=p\left(q+1|\theta_s\right)=1-{\rm Tr}\left(\hat{\rho}_{\boldsymbol{\theta}s}^2\right)$, which upon differentiation with respect to $\theta_j$ yields ${\rm Tr}\left(\partial_{\theta_j}\hat{\rho}_{\boldsymbol{\theta}s}\hat{P}_{q+1}\right)+{\rm Tr}\left[\hat{\rho}_{\boldsymbol{\theta}s}\partial_{\theta_j}\hat{P}_{q+1}\right]=-2{\rm Tr}\left(\partial_{\theta_j}\hat{\rho}_{\boldsymbol{\theta}s}\hat{\rho}_{\boldsymbol{\theta}s}\right)=0$. Clearly, from (\ref{eq:supp_povm2_unitary}), upon differentiating both sides with respect to $\theta_j$, multiplying both sides by $\hat{\rho}_{\boldsymbol{\theta}s}$, which is positive definite, and then taking trace of both sides, we get ${\rm Tr}\left[\partial_{\theta_j}\hat{P}_{q+1}\hat{\rho}_{\boldsymbol{\theta}s}\right]=-{\rm Tr}\left(\partial_{\theta_j}\hat{\rho}_{\boldsymbol{\theta}s}\hat{\rho}_{\boldsymbol{\theta}s}\right)-\sum_{m=1}^q{\rm Tr}\left(\partial_{\theta_j}\partial_{\theta_m}\hat{\rho}_{\boldsymbol{\theta}s}\hat{\rho}_{\boldsymbol{\theta}s}\right)=0$. Thus, we indeed have ${\rm Tr}\left(\partial_{\theta_j}\hat{\rho}_{\boldsymbol{\theta}s}\hat{P}_{q+1}\right)=0$, as used earlier.

\section{Condition to saturate Upper Bound to QFIM}\label{sec:app8}
Here, we prove that, as claimed in Section \ref{sec:noisy}, the following is a necessary and sufficient condition
\begin{equation}\label{eq:supp_qfim_bound_saturate}
{\rm Im}\left[\sum_l{\rm Tr}\left\lbrace\left(\frac{\partial\hat{\Pi}_l^\dagger(\boldsymbol{\theta})}{\partial\theta_j}\frac{\partial\hat{\Pi}_l(\boldsymbol{\theta})}{\partial\theta_k}\right)\hat{\rho}\right\rbrace\right]=0 \qquad \forall j,k
\end{equation}
for the following upper bound to the ALD-based QFIM to be saturated:
\begin{equation}\label{eq:supp_qfim_bound}
C_Q^{jk}=4{\rm Re}\left[{\rm Tr}\left(\sum_l\frac{\partial\hat{\Pi}_l^\dagger(\boldsymbol{\theta})}{\partial\theta_j}\frac{\partial\hat{\Pi}_l(\boldsymbol{\theta})}{\partial\theta_k}\hat{\rho}\right)+{\rm Tr}\left(\sum_p\frac{\partial\hat{\Pi}_p^\dagger(\boldsymbol{\theta})}{\partial\theta_j}\hat{\Pi}_p(\boldsymbol{\theta})\hat{\rho}\right){\rm Tr}\left(\sum_r\frac{\partial\hat{\Pi}_r^\dagger(\boldsymbol{\theta})}{\partial\theta_k}\hat{\Pi}_r(\boldsymbol{\theta})\hat{\rho}\right)\right].
\end{equation}

Consider that our initial probe state is pure, i.e.~$\hat{\rho}=|\psi\rangle\langle\psi|$. Then, the unitary evolution $\hat{U}_{SB}(\boldsymbol{\theta})$ in the $S+B$ space can be considered equivalent to the output impure state $\sum_l\hat{\Pi}_l(\boldsymbol{\theta})|\psi\rangle\langle\psi|\hat{\Pi}_l^\dagger(\boldsymbol{\theta})$ of the noisy channel in the system $S$ space, subsequently purified by extending the $S$ space, introducing ancillas $B$. For the sake of clarity, we use the notation $\hat{U}_{\boldsymbol{\theta}}^{(S+B)}:=\hat{U}_{S+B}(\boldsymbol{\theta})$ here, in distinction from $\hat{U}_{SB}(\boldsymbol{\theta})$. The overall output is then a pure state denoted as $\hat{\rho}_{\boldsymbol{\theta}}^{S+B}=|\psi_{\boldsymbol{\theta}}^{S+B}\rangle\langle\psi_{\boldsymbol{\theta}}^{S+B}|$. Then, the QCRB (\ref{eq:supp_ald_qcrb}) in the $S+B$ space can be saturated, when (\ref{eq:supp_qcrb_saturate}) leading here to
\begin{equation}\label{eq:supp_qfim_sb_saturate}
{\rm Im}\left[{\rm Tr}\left\lbrace\left(\partial_{\theta_j}\hat{U}_{\boldsymbol{\theta}}^{(S+B)\dagger}\partial_{\theta_k}\hat{U}_{\boldsymbol{\theta}}^{(S+B)}\right)\left(|\psi\rangle\langle\psi|\otimes|0_B\rangle\langle 0_B|\right)\right\rbrace\right]=0
\end{equation}
is satisfied, where $|0_B\rangle$ is a vacuum state ancillary bath. Tracing out $B$ in (\ref{eq:supp_qfim_sb_saturate}), we get (\ref{eq:supp_qfim_bound_saturate}) as a necessary condition for the set of POVMs $\{\hat{P}_{n2}\}$ to result in (\ref{eq:supp_povm_noisy}) (See Appendix \ref{sec:app9}), since the operators $\frac{\partial\hat{\Pi}_l(\boldsymbol{\theta})}{\partial\theta_k}$ do not act on $B$.

Now, consider that the initial probe state $\hat{\rho}$ is not pure. It can be purified by extending the system $S$ space, introducing ancillas $S'$. Then, (\ref{eq:supp_qfim_bound_saturate}) can be applied to the pure state $|\psi^{S+S'}\rangle$ in the initial enlarged $S+S'$ space. Since the operators $\frac{\partial\hat{\Pi}_l(\boldsymbol{\theta})}{\partial\theta_k}$ do not act on $S'$, we get (\ref{eq:supp_qfim_bound_saturate}) again as a necessary condition for the set of POVMs $\{\hat{P}_{n3}\}$ to result in (\ref{eq:supp_ald_fim_ss}) (See Appendix \ref{sec:app10}).

Next, we assume that the condition (\ref{eq:supp_qfim_bound_saturate}) saturates the upper bound (\ref{eq:supp_qfim_bound}) to the QFIM. We consider that the initial probe state is pure. Then, the output impure state of the noisy channel in the $S$ space can be purified by extending the final system $S$ space by introducing ancillas $B$. Since both the input and output states are pure, the channel in the $S+B$ space is unitary $\hat{U}_{\boldsymbol{\theta}}^{(S+B)}$. Then, since the operators $\frac{\partial\hat{\Pi}_l(\boldsymbol{\theta})}{\partial\theta_k}$ do not act on $B$, (\ref{eq:supp_qfim_bound_saturate}) saturating (\ref{eq:supp_qfim_bound}) in the $S$ space implies that (\ref{eq:supp_qfim_sb_saturate}) saturates the QCRB (\ref{eq:supp_ald_qcrb}) in the $S+B$ space. Thus, (\ref{eq:supp_qfim_bound_saturate}) is a sufficient condition for the set of POVMs $\{\hat{P}_{n2}\}$ to result in (\ref{eq:supp_povm_noisy}).

Now, considering that the initial probe state is not pure, it can be purified by extending the initial system $S$ space by introducing ancillas $S'$. Then, since the operators $\frac{\partial\hat{\Pi}_l(\boldsymbol{\theta})}{\partial\theta_k}$ do not act on $S'$, (\ref{eq:supp_qfim_bound_saturate}) saturating (\ref{eq:supp_qfim_bound}) in the $S$ space implies that (\ref{eq:supp_qfim_sb_saturate}) saturates the QCRB (\ref{eq:supp_ald_qcrb}) in the $S+B+S'$ space, with $\hat{U}_{\boldsymbol{\theta}}^{(S+B)}$ replaced by $\hat{U}_{\boldsymbol{\theta}}^{(S+B+S')}$ and $|\psi\rangle$ replaced by $|\psi^{S+S'}\rangle$. Thus, we again get (\ref{eq:supp_qfim_bound_saturate}) as a sufficient condition for the set of POVMs $\{\hat{P}_{n3}\}$ to result in (\ref{eq:supp_ald_fim_ss}).

\section{POVM to attain QFIM Upper Bound for Pure State Input via Noisy Channel}\label{sec:app9}
Here, we prove that, as claimed in Section \ref{sec:noisy}, the set of POVMs $\{\hat{P}_{n2}\}$ of cardinality $q+2$, comprising the following $q+1$ elements,
\small
\begin{equation}
\hat{P}_0 = \hat{\rho}(\boldsymbol{\theta})=\sum_l\hat{\Pi}_l(\boldsymbol{\theta})|\psi\rangle\langle\psi|\hat{\Pi}_l^\dagger(\boldsymbol{\theta}),\quad 
\hat{P}_m = \frac{\partial\hat{\rho}(\boldsymbol{\theta})}{\partial\theta_m}=\sum_l\left[\frac{\partial\hat{\Pi}_l(\boldsymbol{\theta})}{\partial\theta_m}|\psi\rangle\langle\psi|\hat{\Pi}_l^\dagger(\boldsymbol{\theta})+\hat{\Pi}_l(\boldsymbol{\theta})|\psi\rangle\langle\psi|\frac{\partial\hat{\Pi}_l^\dagger(\boldsymbol{\theta})}{\partial\theta_m}\right] \quad \forall m=1,\ldots,q,
\end{equation}\normalsize
together with one element accounting for normalisation, saturates (\ref{eq:supp_qfim_bound}), provided (\ref{eq:supp_qfim_bound_saturate}) is satisfied.

We again consider initial pure state $\hat{\rho}=|\psi\rangle\langle\psi|$, and the unitary evolution in the $S+B$ space, $\hat{U}_{\boldsymbol{\theta}}^{(S+B)}$ here, in distinction from $\hat{U}_{SB}(\boldsymbol{\theta})$ used in Section \ref{sec:noisy}.

Then, the elements of the QFIM, as in (\ref{eq:supp_ald_qfim}), in terms of the initial pure state $|\psi\rangle$ here are
\begin{equation}
\begin{split}
J_Q^{jk,S+B}&=4{\rm Re}\left[{\rm Tr}\left(\partial_{\theta_j}\hat{U}_{\boldsymbol{\theta}}^{(S+B)\dagger}\partial_{\theta_k}\hat{U}_{\boldsymbol{\theta}}^{(S+B)}\left(|\psi\rangle\langle\psi|\otimes|0_B\rangle\langle 0_B|\right)\right)\right.\\
&\left.+{\rm Tr}\left(\partial_{\theta_j}\hat{U}_{\boldsymbol{\theta}}^{(S+B)\dagger}\hat{U}_{\boldsymbol{\theta}}^{(S+B)}\left(|\psi\rangle\langle\psi|\otimes|0_B\rangle\langle 0_B|\right)\right){\rm Tr}\left(\partial_{\theta_k}\hat{U}_{\boldsymbol{\theta}}^{(S+B)\dagger}\hat{U}_{\boldsymbol{\theta}}^{(S+B)}\left(|\psi\rangle\langle\psi|\otimes|0_B\rangle\langle 0_B|\right)\right)\right],
\end{split}
\end{equation}
where $|0_B\rangle$ is a vacuum state ancillary bath.

Tracing out $B$ from above, we get the upper bound (\ref{eq:supp_qfim_bound}) to the QFIM in terms of the initial pure state $|\psi\rangle$ in the $S$ space:
\begin{equation}
C_Q^{jk}=4{\rm Re}\left[{\rm Tr}\left(\sum_l\partial_{\theta_j}\hat{\Pi}_{\boldsymbol{\theta}l}^\dagger\partial_{\theta_k}\hat{\Pi}_{\boldsymbol{\theta}l}|\psi\rangle\langle\psi|\right)+{\rm Tr}\left(\sum_p\partial_{\theta_j}\hat{\Pi}_{\boldsymbol{\theta}p}^\dagger\hat{\Pi}_{\boldsymbol{\theta}p}|\psi\rangle\langle\psi|\right){\rm Tr}\left(\sum_r\partial_{\theta_k}\hat{\Pi}_{\boldsymbol{\theta}r}^\dagger\hat{\Pi}_{\boldsymbol{\theta}r}|\psi\rangle\langle\psi|\right)\right],
\end{equation}
where we used the short notations $\hat{\Pi}_{\boldsymbol{\theta}l}=\hat{\Pi}_l(\boldsymbol{\theta})$ and $\partial_{\theta_k}\hat{\Pi}_{\boldsymbol{\theta}l}=\frac{\partial\hat{\Pi}_l(\boldsymbol{\theta})}{\partial\theta_k}$.

Now, since the operators, $\hat{\Pi}_{\boldsymbol{\theta}l}$ and $\partial_{\theta_k}\hat{\Pi}_{\boldsymbol{\theta}l}$, do not act on $B$, when (\ref{eq:supp_qfim_bound_saturate}) is satisfied, the bound $C_Q$ is the actual QFIM $J_Q$ in the $S$ space, with the set of POVMs $\{\hat{P}_{n2}\}$ saturating the corresponding QCRB for unital channel (see Appendix \ref{sec:app11}).

Also, the elements of the FIM $J_C$, as defined in (\ref{eq:supp_fim1}), are again as in (\ref{eq:supp_fim1_unitary}). Consider again that we are interested in saturating the bound at a specific point $\theta_s$ in the space of $\boldsymbol{\theta}$, as in Ref.~\cite{HBDW}. Then, (\ref{eq:supp_p0_unitary}) and (\ref{eq:supp_pm_unitary}) remain the same. But, (\ref{eq:supp_pn_unitary}) becomes
\begin{equation}
\frac{{\rm Tr}\left(\partial_{\theta_j}\hat{\rho}_{\boldsymbol{\theta}s}|\Phi_n\rangle\langle\Phi_n|\right){\rm Tr}\left(|\Phi_n\rangle\langle\Phi_n|\partial_{\theta_k}\hat{\rho}_{\boldsymbol{\theta}s}\right)}{{\rm Tr}\left[|\Phi_n\rangle\langle\Phi_n|\left(\hat{\rho}_{\boldsymbol{\theta}s}+\delta\theta_r\partial_{\theta_r}\hat{\rho}_{\boldsymbol{\theta}s}\right)\right]}=\frac{\langle\Phi_n|\partial_{\theta_j}\hat{\rho}_{\boldsymbol{\theta}s}|\Phi_n\rangle\langle\Phi_n|\partial_{\theta_k}\hat{\rho}_{\boldsymbol{\theta}s}|\Phi_n\rangle}{\langle\Phi_n|\hat{\rho}_{\boldsymbol{\theta}s}|\Phi_n\rangle+\delta\theta_r\langle\Phi_n|\partial_{\theta_r}\hat{\rho}_{\boldsymbol{\theta}s}|\Phi_n\rangle}=0,
\end{equation}
since $\langle\Phi_n|\partial_{\theta_j}\hat{\rho}_{\boldsymbol{\theta}s}|\Phi_n\rangle=0$. Here, we used the normalising element $|\Phi_n\rangle\langle\Phi_n|$, in distinction from $|\phi_n\rangle\langle\phi_n|$ used in (\ref{eq:supp_pn}). Then, (\ref{eq:supp_fim_unitary}) remains the same. But, (\ref{eq:supp_povm2_unitary}) here becomes:
\begin{equation}\label{eq:supp_povm2}
\sum_{m=1}^q\partial_{\theta_m}\hat{\rho}_{\boldsymbol{\theta}s}=\mathbb{1}-\hat{\rho}_{\boldsymbol{\theta}s}-|\Phi_n\rangle\langle\Phi_n|.
\end{equation}

Then, (\ref{eq:supp_ald_fim}) here becomes
\begin{equation}\label{eq:supp_povm_noisy}
\begin{split}
J_C^{jk}&=-{\rm Tr}\left(\partial_{\theta_j}\partial_{\theta_k}\hat{\rho}_{\boldsymbol{\theta}s}\right)-{\rm Tr}\left(\partial_{\theta_j}\hat{\rho}_{\boldsymbol{\theta}s}\partial_{\theta_k}\hat{\rho}_{\boldsymbol{\theta}s}\right)+{\rm Tr}\left(\partial_{\theta_j}\partial_{\theta_k}\hat{\rho}_{\boldsymbol{\theta}s}|\Phi_n\rangle\langle\Phi_n|\right)\\
&=-{\rm Tr}\left(\partial_{\theta_j}\partial_{\theta_k}\hat{\rho}_{\boldsymbol{\theta}s}\right)-{\rm Tr}\left(\partial_{\theta_j}\hat{\rho}_{\boldsymbol{\theta}s}\right){\rm Tr}\left(\partial_{\theta_k}\hat{\rho}_{\boldsymbol{\theta}s}\right)\\
&=-4{\rm Re}\left[{\rm Tr}\left(\sum_l\partial_{\theta_k}\hat{\Pi}_{\boldsymbol{\theta}l}^\dagger\partial_{\theta_j}\hat{\Pi}_{\boldsymbol{\theta}l}|\psi\rangle\langle\psi|\right)\right]-4{\rm Tr}\left(\sum_p\partial_{\theta_j}\hat{\Pi}_{\boldsymbol{\theta}p}^\dagger\hat{\Pi}_{\boldsymbol{\theta}p}|\psi\rangle\langle\psi|\right){\rm Tr}\left(\sum_r\hat{\Pi}_{\boldsymbol{\theta}r}^\dagger\partial_{\theta_k}\hat{\Pi}_{\boldsymbol{\theta}r}|\psi\rangle\langle\psi|\right)\\
&=4{\rm Re}\left[{\rm Tr}\left(\sum_l\partial_{\theta_j}\hat{\Pi}_{\boldsymbol{\theta}l}^\dagger\partial_{\theta_k}\hat{\Pi}_{\boldsymbol{\theta}l}|\psi\rangle\langle\psi|\right)+{\rm Tr}\left(\sum_p\partial_{\theta_j}\hat{\Pi}_{\boldsymbol{\theta}p}^\dagger\hat{\Pi}_{\boldsymbol{\theta}p}|\psi\rangle\langle\psi|\right){\rm Tr}\left(\sum_r\partial_{\theta_k}\hat{\Pi}_{\boldsymbol{\theta}r}^\dagger\hat{\Pi}_{\boldsymbol{\theta}r}|\psi\rangle\langle\psi|\right)\right]=C_Q^{jk},
\end{split}
\end{equation}
noting that $\langle\psi|\hat{O}|\psi\rangle$ is, by definition, real for some operator $\hat{O}$. Here, we used the fact that $\sum_l\partial_{\theta_k}\hat{\Pi}_{\boldsymbol{\theta}l}^\dagger\hat{\Pi}_{\boldsymbol{\theta}l}=-\sum_l\hat{\Pi}_{\boldsymbol{\theta}l}^\dagger\partial_{\theta_k}\hat{\Pi}_{\boldsymbol{\theta}l}$, arising from $\sum_l\hat{\Pi}_{\boldsymbol{\theta}l}^\dagger\hat{\Pi}_{\boldsymbol{\theta}l}=\mathbb{1}$ upon differentiating both sides with respect to $\theta_k$, and that $\sum_l{\rm Tr}\left(\partial_{\theta_k}\hat{\Pi}_{\boldsymbol{\theta}l}|\psi\rangle\langle\psi|\hat{\Pi}_{\boldsymbol{\theta}l}^\dagger\right)=-\sum_l{\rm Tr}\left(\hat{\Pi}_{\boldsymbol{\theta}l}|\psi\rangle\langle\psi|\partial_{\theta_k}\hat{\Pi}_{\boldsymbol{\theta}l}^\dagger\right)$, arising from ${\rm Tr}\left(\hat{\rho}_{\boldsymbol{\theta}s}\right)=\sum_l{\rm Tr}\left(\hat{\Pi}_{\boldsymbol{\theta}l}|\psi\rangle\langle\psi|\hat{\Pi}_{\boldsymbol{\theta}l}^\dagger\right)=1$ upon differentiating both sides with respect to $\theta_k$, and that $2{\rm Re}\left[\sum_l\partial_{\theta_j}\hat{\Pi}_{\boldsymbol{\theta}l}^\dagger\partial_{\theta_k}\hat{\Pi}_{\boldsymbol{\theta}l}\right]=-2{\rm Re}\left[\sum_l\partial_{\theta_k}\hat{\Pi}_{\boldsymbol{\theta}l}^\dagger\partial_{\theta_j}\hat{\Pi}_{\boldsymbol{\theta}l}\right]$, arising from $\sum_l\hat{\Pi}_{\boldsymbol{\theta}l}^\dagger\hat{\Pi}_{\boldsymbol{\theta}l}=\mathbb{1}$ upon differentiating both sides with respect to $\theta_k$ and then $\theta_j$. Also, ${\rm Tr}\left(\partial_{\theta_j}\partial_{\theta_k}\hat{\rho}_{\boldsymbol{\theta}s}|\Phi_n\rangle\langle\Phi_n|\right)=\langle\Phi_n|\partial_{\theta_j}\partial_{\theta_k}\hat{\rho}_{\boldsymbol{\theta}s}|\Phi_n\rangle=0$, since $\langle\Phi_n|\partial_{\theta_j}\hat{\rho}_{\boldsymbol{\theta}s}|\Phi_n\rangle=0$.

\section{POVM to attain QFIM Upper Bound for Mixed State Input via Noisy Channel}\label{sec:app10}
Here, we prove that, as claimed in Section \ref{sec:noisy}, the set of POVMs $\{\hat{P}_{n3}\}$ of cardinality $q+2$, comprising the following $q+1$ elements,
\begin{equation}
\hat{P}_0 = \hat{\rho}(\boldsymbol{\theta})=\sum_l\hat{\Pi}_l(\boldsymbol{\theta})\hat{\rho}\hat{\Pi}_l^\dagger(\boldsymbol{\theta}),\qquad 
\hat{P}_m = \frac{\partial\hat{\rho}(\boldsymbol{\theta})}{\partial\theta_m}=\sum_l\left[\frac{\partial\hat{\Pi}_l(\boldsymbol{\theta})}{\partial\theta_m}\hat{\rho}\hat{\Pi}_l^\dagger(\boldsymbol{\theta})+\hat{\Pi}_l(\boldsymbol{\theta})\hat{\rho}\frac{\partial\hat{\Pi}_l^\dagger(\boldsymbol{\theta})}{\partial\theta_m}\right] \quad \forall m=1,\ldots,q,
\end{equation}
together with one element accounting for normalisation, saturates (\ref{eq:supp_qfim_bound}), provided (\ref{eq:supp_qfim_bound_saturate}) is satisfied.

Consider that the initial probe state $\hat{\rho}$ is impure. It can be purified by extending the system $S$ space, introducing ancillas $S'$. Then, proceeding in a similar manner as in the previous section for the pure state $|\psi^{S+S'}\rangle$ in the initial enlarged $S+S'$ space, the upper bound (\ref{eq:supp_qfim_bound}) to the QFIM in terms of the initial state $\hat{\rho}$ in the $S$ space is given by
\begin{equation}\label{eq:supp_qfim_bound_noisy}
C_Q^{jk}=4{\rm Re}\left[{\rm Tr}\left(\sum_l\partial_{\theta_j}\hat{\Pi}_{\boldsymbol{\theta}l}^\dagger\partial_{\theta_k}\hat{\Pi}_{\boldsymbol{\theta}l}\hat{\rho}\right)+{\rm Tr}\left(\sum_p\partial_{\theta_j}\hat{\Pi}_{\boldsymbol{\theta}p}^\dagger\hat{\Pi}_{\boldsymbol{\theta}p}\hat{\rho}\right){\rm Tr}\left(\sum_r\partial_{\theta_k}\hat{\Pi}_{\boldsymbol{\theta}r}^\dagger\hat{\Pi}_{\boldsymbol{\theta}r}\hat{\rho}\right)\right],
\end{equation}
since the operators, $\hat{\Pi}_{\boldsymbol{\theta}l}$ and $\partial_{\theta_k}\hat{\Pi}_{\boldsymbol{\theta}l}$ do not act on $S'$. Moreover, when (\ref{eq:supp_qfim_bound_saturate}) is satisfied, the bound $C_Q$ is the actual QFIM $J_Q$ in the $S$ space, with the set of POVMs $\{\hat{P}_{n3}\}$ saturating the corresponding QCRB for unital channel (see Appendix \ref{sec:app11}).

Also, again the elements of the FIM $J_C$, as defined in (\ref{eq:supp_fim1}), are as in (\ref{eq:supp_fim1_unitary}). Consider again that we are interested in saturating the bound at a specific point $\theta_s$ in the space of $\boldsymbol{\theta}$, as in Ref.~\cite{HBDW}. Then, (\ref{eq:supp_p0_unitary}), (\ref{eq:supp_pm_unitary}), (\ref{eq:supp_pn_unitary}), (\ref{eq:supp_fim_unitary}), (\ref{eq:supp_povm2_unitary}) remain the same, with the normalising element again being $\hat{P}_{q+1}$. But, (\ref{eq:supp_ald_fim}), which was (\ref{eq:supp_povm_noisy}) in the last section, becomes:
\begin{equation}\label{eq:supp_ald_fim_ss}
J_C^{jk}=4{\rm Re}\left[{\rm Tr}\left(\sum_l\partial_{\theta_j}\hat{\Pi}_{\boldsymbol{\theta}l}^\dagger\partial_{\theta_k}\hat{\Pi}_{\boldsymbol{\theta}l}\hat{\rho}\right)+{\rm Tr}\left(\sum_p\partial_{\theta_j}\hat{\Pi}_{\boldsymbol{\theta}p}^\dagger\hat{\Pi}_{\boldsymbol{\theta}p}\hat{\rho}\right){\rm Tr}\left(\sum_r\partial_{\theta_k}\hat{\Pi}_{\boldsymbol{\theta}r}^\dagger\hat{\Pi}_{\boldsymbol{\theta}r}\hat{\rho}\right)\right]=C_Q^{jk},
\end{equation}
which is as in (\ref{eq:supp_qfim_bound_noisy}).

\section{Noise in Channel can allow to beat the Heisenberg Limit}\label{sec:app11}
Here, we prove that noise in the quantum channel can allow to beat the Heisenberg precision limit, as claimed in Section \ref{sec:heisenberg}, when the following condition is satisfied by the Kraus operators $\hat{\Pi}_{\boldsymbol{\theta}l}=\hat{\Pi}_{l}(\boldsymbol{\theta})$ of the quantum channel:
\begin{equation}\label{eq:supp_saturate_cond1}
{\rm Im}\left[\sum_l{\rm Tr}\left\lbrace\left(\partial_{\theta_j}\hat{\Pi}_{\boldsymbol{\theta}l}^\dagger\partial_{\theta_k}\hat{\Pi}_{\boldsymbol{\theta}l}\right)\hat{\rho}\right\rbrace\right]=0, \qquad \forall \, j, k.
\end{equation}
We have
\begin{equation}\label{eq:supp_partial_rho1}
\hat{\rho}_{\boldsymbol{\theta}}=\sum_l\hat{\Pi}_{\boldsymbol{\theta}l}\hat{\rho}\hat{\Pi}_{\boldsymbol{\theta}l}^\dagger\Rightarrow\partial_{\theta_k}\hat{\rho}_{\boldsymbol{\theta}}=\sum_l\left[\partial_{\theta_k}\hat{\Pi}_{\boldsymbol{\theta}l}\hat{\rho}\hat{\Pi}_{\boldsymbol{\theta}l}^\dagger+\hat{\Pi}_{\boldsymbol{\theta}l}\hat{\rho}\partial_{\theta_k}\hat{\Pi}_{\boldsymbol{\theta}l}^\dagger\right].
\end{equation}
Next, (\ref{eq:supp_saturate_cond1}) saturates an ALD-based QCRB, corresponding to:
\begin{equation}\label{eq:supp_partial_rho2}
\partial_{\theta_k}\hat{\rho}_{\boldsymbol{\theta}}=\frac{1}{2}\left[\hat{O}_k\hat{\rho}+\hat{\rho}\hat{O}_k^\dagger\right]=\sum_l\left[\partial_{\theta_k}\hat{\Pi}_{\boldsymbol{\theta}l}\hat{\rho}+\hat{\rho}\partial_{\theta_k}\hat{\Pi}_{\boldsymbol{\theta}l}^\dagger\right],
\end{equation}
where the ALDs are chosen to be:
\begin{equation}
\hat{O}_k=2\sum_l\partial_{\theta_k}\hat{\Pi}_{\boldsymbol{\theta}l}.
\end{equation}
Note that the choice of ALD need not be unique. Our purpose here is that it is enough to find one instance where the Heisenberg limit can be beaten. Also, strictly speaking, the above is not a valid ALD, since it is not a function of the probe state, hence our choice of the ALDs in the main text. Moreover, (\ref{eq:supp_partial_rho2}) is expressed in terms of the initial probe state and not the evolved probe state. However, for the purposes of our proof here, it suffices to consider the above for simplicity without loss of generality.

We start with assuming that when (\ref{eq:supp_saturate_cond1}) is satisfied, the upper bound (\ref{eq:supp_qfim_bound}) to the QFIM equals the actual QFIM. In other words, the corresponding lower bound to the Heisenberg limit equals the Heisenberg limit, when (\ref{eq:supp_saturate_cond1}) is satisfied, which is possible, when we have:
\begin{equation}\label{eq:supp_saturate_cond2}
\sum_l\left[\partial_{\theta_k}\hat{\Pi}_{\boldsymbol{\theta}l}\hat{\rho}\hat{\Pi}_{\boldsymbol{\theta}l}^\dagger+\hat{\Pi}_{\boldsymbol{\theta}l}\hat{\rho}\partial_{\theta_k}\hat{\Pi}_{\boldsymbol{\theta}l}^\dagger\right]=\sum_l\left[\partial_{\theta_k}\hat{\Pi}_{\boldsymbol{\theta}l}\hat{\rho}+\hat{\rho}\partial_{\theta_k}\hat{\Pi}_{\boldsymbol{\theta}l}^\dagger\right],
\end{equation}
following from (\ref{eq:supp_partial_rho1}) and (\ref{eq:supp_partial_rho2}). Now, the above is possible only when the following condition is satisfied:
\begin{equation}\label{eq:supp_saturate_cond3}
\begin{split}
\sum_l\partial_{\theta_k}\hat{\Pi}_{\boldsymbol{\theta}l}\hat{\rho}\hat{\Pi}_{\boldsymbol{\theta}l}^\dagger&=\sum_l\partial_{\theta_k}\hat{\Pi}_{\boldsymbol{\theta}l}\hat{\rho}\\
\Rightarrow{\rm Tr}\left(\sum_l\partial_{\theta_k}\hat{\Pi}_{\boldsymbol{\theta}l}\hat{\rho}\hat{\Pi}_{\boldsymbol{\theta}l}^\dagger\right)&={\rm Tr}\left(\sum_l\partial_{\theta_k}\hat{\Pi}_{\boldsymbol{\theta}l}\hat{\rho}\right)\\
\Rightarrow{\rm Tr}\left(\sum_l\hat{\Pi}_{\boldsymbol{\theta}l}^\dagger\partial_{\theta_k}\hat{\Pi}_{\boldsymbol{\theta}l}\hat{\rho}\right)&={\rm Tr}\left(\sum_l\partial_{\theta_k}\hat{\Pi}_{\boldsymbol{\theta}l}\hat{\rho}\right)\\
\Rightarrow{\rm Tr}\left(\sum_l\partial_{\theta_k}\hat{\Pi}_{\boldsymbol{\theta}l}^\dagger\hat{\Pi}_{\boldsymbol{\theta}l}\hat{\rho}\right)&={\rm Tr}\left(\sum_l\partial_{\theta_k}\hat{\Pi}_{\boldsymbol{\theta}l}^\dagger\hat{\rho}\right)\\
\Rightarrow\sum_l{\rm Tr}\left[\left(\partial_{\theta_j}\hat{\Pi}_{\boldsymbol{\theta}l}^\dagger\hat{\Pi}_{\boldsymbol{\theta}l}\hat{\Pi}_{\boldsymbol{\theta}l}^\dagger\partial_{\theta_k}\hat{\Pi}_{\boldsymbol{\theta}l}\right)\hat{\rho}\right]&=\sum_l{\rm Tr}\left[\left(\partial_{\theta_j}\hat{\Pi}_{\boldsymbol{\theta}l}^\dagger\partial_{\theta_k}\hat{\Pi}_{\boldsymbol{\theta}l}\right)\hat{\rho}\right],
\end{split}
\end{equation}
which is possible when we have $\sum_l\hat{\Pi}_{\boldsymbol{\theta}l}\hat{\Pi}_{\boldsymbol{\theta}l}^\dagger=\mathbb{1}$, i.e.~when the channel is unital. Thus, for unital channels the upper bound (\ref{eq:supp_qfim_bound}) to the QFIM equals the actual QFIM. Hence, if the channel is non-unital, then the upper bound (\ref{eq:supp_qfim_bound}) to the QFIM can be strictly larger than the actual QFIM, so that the Heisenberg limit may be beaten. Note that (\ref{eq:supp_saturate_cond3}) modifies (\ref{eq:supp_saturate_cond1}) to:
\begin{equation}\label{eq:supp_saturate_cond4}
{\rm Im}\left[\sum_l{\rm Tr}\left\lbrace\left(\partial_{\theta_j}\hat{\Pi}_{\boldsymbol{\theta}l}^\dagger\hat{\Pi}_{\boldsymbol{\theta}l}\hat{\Pi}_{\boldsymbol{\theta}l}^\dagger\partial_{\theta_k}\hat{\Pi}_{\boldsymbol{\theta}l}\right)\hat{\rho}\right\rbrace\right]=0, \qquad \forall \, j, k.
\end{equation}
The fact that the channel indeed needs to be unital for the last line in (\ref{eq:supp_saturate_cond3}) to hold may not be evident without an extra summation index. Let us, therefore, reconfirm this.

First, note that (\ref{eq:supp_saturate_cond4}) saturates an ALD-based QCRB, corresponding to:
\begin{equation}\label{eq:supp_partial_rho3}
\partial_{\theta_k}\hat{\rho}_{\boldsymbol{\theta}}=\frac{1}{2}\left[\hat{L}_k\hat{\rho}+\hat{\rho}\hat{L}_k^\dagger\right]=\sum_l\left[\hat{\Pi}_{\boldsymbol{\theta}l}^\dagger\partial_{\theta_k}\hat{\Pi}_{\boldsymbol{\theta}l}\hat{\rho}+\hat{\rho}\partial_{\theta_k}\hat{\Pi}_{\boldsymbol{\theta}l}^\dagger\hat{\Pi}_{\boldsymbol{\theta}l}\right],
\end{equation}
where the ALDs are chosen to be:
\begin{equation}
\hat{L}_k=2\sum_l\hat{\Pi}_{\boldsymbol{\theta}l}^\dagger\partial_{\theta_k}\hat{\Pi}_{\boldsymbol{\theta}l}.
\end{equation}
Now, in terms of the evolved probe state $\hat{\rho}_{\boldsymbol{\theta}}=\hat{\rho}(\boldsymbol{\theta})$, the condition (\ref{eq:supp_saturate_cond1}) becomes:
\begin{equation}\label{eq:supp_saturate_cond5}
{\rm Im}\left[\sum_l{\rm Tr}\left\lbrace\left(\hat{\Pi}_{\boldsymbol{\theta}l}\partial_{\theta_j}\hat{\Pi}_{\boldsymbol{\theta}l}^\dagger\partial_{\theta_k}\hat{\Pi}_{\boldsymbol{\theta}l}\hat{\Pi}_{\boldsymbol{\theta}l}^\dagger\right)\hat{\rho}_{\boldsymbol{\theta}}\right\rbrace\right]=0, \qquad \forall \, j, k.
\end{equation}
This is obtained from the saturability condition corresponding to $\hat{\rho}^{(S+B)}_{\boldsymbol{\theta}}$ in the $S+B$ space, by tracing out the bath $B$. And this is equivalent to the saturability condition (\ref{eq:supp_saturate_cond1}).

Next, (\ref{eq:supp_saturate_cond5}) saturates an ALD-based QCRB, corresponding to:
\begin{equation}\label{eq:supp_partial_rho4}
\partial_{\theta_k}\hat{\rho}_{\boldsymbol{\theta}}=\frac{1}{2}\left[\hat{Q}_k\hat{\rho}_{\boldsymbol{\theta}}+\hat{\rho}_{\boldsymbol{\theta}}\hat{Q}_k^\dagger\right]=\sum_l\left[\partial_{\theta_k}\hat{\Pi}_{\boldsymbol{\theta}l}\hat{\Pi}_{\boldsymbol{\theta}l}^\dagger\hat{\rho}_{\boldsymbol{\theta}}+\hat{\rho}_{\boldsymbol{\theta}}\hat{\Pi}_{\boldsymbol{\theta}l}\partial_{\theta_k}\hat{\Pi}_{\boldsymbol{\theta}l}^\dagger\right],
\end{equation}
where the ALDs are chosen to be:
\begin{equation}
\hat{Q}_k=2\sum_l\partial_{\theta_k}\hat{\Pi}_{\boldsymbol{\theta}l}\hat{\Pi}_{\boldsymbol{\theta}l}^\dagger.
\end{equation}
Then, (\ref{eq:supp_saturate_cond2}) holds, when the following holds:
\begin{equation}\label{eq:supp_saturate_cond6}
\sum_l\left[\hat{\Pi}_{\boldsymbol{\theta}l}^\dagger\partial_{\theta_k}\hat{\Pi}_{\boldsymbol{\theta}l}\hat{\rho}+\hat{\rho}\partial_{\theta_k}\hat{\Pi}_{\boldsymbol{\theta}l}^\dagger\hat{\Pi}_{\boldsymbol{\theta}l}\right]=\sum_l\left[\partial_{\theta_k}\hat{\Pi}_{\boldsymbol{\theta}l}\hat{\Pi}_{\boldsymbol{\theta}l}^\dagger\hat{\rho}_{\boldsymbol{\theta}}+\hat{\rho}_{\boldsymbol{\theta}}\hat{\Pi}_{\boldsymbol{\theta}l}\partial_{\theta_k}\hat{\Pi}_{\boldsymbol{\theta}l}^\dagger\right],
\end{equation}
that follows from (\ref{eq:supp_partial_rho3}) and (\ref{eq:supp_partial_rho4}).

Now, let us consider both $\hat{\rho}$ and $\hat{\rho}_{\boldsymbol{\theta}}$ to be maximally mixed. Then, (\ref{eq:supp_saturate_cond6}) becomes:
\begin{equation}\label{eq:supp_saturate_cond7}
\begin{split}
\sum_l\left[\hat{\Pi}_{\boldsymbol{\theta}l}^\dagger\partial_{\theta_k}\hat{\Pi}_{\boldsymbol{\theta}l}+\partial_{\theta_k}\hat{\Pi}_{\boldsymbol{\theta}l}^\dagger\hat{\Pi}_{\boldsymbol{\theta}l}\right]&=\sum_l\left[\partial_{\theta_k}\hat{\Pi}_{\boldsymbol{\theta}l}\hat{\Pi}_{\boldsymbol{\theta}l}^\dagger+\hat{\Pi}_{\boldsymbol{\theta}l}\partial_{\theta_k}\hat{\Pi}_{\boldsymbol{\theta}l}^\dagger\right]\\
\Rightarrow\sum_l\partial_{\theta_k}\left[\hat{\Pi}_{\boldsymbol{\theta}l}^\dagger\hat{\Pi}_{\boldsymbol{\theta}l}\right]&=\sum_l\partial_{\theta_k}\left[\hat{\Pi}_{\boldsymbol{\theta}l}\hat{\Pi}_{\boldsymbol{\theta}l}^\dagger\right]\\
\Rightarrow\mathbb{1}&=\sum_l\hat{\Pi}_{\boldsymbol{\theta}l}\hat{\Pi}_{\boldsymbol{\theta}l}^\dagger \qquad \because \, \sum_l\hat{\Pi}_{\boldsymbol{\theta}l}^\dagger\hat{\Pi}_{\boldsymbol{\theta}l}=\mathbb{1},
\end{split}
\end{equation}
where the last line follows from the previous line without an additional constant, since we must also have:
\begin{equation}
\hat{\rho}_{\boldsymbol{\theta}}=\sum_l\hat{\Pi}_{\boldsymbol{\theta}l}\hat{\rho}\hat{\Pi}_{\boldsymbol{\theta}l}^\dagger
\Rightarrow\mathbb{1}=\sum_l\hat{\Pi}_{\boldsymbol{\theta}l}\hat{\Pi}_{\boldsymbol{\theta}l}^\dagger,
\end{equation}
when both $\hat{\rho}$ and $\hat{\rho}_{\boldsymbol{\theta}}$ are maximally mixed. Indeed, both the initial and evolved probe states can be maximally mixed, only if the noisy channel is unital. Thus, the channel indeed needs to be unital for (\ref{eq:supp_saturate_cond6}), and therefore, (\ref{eq:supp_saturate_cond3}) to hold.

\end{document}